\documentclass[a4paper, 11pt]{report}

\usepackage{latexsym}
\usepackage{graphicx}
\usepackage{epsfig}
\newcommand \be {\begin{equation}}
\newcommand \bea {\begin{eqnarray}}
\newcommand \ee {\end{equation}}
\newcommand \eea {\end{eqnarray}}

\newcommand{\bit}{\begin{itemize}}
\newcommand{\eit}{\end{itemize}}
\newcommand{\Z}{\mathbf{Z}}

\newcommand{\C}{\mathbf{C}}
\newcommand{\1}{\mathbf{1}}
\newcommand{\eps}{\epsilon}
\newcommand{\beps}{\bar\epsilon}
\newcommand{\bpsi}{\bar\psi}
\newcommand{\blambda}{\bar\lambda}
\newcommand{\bsigma}{\bar\sigma}
\newcommand{\dalpha}{\dot\alpha}
\newcommand{\dbeta}{\dot\beta}

\newcommand{\lra}{\leftrightarrow}
\newcommand{\la}{\leftarrow}
\newcommand{\ra}{\rightarrow}
\newcommand \dsl {\not\!\partial}

\begin{document}
\vskip 2cm
\thispagestyle{empty}
\begin{center}
{\huge \bf Supersymmetric Noether Currents and Seiberg-Witten Theory}
\vskip .5cm
by
\vskip .5cm
{\Large \bf Alfredo Iorio}
\vskip 1.5cm
A thesis submitted to the School of Mathematics, Trinity College,
University
of Dublin, for the degree of
\vskip1cm
{\Large Doctor of Philosophy}
\vskip3cm
\begin{center}
\begin{figure}[h]
\centering
\epsfig{file = 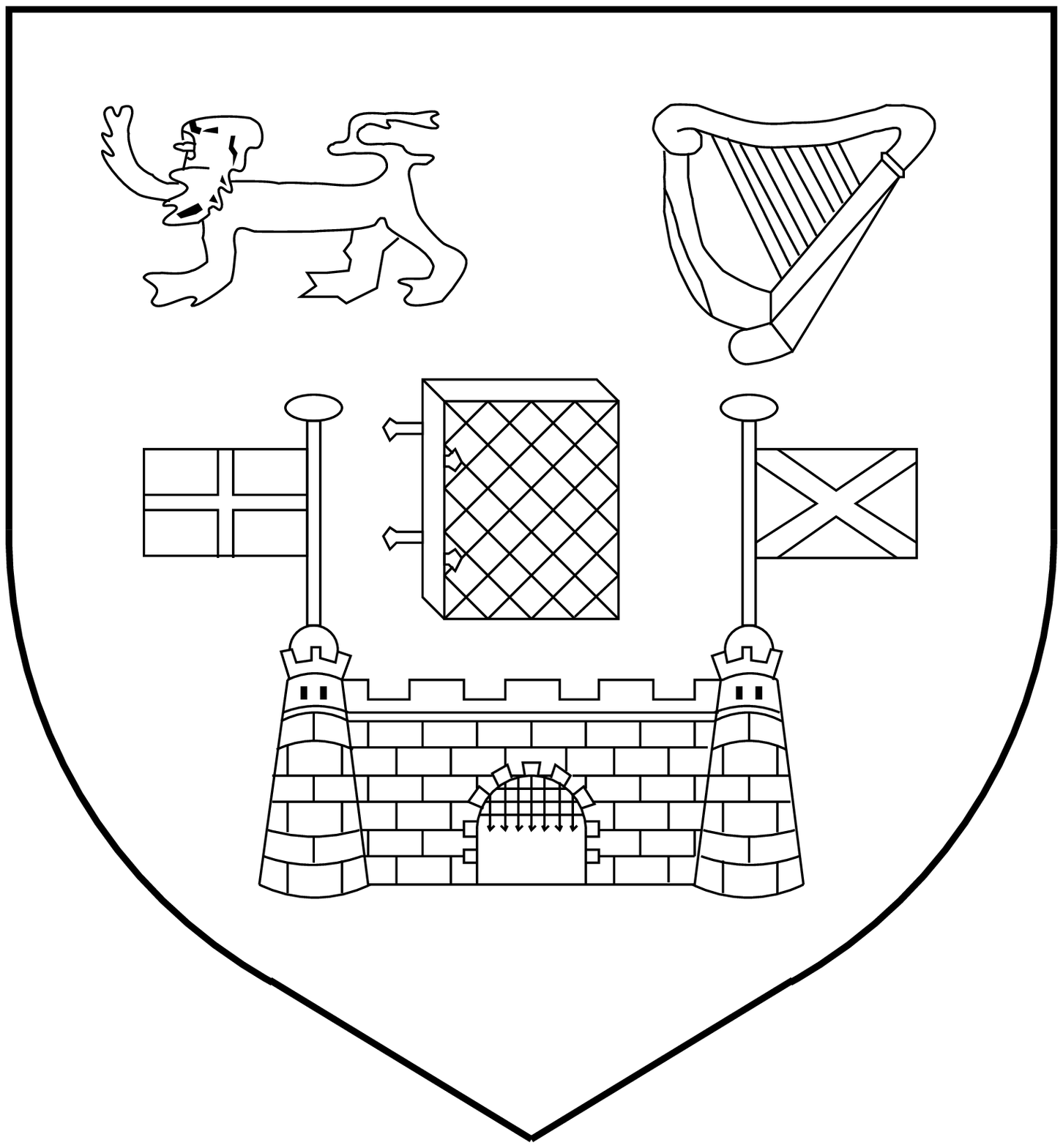 , width = 50mm}
\end{figure}
\end{center} 
\noindent
\vfill December 1999
\end{center}

\newpage

\phantom{Declaration}
\vskip2cm
\begin{center}
{\Large \bf Declaration}  
\end{center}
\vskip1cm
This thesis is entirely my own work. No part of it has been previously
submitted as an exercise for any degree at any university. This thesis is
based on one paper in process of publication that is fully integrated into the 
body of the thesis.

\vskip1cm       
\phantom{EMP}\hskip9.22cm \hrulefill

\hfill {\bf Alfredo Iorio}

\newpage

\phantom{Acknowledgments}
\vskip2cm
\begin{center}
{\Large \bf Acknowledgments}
\end{center}
\vskip1cm

\noindent
I am forever indebted to my supervisor Lochlainn O'Raifeartaigh for his 
patient, enthusiastic and continuous support during my work. 

\noindent
I thank Siddhartha Sen for his advice and for many scientific 
discussions, and Peppino Vitiello for giving me plenty of opportunities to 
begin and carry on this long journey.
A special thank goes to the late Bob Mills, whose joyful lectures I had 
the privilege to attend.

\noindent
I acknowledge the financial support of the Dublin Institute for Advanced 
Studies, the University of Salerno, the Istituto Italiano per gli Studi 
Filosofici-Naples, the FORBAIRT, the Dublin Corporation, and, finally, of 
Alice and John Lee.

\noindent
I should mention too big a number of people to whom I am grateful for  
different reasons, both scientific and personal. The partial list 
of them is: [Trinity] Rossana Caroni, Stephanie Lacolley, Leticia 
Merin, Alison Musgrave, Beatrice Paladini,
David Adams, Alan Bates (Topol), Colm Connaughton, James Harris,
Conall Kennedy, Oliver Mason, Tom Philbin,  Carlos Pinto, Emil Prodanov,
Alex Shimabukuro (Buendia), Fabian Sievers, Samik Sen; 
[DIAS] Margaret Matthews, Brian Dolan, Jan Pawloski, Ivo Sachs,  Michael Tuite,
Shreedar Vinnakota, Paul Watts, Sylvain Wolf;
[Salerno] Eleonora Alfinito, Massimo Blasone, Fabrizio Bobba, 
Michele De Marco, Luciano Di Matteo, Gaetano Lambiase.

\noindent
To Marco Nicastro goes a special mention for his unbelievable patience in 
listening to my philosophical nonsense. Another special mention goes to David 
Malone for ``building up'' and lending to me an everlasting PC!

\noindent
Thanks for their friendship to  Anna Rita Nicastro, Daniela Prodanova,
Salvatore Fabrizio, Christoph Helmig, Alfredo Iorio (senior), Rocco La Corte, 
Francesco Monticelli, Giovanni Persano, Ivan Rancati, Domenico Recinto, Mauro 
Rizzi and a special thank to Mita Petitti and Marie Anne Tissot.

\noindent
Finally, my deepest gratitude goes to Giuliana, for being my sun and my moon,
and to my sister Giovanna, to whom I can only say: without you I could never
make it.

\newpage 

\topskip 2cm

\rightline{\Large {\it A mio padre e mia madre}}

\smallskip

\rightline{\Large {\it To my father and my mother}}

\newpage

\tableofcontents

\newpage

\topskip 2cm

\chapter*{General Introduction}
\addcontentsline{toc}{chapter}{General Introduction}

\begin{quote}
``{\it Credo di essere semplicemente un uomo medio, 

che ha le sue convinzioni profonde, 

e che non le baratta  per niente al mondo}''. 

A. Gramsci, Lettera dal carcere del 12.XI.1927
\end{quote}

\noindent
The impact of Noether theorem \cite{emmy} in physics could be the 
subject of more than one thesis of philosophy of science. The motivations 
behind it are at the core of the contemporary approach to theoretical 
physics based on various versions of the symmetry principle. We suggest here
to the reader some historical and philosophical references \cite{weyl} for 
those who like these topics as well as {\it hard core} physics.

\noindent
This thesis has been devoted to the construction of the Noether 
supercharges for the Seiberg-Witten (SW) model \cite{sw}. One of our most 
important results is the first complete and direct derivation of the SW 
version of the mass formula \cite{io}. 

\noindent
The astonishing results obtained by Seiberg and Witten in their seminal 
papers are commented in a variety of review papers since their work was 
published in 1994.
Their most exciting achievement is the exact solution of a 
{\it quasi}-realistic quantum Yang-Mills model in four dimensions which leads 
to the explanation, within the model, of the confinement of electric charge 
along the lines of the long suspected {\it dual} Meissner effect. 
The spin-off's are various and in a wide range of related fields, among others 
surely there is Supersymmetry (Susy) itself, their model being strongly
based on the very special features of N=2 Susy. 

\noindent
Although Susy has been extensively developed this is not 
the case for Susy Noether currents and charges and this is regrettable
because many aspects of Susy theories could be clarified by the currents. 
A case in point is the SW model where the occurrence of a 
non-trivial central charge $Z$ is vital. In a nutshell the important features 
of $Z$ are:
\begin{itemize}
\item{It allows for SSB of the gauge symmetry within the supersymmetric 
theory.}
\item{It produces the complete and exact mass spectrum given 
by\footnote{$n_e$ and $n_m$ are the electric and magnetic charges 
respectively and $a$ and $a_D$ are the v.e.v.'s of the scalar field and its 
dual surviving the Higgs phenomenon in the spontaneously broken phase 
SU(2) $\rightarrow$ U(1).} $M = |Z| = \sqrt2 |n_e a + n_m a_D|$.}
\item{It exhibits an explicit SL(2, $\Z$) duality symmetry whereas this 
symmetry is not a symmetry of the theory in the Noether sense.}
\item{It is the most important global piece of information at our disposal, 
therefore it is vital for the exact solution of the model.}
\end{itemize}

\noindent
Susy Noether currents present quite serious difficulties due to the following 
reasons. First Susy is a space-time symmetry therefore the standard procedure 
to find Noether currents does not give a unique answer. A term,
often called {\it improvement}, has to be added to the term one would obtain 
for an internal symmetry. 
The additional term is not unique, it can be fixed only by requiring the
charge to produce the Susy transformations one starts with, and  for non 
trivial theories it is by no means easy to compute.
Second the linear realization of Susy involves Lagrange multipliers
called dummy-fields, which of course have no canonical conjugate momenta.
On the other hand, if dummy-fields are eliminated to produce a standard
Lagrangian, then the variations of the fields are no longer linear and 
the Noether currents are no longer bilinear.
A further problem is that the variations of the fields involve 
space-time derivatives and this happens in a {\it fermi-boson asymmetric} 
way (the variations of the fermions involve derivatives of 
the scalars but not conversely). This implies some double-counting 
solved only by a correct choice of the current.
Besides these problems we also had to deal with an effective Lagrangian.
In this case the Lagrangians are not constrained by renormalizability 
requirements, as it is the case for SW effective Lagrangian.
For that theory we deal with terms quartic in the fermions and coefficients
of the kinetic terms non-polynomial functions of the scalar field. Because
of this, the Noether procedure requires a great deal of care.

\noindent
We have solved all those problems by implementing a canonical formalism 
in the different cases under consideration. Firstly we construct the Noether
currents for the classical limit of the U(1) sector of the theory. 
In this case Susy is linearly realized 
regardless of the dummy fields, no complications arising in the effective case
are present and the fields are non-interacting. When the procedure is clearly
stated in this case we move to the next level, the effective U(1) sector
and we see what is left from the classical case and what is new. Now the 
currents are very different and, for instance, we cannot use a
formula one finds classically to overcome the above mentioned 
fermi-bose asymmetry in the transformations of the fields.
Nevertheless the constraints imposed by Susy are 
strong enough to force the effective centre to an identical {\it form} as 
the classical one\footnote{Of course this does {\it not} mean that 
quantum corrections are not present, as is expected to be the case for N=4,
but only that having a {\it dictionary} we could replace classical quantities
by their quantum correspondents with no other changes.} proving the SW
conjecture that $Z = n_e a + n_m a_D$. The last step 
is to consider the SU(2) sector. There we find that the canonical 
procedure implemented in the U(1) sector does not need any further 
change and our analysis confirms that U(1) is the only sector that 
contributes to the centre.

\noindent
Naturally the future work will be the generalization of our 
results to any Susy theory, possibly to obtain a {\it Susy-Noether Theorem}.
The task is by no means easy due to the above mentioned problems
 and other difficulties. For instance, as well known, for 
ordinary space-time symmetries the energy momentum tensor $T_{\mu \nu}$ 
can be obtained by embedding the theory in a curved space-time with metric 
$g_{\mu \nu}$, defining $T_{\mu \nu} = \frac{\delta S}{\delta g_{\mu \nu}}$
and then taking the flat-space limit. In Susy the situation is much more 
complicated because the embedding has to be in a curved {\it super}space 
which only has a quasi-metrical structure.

\noindent
One may also want to investigate the (non-holomorphic) 
next-to-leading order term in the superfield expansion of the SW effective 
Action. The presence of derivatives higher than second spoils the canonical 
approach and Noether procedure cannot be trivially applied. The interest here 
is to understand how the lack of canonicity and holomorphy (a crucial 
ingredient for the solution of SW model) affects the currents and charges, 
and therefore the whole theory itself. Of course this analysis is somehow more 
general and it could help to understand how to handle the symmetries of full 
effective Actions.

\newpage

\chapter{Noether Theorem and Susy}\label{c1}

\noindent
In this Chapter we want to review the difficulties of the Noether standard
procedure in relation to Susy. In the first two Sections we shall state 
Noether theorem and we shall discuss in particular its application to 
space-time symmetries and Susy. The aim is to clarify some of the points we 
found either uncovered or obscure or even wrong in literature. We show a 
recipe we have found to deal with Susy Noether charges, also in the context 
of an effective field theory. The Chapter ends with the application of this 
recipe to a supersymmetric toy model, namely the massive Wess-Zumino model.

\section{Noether Theorem}\label{1.1}

\noindent
Given a theory described by an Action 
${\cal A} = \int d^4 x {\cal L} (\Phi_i , \partial \Phi_i)$,
where $\Phi_i$ are fields of arbitrary spin,
we define a {\it symmetry} of the theory 
a transformation of the coordinates and/or of the fields that leaves 
$\cal A$ invariant {\it without} the use of the equations of motion for the 
fields (off-shell). 
The last requirement is crucial because any transformation leaves $\cal A$ 
invariant when the fields obey the equations of motion (on-shell). 
Noether theorem for classical fields states that 

\noindent
``{\it To any continuous symmetry of the Action corresponds a conserved 
charge''.}

\noindent
The invariance of the Action only ensures the invariance of the Lagrangian 
density up to a total divergence 
\be
\delta {\cal A} = 0 \Rightarrow \delta {\cal L} = \partial_\mu V^\mu
\ee
As we shall see $V^\mu$ plays a major role in Susy.

\noindent
There are different ways to prove this theorem\footnote{We leave to Appendix 
\ref{mercaldo} a detailed discussion of one proof.},
the simplest one is obtained in Quantum Mechanics in Hamiltonian formalism 
\cite{lor1}. It consists in the observation that $[H , Q] = 0$, where 
$H$ is the Hamiltonian and $Q$ the charge, tells us at once that 
time-conserved charges are symmetries of the theory! The classical derivation 
of the theorem is based on Lagrangian rather than Hamiltonian formulation. 
For instance, one could prove the theorem by comparing the variation off-shell
to the variation on-shell of the Lagrangian density $\cal L$. 
On the one hand, by the above given definition of symmetry, one has that 
off-shell 
\be\label{+}
\delta {\cal L} = \partial^\mu V_\mu 
\ee
On the other hand the same transformation (and any other transformation) 
on-shell gives
\be
\delta {\cal L} = \partial^\mu N_\mu
\ee
where 
\be
N_\mu \equiv \Pi^i_\mu \delta \Phi_i \quad{\rm and}\quad
\Pi^i_\mu = \frac{\partial {\cal L}}{\partial \partial^\mu \Phi_i} 
\ee 

\noindent
Therefore one can write a current given by
\be\label{N-V} 
J_\mu = N_\mu - V_\mu
\ee
that obeys
\be
\partial^\mu J_\mu = ({\rm E.L.})_i \delta \Phi^i
\ee
where $({\rm E.L.})_i$ stands for the Euler-Lagrange constraint for the field
$\Phi^i$, given by 
$\partial_\mu \Pi_i^\mu - \frac{\partial \cal L}{\partial \Phi^i}$.
Thus the Noether  current is conserved on-shell.

\noindent 
We shall call $N_\mu$ the {\it rigid} current
as this is the only contribution to the Noether  current when 
rigid internal symmetries are concerned\footnote{It is a well known fact that 
{\it local} gauge symmetries only fix the form of the interaction but do not
introduce new charges. For instance, in Quantum Electrodynamics we have 
${\cal L}_{QED} = -\frac{1}{4} v_{\mu \nu} v^{\mu \nu} 
+ i \bpsi \gamma_\mu (\partial^\mu - e v^\mu) \psi$,  which is 
U(1) locally invariant. This means that $\delta v_\mu = \partial_\mu \theta$
and $\delta \psi = i \theta \psi$, where $\theta$ is the $x$ dependent gauge 
parameter. From Noether theorem it follows that on-shell
$J^{\theta \nu} = \partial_\mu (v^{\mu \nu} \theta)$, therefore 
$Q^\theta \equiv \int d^3 x \vec{\nabla} \cdot (\vec{E} \theta)
\to 0$ for $\theta \to 0$ at infinity. Therefore the only conserved charge of
this theory is $e \int d^3 x \bpsi \gamma^0 \psi$}.

\noindent  
The other part $V_\mu$ is never zero for space-time symmetries and in general 
is not unique. As a matter of fact it is obtained from (\ref{+}) thus 
an improvement term $\partial_\nu W^{[\mu \nu]}$ could be 
added to it with no effects on the theorem.
For ordinary space-time transformations it could be written 
as\footnote{See Appendix \ref{mercaldo}} \cite{lop}
\be\label{ldx}
V_\mu =  - {\cal L} \delta x_\mu
\ee
For instance, if one considers the translation symmetry of a scalar field 
theory, for which $\delta_\mu \phi = \partial_\mu \phi$ and 
$\delta_\mu x_\nu = \eta_{\mu \nu}$,
the Noether  current is the {\it canonical} energy-momentum tensor given by
\be\label{t}
T_{\mu \nu} = \Pi_\mu \partial_\nu \phi - \eta_{\mu \nu} {\cal L} 
\ee
The reader could wonder about the sign of $V_\mu$ entering the canonical 
expression (\ref{t}). The point is that one may also obtain the 
energy-momentum tensor $T_{\mu \nu}$ in a slightly different way, namely 
by not explicitly making use of the $V_\mu$. This is done by considering 
$x$-dependent and $x$-independent variations of the fields, and identifying 
the $T_{\mu \nu}$ as the coefficient of $\delta x_\nu$. This is explained in 
some details in Appendix \ref{mercaldo} (see in particular Eq.(\ref{epsx})).

\noindent
A different way to produce the $T_{\mu \nu}$ is by embedding the Action
$\cal A$ in a curved space-time, computing its derivative with respect
to the metric tensor 
\be\label{ti}
T_{\mu \nu} = \frac{1}{\sqrt g} \frac{\delta {\cal A}}{\delta g^{\mu \nu}}
\ee
and taking the flat space-time limit. This gives the {\it symmetric} 
(Belinfante) energy-momentum tensor but for instance this 
$T_{\mu \nu}$ is not improved to give $T_\mu^\mu = 0$, as required 
by the scale symmetry. Equation (\ref{ti}) may also give the improved
energy-momentum tensor provided that a suitable extra coupling of 
the fields to the Ricci scalar is introduced \cite{lor3}.

\noindent
Even if we do not require any improvement there is another problem with 
space-time symmetries, namely how to express $V_\mu$ in terms of 
canonical momenta $\Pi^i_\mu$ and transformations $\delta \Phi_i$ \cite{wei}.

\noindent
From the previous discussion it is clear that many difficulties arise
in the computation of the currents when space-time symmetries are involved.
Susy is a very special case of space-time symmetry and we shall see in the
next Section that extra complications appear.

\section{Susy-Noether Theorem}\label{1.2}

\noindent
We follow the Weyl notation and the conventions of Wess and Bagger \cite{wb},
explained in some details in Appendix \ref{not}. For what follows let us 
introduce the {\it $N$ extended} Susy algebra, given by
\bea
[ Q^L_\alpha \; , \; \bar{Q}_{M \dalpha} ]_{+} &=& 
2 \delta^L_M \sigma^\mu_{\alpha \dalpha} P_\mu \label{susy1} \\
{[} Q^L_\alpha \; , \; Q^M_\beta {]}_{+} &=&
\eps_{\alpha \beta} Z^{[LM]} \label{susy2} \\
{[} \bar{Q}_{L \dalpha} \; , \; \bar{Q}_{M \dbeta} {]}_{+} &=&
\eps_{\dalpha \dbeta} Z^*_{[LM]} \label{susy3}
\eea
where $[,]_{+}$ is the anticommutator, $L,M = 1, ..., N$, 
$\alpha , \dalpha = 1,2$, the $Q_\alpha$'s are the supersymmetry charges, 
$P_\mu$ is the four-momentum and $Z^{[LM]}$ are central terms.

\noindent
It is beyond the scope of this thesis to discuss Susy in all details. 
A partial list of references on Susy is \cite{susy}, \cite{lik}, \cite{shif},
\cite{muller}, \cite{west}, \cite{1001}. We shall explain some of its nice 
features in the following Chapters. In particular Chapter 2 is a pedagogical 
introduction to some of the more advanced applications. What we want to say 
here is that Susy is the only known way to non trivially combine space-time 
(Poincar\`e) and internal symmetries of the $S$ matrix, according to the 
Haag-Lopuszanski-Sohnius generalization \cite{hls} of the Coleman-Mandula 
theorem \cite{cm}.

\noindent
The algebra (\ref{susy1})-(\ref{susy3}) is only the part of the full 
Susy algebra we shall be interested in.
Together with the ordinary Poincar\`e algebra it is referred to as
the Super-Poincar\`e (SP) algebra\footnote{The rest of the algebra contains 
the internal symmetry algebra and the non trivial commutations between the 
$Q_\alpha$'s and the internal symmetry generators.}.
From (\ref{susy1})-(\ref{susy3}) it is evident that Susy is a (special kind!) 
of space-time symmetry. This can be seen for instance by looking at the r.h.s.
of Eq. (\ref{susy1}) where we find $P_\mu$, the generator of 
translations\footnote{The space-time nature of Susy becomes more evident 
in superspace language. Let us consider N=1 for simplicity.
The generic group element of the SP group is given by \cite{ivo}  
\be
g = \exp\{ a_\mu P^\mu + \eps^\alpha Q_\alpha 
+ \beps_{\dalpha} \bar{Q}^{\dalpha}\} 
\exp \{ \omega_{\mu \nu} L^{\mu \nu} \} \nonumber 
\ee
where $L^{\mu \nu}$ are the Lorentz generators, then the coset space of 
SP/Lorentz is parameterized by the supercoordinates
$z^A \equiv (x^\mu , \theta^\alpha , \bar{\theta}^{\dalpha})$ 
corresponding to the group element 
\be
\exp\{ x_\mu P^\mu + \theta^\alpha Q_\alpha 
+ \bar{\theta}_{\dalpha} \bar{Q}^{\dalpha}\} \nonumber 
\ee 
thus the left action by a $g_o \in {\rm SP}$ is equivalent to a 
transformation of the supercoodinates given by
\bea
x^\mu &\to& x^\mu 
+ i \eps_o \sigma^\mu \bar{\theta} - i \theta \sigma^\mu \beps_o 
+ \omega_o^{\mu \nu} x_\nu \nonumber \\
\theta^\alpha &\to& \theta^\alpha + \eps_o^\alpha 
+ \frac{1}{4} (\sigma_{\mu \nu} \theta)^\alpha \omega_o^{\mu \nu} \nonumber \\
\bar{\theta}^{\dalpha} &\to& \bar{\theta}^{\dalpha} + \beps_o^{\dalpha} 
+ \frac{1}{4} (\bar\theta \bsigma_{\mu \nu})^{\dalpha} \omega_o^{\mu \nu} 
\nonumber
\eea}. 
Therefore Susy currents share all the difficulties mentioned in the 
previous Section with respect to space-time currents. The situation is 
even more complicated now due to the special nature of this symmetry. 
Following the same approach described in the last Section, for ordinary 
space-time symmetries, we shall work with {\it rigid} Susy, namely we shall 
take the parameters $\eps_\alpha$'s to be $x$-independent. Thus, in our 
approach, being $N_\mu$ the part of the current with no ambiguities, the 
problem amounts to find a suitable $V_\mu$. Of course, one could also obtain 
the Susy currents by letting the $\eps^\alpha$'s become 
local\footnote{This is in the same spirit of what discussed earlier 
for standard local internal symmetries. But in that case no ambiguities due
to improvements arise and the current is once and for all given by the rigid 
one $N^\mu$.}
and then identifying the currents as the coefficients of the
$\partial_\mu \eps_\alpha$'s after variation of the Action and partial 
integration
\be\label{jsylv}
\delta_{\rm local} {\cal A} = \int d^4 x 
\eps^\alpha (\partial_\mu \hat{J}^\mu_\alpha) 
= \int d^4 x 
(\partial_\mu \eps^\alpha)  \hat{J}^\mu_\alpha + {\rm surface \; terms}
\ee
(see also the discussion following Eq.(\ref{t})).

\noindent
The point is that one wants to produce the right (improved) Susy-Noether 
charges $Q^L_\alpha$ that correctly generate the Susy transformations of the 
fields, and this is not straightforward. For instance the charges obtained 
from the currents $\hat{J}^\mu$ in (\ref{jsylv}) need to be improved 
\cite{syl}.

\noindent
Furthermore, although $V_\mu$ could be obtained as related to the second-last 
term in the superfield expansion \cite{1001}, \cite{lik}, this $V_\mu$ in 
general does not correspond to the one required. If one also demands  
the supercurrent to enter a supermultiplet with the $R$-current and 
$T_{\mu \nu}$ we have a $V_\mu$ different from the one obtained from 
superfield expansion and again we cannot produce the Susy transformations. 
This problem cannot be cured by a simple analogue of Eq.(\ref{ti}), to obtain 
an improved supercurrent from the Action embedded in a curved superspace 
because only a quasi-metrical structure is given. Note also that there is no 
simple analogue of Eq.(\ref{ldx}).

\noindent
A second point is that the linear realization of Susy involves bosonic 
Lagrange multipliers called {\it dummy fields} to which we cannot associate
a conjugate momentum and the standard Noether  procedure, based on 
such conjugates, breaks down. Of course the dummy fields can be eliminated
by using their Euler-Lagrange equations but this introduces other 
ambiguities, unsolved in literature. Namely: {\it when} to put the 
dummy fields on-shell, before or after the computation of $V_\mu$? 
Does that mean that all the fields have to be on-shell? Note that this last 
point is vital since the definition of symmetry in the first place is based 
on the variation off-shell of the Action.

\noindent
Finally, the probably best known feature of Susy is that it 
transforms fermions into bosons and {\it vice versa}. It does so by
transforming fermions into derivatives of the bosons and bosons into fermions.
Therefore the conjugate momenta of the bosons appear in the Susy 
transformations of the fermions but the contrary is not true in general. 
This makes even more difficult to express the full current in terms of 
canonical momenta and transformations.

\noindent
In a nutshell the difficulties of Susy-Noether  currents are

\begin{itemize}
\item{Susy is a {\it super}space-time symmetry;}
\item{Susy involves dummy-fields;}
\item{Susy variations involve space-time derivatives in a way not 
symmetrical with respect to fields of different spin.}
\end{itemize}

\noindent
For the application in which we shall be more interested in, the SW model 
\cite{sw}, the situation is even more complicated due to the following problem 
that closes the list of difficulties encountered in the computations 
illustrated later: 

\begin{itemize}
\item{Effective Lagrangians, even non-Susy.}
\end{itemize}

\noindent
Namely, as we shall see, in SW theory we have to deal with effective 
Lagrangians and 
renormalization does not constraint the fermionic terms to be bilinear and 
the coefficients of the kinetic terms to be constant and in general this 
is not true. As a matter of fact, the SW effective Lagrangian is quartic in 
the fermionic fields and has coefficients of the kinetic terms that are 
non-polynomial functions of the scalar field. Because of this, the Noether 
procedure requires a great deal of care\footnote{Generally speaking, the 
Noether procedure has always to be handled with care when applied to quantum 
theories. On this point see \cite{lop}.}. 
For example we shall encounter equal 
time commutations (Poisson brackets\footnote{See Appendix \ref{poisson}}) 
between fermions and bosons such as
\be
\{ \psi \; , \; \pi_\phi \} = f(\phi) \psi
\quad{\rm from}\quad \{ \pi_{\bpsi} \; , \; \pi_\phi \} = 0 
\ee
where $f(\phi)$ is a non-polynomial function of the scalar field related 
to the coefficient of the kinetic terms. This reflects the difficulty of 
treating Noether currents in a quantum context \cite{lop}.

\noindent
All these problems are solved in our analysis and we give here the recipe
we have found:

\begin{itemize}
\item{The Susy-Noether  charge that correctly reproduces the Susy 
transformations is the one obtained from $J^\mu = N^\mu - V^\mu$ where
$\delta {\cal L} = \partial_\mu V^\mu$ and $V^\mu$ has to be extracted 
as it is, i.e. no terms like $\partial_\nu W^{[\nu \mu]}$ have to be 
added.}
\item{The variation $\delta {\cal L}$ has to be performed off-shell by the
definition of symmetry. Nevertheless the dummy fields, and {\it only} them,
automatically are projected  on-shell.}
\item{The full current $J^\mu$ contains terms of the form 
$\pi_\psi \delta\psi$, that generate the fermionic transformations. The 
{\it same} term can be written as $\pi_\phi \delta \phi \; + \cdots$ therefore 
it also generates the bosonic transformations. The situation is more 
complicated for effective theories.}
\item{The canonical commutation relations are preserved also at the effective 
level, even if some of the usual assumptions, such as that at equal time all 
fermions and bosons commute, are incorrect. Noether currents at the effective 
level do not exhibit the same simple expressions as at the classical level.}
\end{itemize}

\noindent
Of course a recipe is not a final solution and lot of work has to be done to 
fully understand the issue of Susy-Nother currents or more generally 
space-time Noether  currents. Nevertheless our work surely is a guideline in 
this direction and successfully solved the problem of the SW Susy currents 
that we intended to study.

\section{Wess-Zumino model}\label{1.3}

\noindent
Before starting our journey to the analysis of SW theory, we want to apply 
the above outlined recipe to the simplest N=1 supersymmetric model where the 
dummy fields couple to dynamical fields: the Wess-Zumino massive model.

\noindent
The Lagrangian density and supersymmetric transformations of the fields
for this model are given by \cite{wb}
\be\label{wz}
{\cal L} = -\frac{i}{2} \psi \not\!\partial {\bar\psi}
-\frac{i}{2} {\bar\psi} \not\!{\bar \partial} \psi
- \partial_\mu A \partial^\mu A^\dagger + F F^\dagger
+ m A F + m A^\dagger F^\dagger - \frac{m}{2} \psi^2 
-\frac{m}{2} {\bar \psi}^2
\ee
and
\bea
\delta A = \sqrt2 \epsilon\psi & \delta A^\dagger = 
\sqrt2 \bar\epsilon \bar\psi \\
\delta \psi_\alpha  = i \sqrt2 (\sigma^\mu \beps)_\alpha \partial_\mu A 
+ \sqrt2 \eps_\alpha F & \delta \bpsi^{\dalpha} = 
i \sqrt2 (\bsigma^\mu \eps)^{\dalpha} \partial_\mu A^\dagger 
+ \sqrt2 \beps^{\dalpha} F^\dagger \\
\delta F = i \sqrt2 {\bar\epsilon}\not\!{\bar \partial} \psi &
\delta F^\dagger = i \sqrt2 \epsilon \not\!\partial {\bar\psi}
\eea
where $A$ is a complex scalar field, $\psi$ is its Susy fermionic partner in 
Weyl notation and $F$ is the complex bosonic dummy field.

\noindent
A note on partial integration in the fermionic sector of (\ref{wz}) is now
in order. We see that $\psi$ and $\bpsi$ play the double role of 
{\it fields} and {\it momenta} at the same time. It is just a 
matter of taste to choose Dirac brackets for this second class 
constrained system \cite{dir} or to partially integrate to fix a proper 
phase-space and implement the canonical Poisson brackets.

\noindent
If one chooses the canonical Poisson brackets, as we did, then it is only a  
matter of convenience when to partially integrate the fermions. In 
fact, even if $N^\mu$ and $V^\mu$ both change under
partial integration, the total current $J^\mu$ is {\it formally} invariant,
namely its expression in terms of fields and their derivatives is 
invariant but the interpretation in terms of momenta and variations of 
the fields is different. Of course both choices give the same results, 
therefore one could either start by fixing the proper phase space since
the beginning or just do it at the end.

\noindent
Let us keep (\ref{wz}) as it stands, define the following non canonical momenta
\bea
\pi^\mu_{\psi \alpha} = \frac{i}{2} (\sigma^\mu \bpsi)_\alpha &
\pi^{\mu \dalpha}_{\bpsi} = \frac{i}{2} (\bsigma^\mu \psi)^{\dalpha} 
\label{momwz}\\
\pi^\mu_A = -\partial^\mu A^\dagger & \pi^\mu_{A^\dagger} = -\partial^\mu A
\eea
and use Eq. (\ref{N-V}) to obtain the supersymmetric current $J^\mu$. 

\noindent
We use the first two ingredients of the recipe to compute $V^\mu$ by 
varying (\ref{wz}) off-shell, under the given transformations, obtaining
\bea
V^\mu &=& \delta A \pi^\mu_A + \delta A^\dagger \pi^\mu_{A^\dagger} 
        - \delta^A \psi \pi^\mu_\psi 
        - \delta^{A^\dagger} \bar\psi \pi^\mu_{\bar\psi} 
        + \delta^F \psi \pi^\mu_\psi
        + \delta^{F^\dagger} \bar\psi \pi^\mu_{\bar\psi} \nonumber \\
      && -2 \delta^{F_{\rm on}}\psi \pi^\mu_\psi
         -2 \delta^{F^\dag_{\rm on}}\bpsi \pi^\mu_{\bpsi}
\eea
where $\delta^X Y$ stands for the part of the variation of $Y$ which 
contains $X$ (for instance $\delta^F \psi$ stands for $\sqrt2 \epsilon F$) 
and $F_{\rm on}$, $F^\dag_{\rm on}$ are the dummy fields 
given by their expressions on-shell ($F = - m A^\dag$, $F^\dag = - m A$). 
Note here that we succeeded in finding an expression for $V^\mu$ in terms of 
$\pi^\mu$'s and variations of the fields. Note also that the terms involving 
$F_{\rm on}$ and  $F^\dag_{\rm on}$ were obtained without any request but 
they simply came out like that.

\noindent
Then we write the rigid current
\be
N^\mu = 
\delta A \pi^\mu_A + \delta A^\dagger \pi^\mu_{A^\dagger}
+ \delta \psi \pi^\mu_\psi + \delta \bar\psi \pi^\mu_{\bar\psi}
\ee
and the full current is given by 
\be\label{fullwz}
J^\mu = N^\mu - V^\mu 
      = 2 {\big (}\delta^{\rm on} \psi \pi_{\psi}^\mu +
	  \delta^{\rm on} \bar\psi \pi_{\bar\psi}^\mu {\big )}
\ee
therefore $J^\mu = 2 (N^\mu)_{\rm fermi}^{\rm on}$, with obvious 
notation.
In the bosonic sector $N^\mu$ completely cancels out against the 
correspondent part of $V^\mu$. In the fermionic sector
$\delta^F \psi \pi_{\psi}^\mu$  in $N^\mu$
cancels out against the term coming from 
$V^\mu$, $\delta^A \psi \pi_{\psi}^\mu$ in $N^\mu$ and in $V^\mu$ add up 
and combined with the $2 \delta^F_{\rm on} \psi \pi_{\psi}^\mu$ in $V^\mu$
gives $2 \delta_{\rm on} \psi \pi_{\psi}^\mu$ in the full current $J^\mu$. 
Similarly for $\bar\psi$. This illustrates the third difficulty.

\noindent
Therefore we conclude that:  
{\bf a} the dummy fields are on-shell automatically and,
if we keep the fermionic non canonical momenta given in (\ref{momwz}),
{\bf b} the full current is given by {\it twice} the fermionic rigid 
current $(N^\mu)_{\rm fermi}^{\rm on}$.

\noindent 
The result {\bf a} is the second ingredient of the recipe given above. We 
shall see in the highly non trivial case of the SW effective Action that this 
result still holds and it seems to be a general feature of Susy-Noether 
currents. 


\noindent 
The result {\bf b} instead is only valid for simple Lagrangians and 
it breaks down for less trivial cases. There are two reasons for that 
curious result: the fictitious double counting of the fermionic 
degrees of freedom and the third difficulty explained above. 
Nevertheless, when applicable, Eq.(\ref{fullwz}) remains a labour saving 
formula. All we have to do is to rewrite $J^\mu$ in terms of fields 
and their derivatives
\be
J^\mu = \sqrt2 (\bar\psi \bar\sigma^\mu \sigma^\nu \bar\epsilon \partial_\nu A
+ i \epsilon \sigma^\mu \bar\psi F_{\rm on} + {\rm h.c.})
\ee
then choose one partial integration 
\bea
J^\mu &=& \delta_{\rm on}\psi {\pi^\mu}^I_\psi +
\sqrt2 \psi \sigma^\mu \bar\sigma^\nu \epsilon \partial_\nu A^\dagger
+ i \sqrt2 \bar\epsilon \bar\sigma^\mu \psi F^\dagger_{\rm on} \\
{\rm or} &=& 
\sqrt2 \bar\psi \bar\sigma^\mu \sigma^\nu \bar\epsilon \partial_\nu A
+ i \sqrt2 \epsilon \sigma^\mu \bar\psi F_{\rm on} 
+ \delta_{\rm on}\bar\psi {\pi^\mu}^{II}_{\bar\psi}
\eea
where ${\pi^\mu}^I_\psi = i \sigma^\mu \bar\psi$ 
(${\pi^\mu}^{I}_{\bar\psi} = 0$) and 
${\pi^\mu}^{II}_{\bar\psi} = i {\bar\sigma}^\mu \psi$ 
(${\pi^\mu}^{II}_\psi = 0$) are the canonical
momenta obtained by (\ref{wz}) conveniently integrated by parts, 
and perform our computations using canonical Poisson brackets. To integrate
by parts in the effective SW theory a greater deal of care is needed due to 
the fact that the coefficients of the kinetic terms are functions of the 
scalar field.

\noindent
Choosing the setting $I$, for instance, what is left is to check that the 
charge
\be\label{qwz}
{\cal Q} \equiv \int d^3 x J^0 (x) = \int d^3 x {\Big (}
\delta_{\rm on}\psi \pi^I_\psi +
\sqrt2 \psi \sigma^0 \bar\sigma^\nu \eps \partial_\nu A^\dagger
+ i \sqrt2 \beps \bsigma^0 \psi F^\dagger_{\rm on} {\Big )}
\ee
correctly generates the transformations. This is a trivial task in this 
case since the current and the expression of the dummy 
fields on-shell are very simple and the transformations can be read off 
immediately from the charge (\ref{qwz}). We shall see that this is not always 
the case. It is worthwhile to notice at this point that to generate
the transformations of the scalar field $A^\dag$ one has to use 
\be
\delta^A \psi {\pi^\mu}^I_\psi = \delta A^\dag \pi_{A^\dag} 
+ \sqrt2 \bpsi \bsigma^0 \sigma^i \beps \partial_i A 
\ee
Notice also that the transformation of $\bpsi$ is obtained by acting with the 
charge on the conjugate momentum of $\psi$: 
$\{ {\cal Q} \; , \; \pi^I_\psi \}_{-}$.

\chapter{SW Theory}

\noindent
In this Chapter we want to introduce the model discovered by Seiberg and
Witten in \cite{sw}, focusing on the aspects we are more interested in. For a 
complete review we leave the reader to the excellent literature \cite{bil}, 
\cite{gom}, \cite{ket1}, \cite{lor8}, \cite{ler}, and of course to their 
beautiful seminal paper.

\noindent 
The solution of this model essentially consists in the computation of a 
complex function ${\cal F}$. This amounts to find {\it singularities} and 
{\it monodromies} and to construct the relative differential equation.
We intend to describe this strategy here, by stressing on the vital role 
of the quantum corrected mass formula, descending from the N=2 Susy.

\noindent
In the first Section we introduce the model and make clear the mathematical 
side of the problem. In the second Section we describe in greater detail the 
physics, showing how the mass formula allows for a very intuitive
interpretation of a singularity. In the third Section we introduce 
electromagnetic (e.m.) duality, again by analyzing the mass formula, and we 
show how the monodromies around the above mentioned singularities identify 
a unique ${\cal F}$. In the last Section we collect the arguments presented
and motivate the interest in the computation of the central charge of the SW 
model.

\section{Introduction} \label{2.intro}

\noindent
SW model is a N=2 supersymmetric version of a SU(2) Yang-Mills theory 
in four dimensions.

\noindent
This is the first and only example of exact solution of a non-trivial 
four dimensional quantum field theory. The task was achieved by cleverly 
combining together the following ingredients:
\begin{description}
\item{\it N=2 Susy:}  holomorphy of the effective Action, non trivial 
non-renormalization properties, central charge $Z$; 
\item{\it Spontaneous Symmetry Breaking (SSB):} the space of gauge 
inequivalent vacua in the quantum theory, ${\cal M}_q$, exhibits 
singularities defined in terms of the Higgs vacuum expectation values
(v.e.v.'s);
\item{\it E.M. Duality:} electrically charged elementary particles in the 
asymptotically free sector and magnetically charged topological excitations 
in the infrared slave sector are exchanged by means of duality.
\end{description}
The N=2 supersymmetric, SU(2) gauged, Wilsonian effective 
Action\footnote{The Wilsonian effective Action differs from the standard 
one particle irreducible effective Action when massive and massless modes are 
both present. The Wilsonian effective Action allows for the description of 
the strong coupling regimes in terms of massless (or light) modes only. 
We shall not enter into details here. For a lucid introduction see \cite{gom}.}
in N=2 superfield language is given by \cite{bil}
\be\label{a1su2}
{\cal A} = \frac{1}{4\pi} {\rm Im} \int d^4x d^2\theta d^2\tilde{\theta} 
{\cal F} (\Psi^a \Psi^a)
\ee
where $\theta$ and $\tilde{\theta}$ are the grassmanian coordinates of the 
N=2 superspace\footnote{See note on superspace in Chapter 1} and
$\Psi^a \Psi^a$, $a = 1,2,3$, is the SU(2) gauge Casimir.  
$\Psi^a$ is the N=2 superfield that combines a scalar field $A^a$, a vector 
field $v^a_\mu$, two Weyl fermions $\psi^a$ and $\lambda^a$ (and possibly 
dummy fields) into a single Susy multiplet. Thus all the fields are in the 
same representation of the gauge group SU(2) as $v^a_\mu$, i.e. the 
adjoint representation.
${\cal F}$ is a holomorphic\footnote{${\cal F}$ is not a function of 
$\bar\Psi$ and this only happens if we stop at the leading order term in the 
expansion in $p_\mu$ of the effective Action. For instance the next-to-leading
order term ${\cal H}(\Psi , \bar\Psi)$ is no longer holomorphic \cite{ket} 
\cite{weir}.} and analytic\footnote{By analytic, we mean 
that it can have branch cuts, poles etc., but no essential singularities.} 
function. 

\noindent
The point we want to make here is that the knowledge of the function $\cal F$,
sometimes called {\it prepotential}, completely determines the 
theory.

\noindent
The key idea of Seiberg and Witten is to compute ${\cal F}$ by first posing 
and then solving what mathematicians call a ``Riemann-Hilbert (RH) 
problem\footnote{In a paper published in 1900 \cite{hilbert} Hilbert presented
a list of 23 problems. The statement we are describing here appears as the 
$21^{st}$ one in the list. The RH problem seems to be very fruitful in physics.
Recently it has been applied to renormalization in Quantum Field Theory 
\cite{kreimer}.}'' \cite{anosov}, namely: {\it given as initial data 
singularities and monodromies, does there exist a Fuchsian system having 
these data}?

\noindent
A Fuchsian system is a system of differential equations in the complex domain,
given by 
\be\label{fuchs}
\frac{d f^i (z)}{d z} = A^{ij} (z) f^j(z) \quad i,j = 1 , ... , p 
\ee
where the $f^i(z)$'s are in general multi-valued complex functions and the 
matrix $A(z)$ is holomorphic in $S = \C - \{z_1, ... , z_n\}$  and 
$z_1, ... , z_n$ are poles of $A(z)$ of order at most one.
We can naturally associate a group structure to a fundamental system of 
solutions of (\ref{fuchs}), say\footnote{This corresponds to the 
simple request to have $p$ linearly independent solutions combined together 
into an invertible $p \times p$ complex matrix, say $F(z) \in GL(p, \C)$.} 
$GL(p, \C)$. We shall see that in SW theory this group turns out to be a 
subgroup of $SL(2, \Z)$, namely 
\be\label{gamma2}
\Gamma_2 \equiv {\Big \{} \gamma \in {\rm SL(2, \Z)}: 
\gamma = \1 + \left( \begin{array}{cc} l & m  \\ n & p \end{array} \right)
l,m,n,p \in \Z {\Big \}}
\ee

\noindent
If we now consider the universal covering surface\footnote{This simply 
means that we are considering all the Riemann sheets obtained by winding 
around the singularities $z_1, ... , z_n$. For instance, in the case of a 
logarithmic function of one complex variable, $\tilde{S}$ represents the 
infinite copies of the complex plane.} of $S$, say $\tilde{S}$, we can define 
maps $\delta : \tilde{S} \to S$. The {\it monodromy} representation of 
$GL(p, \C)$ is then defined as $M : \delta \to M(\delta) \in GL(p, \C)$.
More practically the monodromy constant matrices $M$ are obtained by winding 
around the singularities $z_i$'s of $A(z)$ with loops $\alpha_i$'s in 
one-to-one correspondence with the $z_i$'s. 

\noindent
Therefore the RH problem consists in finding a system of the type (\ref{fuchs})
starting from the knowledge of the singularities $z_1, ... , z_n$ and the 
monodromies around them. If at least one of the matrices 
$M(\alpha_1), ... , M(\alpha_n)$ is diagonalizable then the RH problem has a 
positive answer\cite{anosov}.

\noindent
We want to show in the following how these singularities arise in SW model, 
their physical meaning and the vital role of the central charge $Z$ of the 
underlying N=2 Susy.

\section{SSB and mass spectrum} \label{2.ssb}

\noindent
The Action (\ref{a1su2}) is obtained  in component fields in the following 
Chapters and is given by 
\bea\label{asu2}
{\cal A} &=& \frac{1}{4\pi} {\rm Im} \int d^4x 
{\Big [} {\cal F}^{ab} {\Big (} - \frac{1}{4} v^{a \mu \nu} \hat{v}^b_{\mu \nu}
- {\cal D}_\mu A^a {\cal D}^\mu A^{\dag b}
- i \psi^a \not \!\!{\cal D} \bpsi^b  
- i \lambda^a \not\!\!{\cal D} \blambda^b \nonumber \\
&& - \frac{1}{\sqrt2} \eps^{adc}  
(A^c \bpsi^b \blambda^d + A^{\dag c} \psi^b \lambda^d) 
+ \frac{1}{2} \eps^{acd}  \eps^{bfg}
A^c A^{\dag d} A^f A^{\dag g} {\Big )} {\Big ]} \\
&+& {\cal A}_{\rm quantum} \nonumber
\eea
where ${\cal F}$ is now a function of the scalar fields only, 
${\cal F}^{a_1 \cdots a_n} \equiv 
\partial^n {\cal F} / \partial A^{a_1} \cdots \partial A^{a_n}$,
$v^{a \mu \nu}$, $\hat{v}^a_{\mu \nu}$  and ${\cal D}_\mu$ are the vector 
field strength, its self-dual projection and the covariant derivative 
respectively\footnote{These quantities and our SU(2) conventions are all 
given later in greater detail.}.

\noindent
The Action (\ref{asu2}) is immediately recognized as (an effective version of)
a Georgi-Glashow type of Action. It has: self-coupled gauge fields, 
topological excitations (instantons and monopoles), gauge fields coupled to 
matter, a Yukawa potential, and a Higgs potential to spontaneously break the 
gauge symmetry.
The purely quantum term contains third and fourth derivatives of $\cal F$, 
vertices with two fermions coupled to the gauge fields and vertices with 
four fermions.
The SU(2) gauge symmetry can be spontaneously broken down to U(1) preserving 
the N=2 Susy.

\noindent
This is possible since the Higgs potential 
$
{\rm Tr} ([\vec{A}, \vec{A}^\dag])^2
$, 
where $\vec{X} = \frac{1}{2} X^a\sigma^a$ and the $\sigma^a$'s are the 
generators of SU(2)\footnote{See previous Note.},
admits {\it flat directions}, i.e. directions in the 
group that cost no energy. This is the first requirement to spontaneously 
break SU(2) down to U(1), but preserve Susy at the same time, since
the Hamiltonian of a supersymmetric theory is always bounded below. In 
particular the Higgs potential must be zero on the vacuum\cite{lor5}. By 
choosing a direction, say $<0|A^a|0> = \delta^{a 3} a$, the potential is 
indeed still zero on the vacuum  preserving Susy but spontaneously breaking 
the gauge symmetry.

\bigskip

\noindent
We now want to show that the algebraic structure of N=2 Susy indeed allows for 
a SSB of the gauge symmetry, but only for non-vanishing central charge. The 
problem is how to handle the jump in the dimension $d$ of the representation 
of Susy when the Higgs mechanism switches the masses on, but the number of 
degrees of freedom is left invariant.

\noindent
The irreducible representations of extended Susy are easily found in terms of 
suitable linear combinations of the supercharges $Q_\alpha^L$, 
$L = 1, ... , N$, to obtain creation and annihilation operators acting on a 
Clifford vacuum \cite{wb}. On general grounds one finds that the dimension of 
the representation of the Clifford algebra corresponding to {\it massless} 
states is given by 
\be
d = 2^{N}
\ee
while for the {\it massive} case this number becomes 
\be
d = 2^{2 N}
\ee
As well known, the number at the exponent is the number of the anticommuting
creation and annihilation operators mentioned above\footnote{There are 
N (2N) creation and N (2N) annihilation operators in the massless (massive) 
case.}. Thus we have a problem if we want to keep Susy in both phases, 
massless and massive.

\noindent
The way out was found in \cite{wo}.
Let us consider the algebra given in (\ref{susy1})-(\ref{susy3}) for N=2,
our case. In the rest frame we can write\cite{wb}
\bea
[ Q^L_\alpha \; , \; (Q^M_\beta)^\dag ]_{+} &=& 
2 M \; \delta^L_M \delta_\alpha^\beta   \\
{[} Q^L_\alpha \; , \; Q^M_\beta {]}_{+} &=&
\eps_{\alpha \beta} \; Z^{[LM]}  \\
{[} (Q_\alpha^L)^\dag \; , \; (Q_\beta^M)^\dag {]}_{+} &=&
\eps^{\alpha \beta} \; Z^*_{[LM]} 
\eea
where $L,M = 1,2$.

\noindent
By performing a unitary transformation on the $Q^L_\alpha$ we can 
introduce new charges $\tilde{Q}^L_\alpha = U^L_M Q^M_\alpha$ that 
obey\footnote{In this basis $Z^{[LM]}$ is mapped to $\eps^{LM} 2 |Z|$, where 
$Z = |Z| e^{i\zeta}$ and $|Z| \geq 0$.}
\bea
[ \tilde{Q}^L_\alpha \; , \; (\tilde{Q}^M_\beta)^\dag ]_{+} &=& 
2 M \; \delta^L_M \delta_\alpha^\beta   \\
{[} \tilde{Q}^L_\alpha \; , \; \tilde{Q}^M_\beta {]}_{+} &=&
2 |Z| \; \eps_{\alpha \beta} \eps^{LM} \\
{[} (\tilde{Q}_\alpha^L)^\dag \; , \; (\tilde{Q}_\beta^M)^\dag {]}_{+} &=&
2 |Z| \; \eps^{\alpha \beta} \eps_{LM} 
\eea
where $\eps^{LM} = - \eps^{ML}$, $\eps^{12}=1=-\eps_{12}$.

\noindent
We can now define the following annihilation operators
\bea
a_\alpha &=& \frac{1}{\sqrt2} 
{\big (} \tilde{Q}^1_\alpha + 
\eps_{\alpha \gamma} (\tilde{Q}_\gamma^2)^\dag {\big )} \\
b_\alpha &=& \frac{1}{\sqrt2} 
{\big (} \tilde{Q}^1_\alpha -
\eps_{\alpha \gamma} (\tilde{Q}_\gamma^2)^\dag {\big )}
\eea
and their conjugates $a_\alpha^\dag$ and $b_\alpha^\dag$, in terms of 
which we can write the algebra as
\bea
{[}a_\alpha \; , \; a_\beta {]}_{+} &=& {[}b_\alpha \; , \; b_\beta {]}_{+}
= {[}a_\alpha \; , \; b_\beta {]}_{+} = 0 \\
{[}a_\alpha \; , \; a^\dag_\beta {]}_{+} &=& \delta_{\alpha \beta} 2(M + |Z|) 
\label{aadag}\\
{[}b_\alpha \; , \; b^\dag_\beta {]}_{+} &=& \delta_{\alpha \beta} 2(M - |Z|)
\label{bbdag}
\eea
For $\alpha = \beta$ the anticommutators (\ref{aadag}) and (\ref{bbdag}) are 
never less than zero on any states. Therefore from (\ref{aadag}) follows 
$M + |Z| \geq 0$ and from (\ref{bbdag}) follows $M - |Z| \geq 0$. By 
multiplying these two inequalities together we obtain
\be\label{bog}
M \ge |Z|
\ee
Thus, for non-vanishing central charge, the saturation of this inequality,
$M = |Z|$, implies that the operators $b_\alpha$ must vanish. This reduces 
the number of creation and annihilation operators of the Clifford algebra 
from 4 to 2. Therefore the dimension of the massive representation reduces 
to the dimension of the massless one: from $2^4 = 16$ to $2^2 = 4$. We have 
a so-called {\it short} Susy multiplet.

\noindent
States that saturate (\ref{bog}) are called Bogomolnyi-Prasad-Sommerfield
(BPS) states\cite{bps}. They are the announced way out from the problem 
posed by the Higgs mechanisms: the fields in the massive phase have to belong 
to the short Susy multiplet, i.e. they have to be BPS states. 
It is now matter to give physical meaning to the central charge $Z$ arising 
from the algebra. 

\bigskip

\noindent
We shall concentrate first on the classical case.
In \cite{wo} the authors considered the classical N=2 
supersymmetric Georgi-Glashow Action with gauge group O(3) spontaneously 
broken down to U(1) and its supercharges. In Chapter 4 we shall compute the 
quantum central charge for the SU(2) Action (\ref{asu2}), for the moment let 
us just write down the classical limit of it that gives back the result 
obtained in \cite{wo}
\be\label{zi}
Z = i \sqrt2 \int d^2 \vec{\Sigma} \cdot (\vec{\Pi}^a A^a + 
\frac{1}{4\pi} \vec{B}^a A^a_D) \quad \quad a = 1, 2, 3
\ee
where $d^2 \vec{\Sigma}$ is the measure on the sphere at spatial infinity 
$S^2_\infty$, the $A^a$'s are the scalar fields, the $\vec{B}^a$'s are the 
magnetic fields, $\vec{\Pi}^a$ is the conjugate momentum of the vector field 
$\vec{v}^a$ and $A^a_D \equiv \tau A^a$ where\footnote{The $\theta$ 
contribution to the complex coupling $\tau$ was discovered by Witten in 
\cite{wit} shortly after.}
\be
\tau = \frac{\theta}{2\pi} + i \frac{4\pi}{g^2}
\ee
is the classical {\it complex} coupling constant, $g$ is the SU(2) coupling
constant and $\theta$ the CP violating vacuum angle\cite{ryder}.
In the classical case $A^a_D$ is merely a formal quantity with no precise 
physical meaning. On the contrary, in the low-energy sector of the quantum 
theory, it becomes the {\it e.m. dual} of the scalar field.

\noindent
In the unbroken phase $Z=0$, but, as well known, in the broken phase 
this theory admits 't Hooft-Polyakov monopole 
solutions \cite{tho}. In this phase the scalar fields (and the vector 
potentials) tend to their vacuum value
$A^a \sim a \frac{r^a}{r}$ ($v^{a i} \sim \eps^{i a b} \frac{r^b}{r^2}$,
$v^{a 0} = 0$), where $a \in \C$, as $r \to \infty$. This behavior gives
rise to a magnetic charge. By performing a SU(2) gauge transformation on this 
radially symmetric (``hedgehog'') solution we can align $<0|A^a|0>$ along one
direction (the Coulomb branch), say $<0|A^a|0> = \delta^{a 3} a$, and the 
't Hooft-Polyakov monopole becomes a U(1) Dirac-type monopole \cite{ryder},
\cite{gom}, \cite{ket1}.

\noindent
In this spirit we can define the electric and magnetic charges as \cite{wo}
\bea
q_e &\equiv& \frac{1}{a} \int d^2 \vec{\Sigma} \cdot \vec{\Pi}^3 A^3 \\ 
q_m &\equiv& \frac{1}{a} \int d^2 \vec{\Sigma} \cdot \frac{1}{4\pi} 
\vec{B}^3 A^3 
= \frac{1}{a_D} \int d^2 \vec{\Sigma} \cdot \frac{1}{4\pi} 
\vec{B}^3 A^3_D
\eea
where $a_D = \tau a$ and only the U(1) fields remaining massless after SSB 
appear.
These quantities are quantized, since\footnote{We say that the 
't Hooft-Polyakov magnetic charge is the winding number of the 
map $SU(2) \sim S^2 \to S^2_\infty$, that identifies the homotopy class of the 
map. By considering the maps $U(1) \sim S^1 \to S^1_\infty$, where $S^1_\infty$
is the equator of $S^2_\infty$, it is clear that a similar comment holds 
for the U(1) Dirac type magnetic charge. It turns out that the two homotopy 
groups, $\pi_2(S^2)$ and $\pi_1(S^1)$, are isomorphic to $\Z$.
For an enjoyable and pedagogical introduction to topological objects and 
their role in physics I recommend \cite{col}.} 
$q_m \in \pi_1(U(1)) \sim \pi_2(SU(2)) \sim \Z$ and $q_e$ is quantized due to 
Dirac quantization of the electric charge in presence of a magnetic 
charge\cite{gom}.

\noindent
Thus, after SSB, the central charge becomes
\be\label{zizi}
Z = i\sqrt2 (n_e a + n_m a_D)
\ee
The mass spectrum of the theory is then given by
\be\label{mass}
M = \sqrt2 |n_e a + n_m a_D|
\ee 
We shall call this formula the Montonen-Olive mass formula\footnote{
In our short-cut to write down the classical version of the mass formula,
we did not follow the chronological order of the various discoveries that 
led to it.

\noindent
First Bogomolnyi, Prasad and Sommerfield \cite{bps} showed that, for a theory
admitting monopole solutions, the formula
\be\label{mass2}
M = a (q_e^2 + q_m^2)^{1/2} 
\ee
holds classically for monopoles and dyons (topological excitations carrying 
electric and magnetic charge). Then Montonen and Olive \cite{mo} showed that 
it is true classically for all the states, elementary particles included. 
Finally Witten and Olive \cite{wo} obtained it, again classically, from 
the N=2 Susy.

\noindent
The formula (\ref{mass2}) can be written in the following form
\be\label{mass3}
M = |a g (n_e + \tau_0 n_m)| 
\ee
where $q_e \equiv g n_e$, $q_m = (- 4\pi / g) n_m$ and
$\tau_0 \equiv i 4\pi / g^2$. This is the formula we are showing here, 
provided $a g \to a$ and $\tau_0$ is {\it improved} to $\tau$.}. 
It is now crucial to notice that this formula holds for the whole spectrum
consisting of {\it elementary} particles, two $W$ bosons and two 
fermions\footnote{We work with Weyl (chiral) components $\psi^a$ and 
$\lambda^a$, whose masses are generated by the Yukawa potential in 
(\ref{asu2}).}, and {\it topological} excitations, monopoles and 
dyons. For instance the mass of the $W$ bosons and the two fermions 
can be obtained by setting $n_e = \pm 1$ and $n_m = 0$, which gives
$m_W = m_{\rm fermi} = \sqrt2 |a|$, whereas the mass of a monopole 
($n_e =  0$ and $n_m = \pm 1$) amounts to $m_{\rm mon.} = \sqrt2 |a_D|$. 
This establishes a democracy between particles and topological excitations 
that becomes more clear when e.m. duality is implemented.

\bigskip

\noindent
After this long preparation we are now in the position to introduce the most 
important tool to reduce the solution of SW model to that of a RH problem 
in complex analysis: the singularities.

\noindent
Since the Higgs v.e.v. $a \in \C$, we can think of $\C$ as the space of 
gauge inequivalent vacua, namely to $a, a'$, with $a \ne a'$, correspond two 
vacua not related by a SU(2) gauge transformation (but only by a transformation
in the little group U(1)). To be more precise we have to introduce the SU(2) 
invariant parameter
\be\label{ua2} 
u(a) = {\rm Tr} <0|\vec{A}^2|0>  = \frac{1}{2} a^2
\ee
to get rid of the ambiguity due to the discrete Weyl group of SU(2), which 
still acts as $a \to -a$ within the Cartan subalgebra. 
This is now a good coordinate on the complex manifold of gauge 
inequivalent vacua. We shall call this manifold $\cal M$, for {\it moduli} 
space.

\noindent
Eventually we can define a singularity of ${\cal M}$ as {\it a value 
of $u$ at which some of the particles of the spectrum, either elementary or 
topological, become massless}. Classically there is only one of such values, 
namely $u=0$ where the SU(2) gauge symmetry is fully restored and $\cal M$ 
looses its meaning. It is worthwhile to notice that the classical moduli 
space is merely a tool to introduce the idea of a singularity, since the 
running of the coupling is a purely quantum effect, therefore there is no 
physical reason to vary $u$ classically. Nevertheless the crucial point is to 
keep the idea of a singularity of $\cal M$ as a point where {\it some} 
particles become massless.

\bigskip

\noindent
The big step is to go to the quantum theory (\ref{asu2}) where non trivial 
renormalization leads to a non vanishing beta function.
The running of the effective coupling $g_{\rm eff} (\mu)$, where $\mu$ is the 
renormalization scale, presented in Figure 2.1, explains why the 
physics changes dramatically from the high energy regime to the low energy one.
In fact at low energies the coupling is expected to become strong and we 
cannot make reliable predictions based on perturbative analysis as at high 
energies where the coupling is weak.
The masses of the elementary fields in SU(2)/U(1) become big in the low energy 
sector and the effective theory can be described all in terms of the massless 
U(1) fields (the heavy fields can be integrated out form the effective Action).

\begin{center}
\begin{figure}
\epsfig{file = 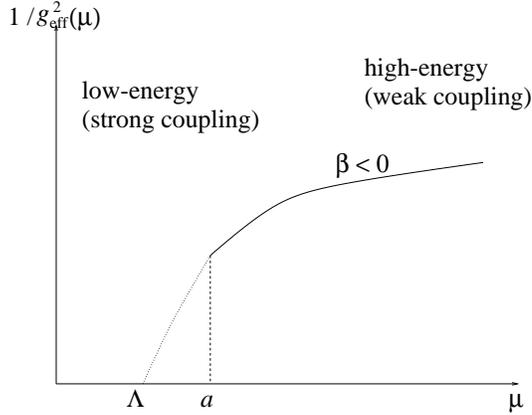 , width = 70mm}
\caption{Running of the effective coupling. $a$ is the Higgs v.e.v. and 
$\Lambda$ is the dynamically generated mass scale at which the $W$ bosons and 
two fermions are expected to become infinitely heavy.}
\end{figure}
\end{center} 
\noindent

\noindent
We can replace $\mu$ in 
$
\tau_{\rm eff} (\mu) = \frac{\theta_{\rm eff} (\mu)}{2\pi} 
+ i \frac{4\pi}{g^2_{\rm eff} (\mu)}
$ 
by the Higgs v.e.v.. Thus now is $a$ (therefore $u$) that varies and 
$\tau_{\rm eff} (a)$ becomes a field-dependent coupling as often happens in 
effective theories. Therefore in the quantum theory we can define a proper
moduli space ${\cal M}_q$.

\noindent
What happens to the mass formula (\ref{mass})? Seiberg and Witten conjectured 
that, due to the preservation of the N=2 Susy, the formula is {\it formally} 
unchanged: quantum corrections play a major role since now \cite{sw}
\be
a^{\rm class}_D = \tau_{\rm class} a \to
a^{\rm eff}_D \equiv \frac{\partial {\cal F}(a)}{\partial a}
\ee
where ${\cal F}(a)$ is the prepotential in the low energy sector, evaluated 
at $a$ (see more on this in the next Section), but no other changes are 
expected. 
Therefore the quantum improvement of the Montonen-Olive mass formula 
(\ref{mass}) is simply given by
\be\label{massq}
M = \sqrt2 |n_e a + n_m {\cal F}'(a)|
\ee
This statement is vital for the whole theory. Nevertheless no direct proof
from the N=2 Susy algebra was presented. In the following Chapters we shall
dedicate most of our attention to this point.

\noindent
The vital importance of the mass formula is immediately seen if one wants to
define the singularities of the quantum theory in the spirit outlined above.
In fact Seiberg and Witten conjectured that, in the quantum theory, 
the singularity at $u=0$ splits into $u = \pm \Lambda^2$, where monopoles and 
dyons, and {\it not} $W$ bosons and fermions (as in the classical case) 
are supposed to become massless. 
This makes sense if the $W$ bosons in the low energy sector
can decay into a monopole + dyon pair. Since all the states are BPS, one can 
show \cite{ler} that the mass formula (\ref{massq}) indeed allows for this 
decay. Thus, if some particles have to become massless in the low energy 
sector, these cannot be the $W$ bosons, whose mass is frozen at low energies, 
but only the topological excitations. 
Of course this is only a sufficient but not necessary condition for this to 
happen. Furthermore one should explain why only two singularities and why at 
$\pm \Lambda^2$ and there is no rigorous proof of these points.
 
\noindent
In Figure 2.2 we present a pictorial summary of this Section. We see from the 
picture that also a third singular point appears at $u = \infty$.
We could say that, due to the asymptotic freedom, at that point all the 
elementary particles become massless. As we shall see in the following 
Section, this point is somehow on a different footing respect to the other two.

\noindent
In the following we shall remove the suffix ``eff'' from the effective 
quantities, since their field dependence clearly identifies them.

\begin{center}
\begin{figure}
\epsfig{file = 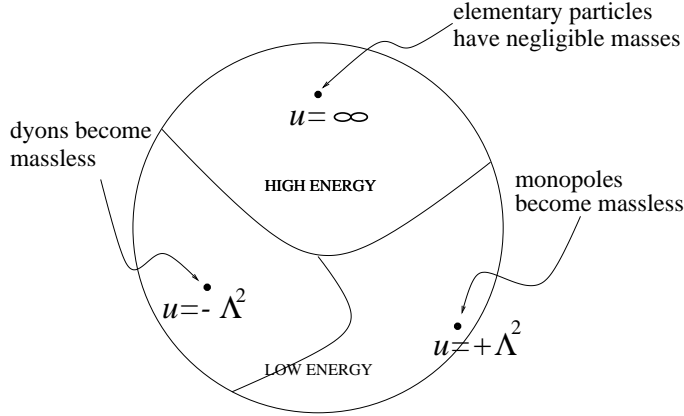 , width = 90mm}
\caption{The quantum moduli space ${\cal M}_q$. The singularities and the 
different corresponding {\it phases} are shown.}
\end{figure}
\end{center}

\section{Duality and the solution of the model} \label{2.dual}

\noindent
It is now matter to associate these singularities to the function $\cal F$
we are looking for and determine the monodromies around them.

\noindent
For large values of $a$ ($a >> \Lambda$) the theory (\ref{asu2}) is 
weakly coupled, thus a perturbative computation to evaluate $\cal F$ leads
to a reliable result. This computation was performed in \cite{sei2} and 
it turns out that 
\be\label{fu1}
{\cal F} (a) = \frac{1}{2} \tau a^2 
+ a^2 \frac{i \hbar}{2 \pi} \ln (\frac{a^2}{\Lambda^2})
+ a^2 \sum_{k=1}^{\infty} c_k \frac{\Lambda^{4k}}{a^{4k}}
\ee
where the function is parameterized by the Higgs v.e.v. $a$ (i.e. evaluated on 
the vacuum). The first two terms are the perturbative contributions: 
tree level and one loop terms respectively\footnote{For N=2 Susy these 
are the only two contributions to the perturbative $\cal F$ 
(non-renormalization) whereas for N=4 the tree level (classical) term is 
enough (super-renormalization).}, and the last term is the non-perturbative 
instanton contribution. From this expression we see that the classical limit 
consists in the substitution ${\cal F}(a) \to \frac{1}{2}\tau a^2$.

\noindent
${\cal F} (a)$ is well defined only in the region of ${\cal M}_q$ near 
$u=\infty$ since the instanton sum converges there. If we try to globally 
extend ${\cal F}(a)$ to the whole ${\cal M}_q$ this is not longer the case.
This can be seen from another perspective. If one requires the mass formula 
(\ref{massq}) to hold on the whole ${\cal M}_q$, since at $u=0$ 
there are no singularities, 
$Z|_{u=0} = i \sqrt2 (a(u) n_e + a_D(u) n_m)|_{u=0} \ne 0$. If we use the 
relation (\ref{ua2}) to write $a = \sqrt{2u}$, we expect the elementary 
particles ($n_e \ne 0$ and $n_m=0$) to become massless, but this implies 
$Z|_{u=0} = i \sqrt2 a(u)|_{u=0} = 0$, which contradicts the hypothesis. 
Therefore $u= 0 \not\!\Rightarrow a=0$ and $a$ is not a good coordinate
to evaluate $\cal F$ in the low-energy sector. We learn here that the functions
$a(u)$ and $a_D(u)$ are very different in the three sectors of ${\cal M}_q$.
All these are clear signals that we need different local descriptions in the 
weak coupling and strong coupling phases of the quantum theory.

\bigskip

\noindent
There is a peculiar symmetry, well known in physics, that exchanges weak and 
strong coupling regimes: $G \to 1 / G$, where $G$ indicates a 
generic coupling. This symmetry is called {\it duality}\footnote{This is 
referred to as $S$ duality. Shortly we shall see that in SW theory this 
duality is represented by only one of the generators of the whole duality 
group $SL(2, \Z)$, the other one corresponding to the $T$ duality \cite{sw}.} 
and is the way out of our dilemma. 
Well known examples are certain two-dimensional theories, where duality may 
exchange different phases of the same theory, as for the Ising model 
\cite{ket1}, or map solutions of a theory into solutions of a different 
theory, as for the bosonic Sine-Gordon and fermionic Thirring models 
\cite{col1}. In the latter case duality exchanges the solitonic solutions of 
the Sine-Gordon model with the elementary particles of the Thirring model.

\noindent
As explained above this is not a symmetry in the Noether sense, but rather a 
transformation that connects different phases. To see how this applies to our 
problem let us look again at the central charge $Z$
\be
Z= i \sqrt2 (n_e a + n_m a_D) = i \sqrt2 (n_m , n_e)
\left( \begin{array}{c} a_D\\ a \end{array} \right)
\ee
If we act with 
$S^{-1} \equiv \left( \begin{array}{cc} 0 & - 1  \\ 1 & 0 \end{array} \right)$
on the row vector $(n_m , n_e)$, we exchange electric charge
with magnetic charge and {\it vice-versa}. This is the e.m. duality 
transformation: it maps electrically charged elementary particles to 
magnetically charged collective excitations, giving meaning to the democracy
announced above between all the BPS states. In SW theory this is an exact 
symmetry of the low energy  Action, as well explained in \cite{bil} and it 
corresponds to the mapping
\be
\tau (a) \to - \frac{1}{\tau_D(a_D)}
\ee
where $\tau (a)$ is the effective coupling introduced in the last Section, 
and $\tau_D$ its dual. Thus by means of this transformation we map the strong 
coupling regime to the low coupling one and 
{\it vice-versa}.

\noindent
The mass of all the particles, regardless to which phase of ${\cal M}_q$
one considers, has to be given by the mass formula (\ref{massq}). Therefore to
$S^{-1}$ acting on $(n_m , n_e)$ it has to correspond $S$ acting on the 
column vector, namely
\be
\left( \begin{array}{c} a_D\\ a \end{array} \right) \to
S \left( \begin{array}{c} a_D\\ a \end{array} \right)
= \left( \begin{array}{c} a\\ -a_D \end{array} \right)
\ee
so that $Z$ is left invariant. 
Thus the $S$ duality invariance of $Z$ suggests that the good 
parameter for ${\cal F}$ (or better, its dual ${\cal F}_D$) near $u=0$ is 
$a_D$ rather than $a$. As already noted the functions $a_D(u)$ and $a(u)$ are
now different from the ones obtained near $u=\infty$, and the task is to find 
them. 
The mass formula is actually invariant under the full group $SL(2, \Z)$ of 
$2 \times 2$ unimodular matrices with integer entries, generated by
\be
S = \left( \begin{array}{cc} 0 & 1  \\ - 1 & 0 \end{array} \right)
\quad {\rm and} \quad
T = \left( \begin{array}{cc} 1 & b  \\ 0 & 1 \end{array} \right)
\ee
where $b \in \Z$.

\noindent
We now have to make this symmetry compatible with the singularities by 
considering the monodromies of $a_D(u)$ and $a(u)$ around them. This will
restrict the group of dualities to a subgroup of SL(2, $\Z$) containing 
the monodromies.

\noindent
The monodromy at $u=\infty$ can be easily computed, since here we can trust 
the perturbative expansion (\ref{fu1}) and we have $a = \sqrt{2 u}$ and 
$a_D (u) \sim i \frac{\hbar}{\pi} \sqrt{2 u} 
{\Big (} \ln (\frac{2 u}{\Lambda^2}) + 1 {\Big )}$. By winding around 
$u=\infty$, the branch point of the logarithm, we obtain
\be
\left( \begin{array}{c} a_D (u)\\ a (u) \end{array} \right) \to 
\left( \begin{array}{c} a_D (e^{i 2 \pi}u)\\ a(e^{i 2 \pi}u) \end{array} 
\right)
= \left( \begin{array}{c} -a_D(u) + 2a(u)\\ -a(u) \end{array} \right)
\ee
or
$
\left( \begin{array}{c} a_D\\ a \end{array} \right) \to
M_\infty \left( \begin{array}{c} a_D\\ a \end{array} \right)
$
where
\be
M_\infty = \left( \begin{array}{cc} -1 & 2  \\ 0 & -1 \end{array} \right)
\ee
This matrix is diagonalizable, therefore the RH problem has a positive 
solution \cite{anosov}. We are on the right track!

\noindent
To find the other two monodromies we require the state of vanishing
mass responsible for the singularity to be invariant under the 
monodromy transformation:
\be
(n_m , n_e) M_{(n_m , n_e )} = (n_m , n_e) 
\ee
This simply means that, even if SL(2, $\Z$) maps particles of one phase to 
particles of another phase, once we arrive at a singularity the monodromy 
does not change this state into another state.
From this it is easy to check that the form of the monodromies around the 
other two points $\pm \Lambda^2$ has to be
\be\label{genmon}
M_{(n_m , n_e )} = \left( \begin{array}{cc} 
1 + 2 n_m n_e & 2 n_e^2  \\ -2 n_m^2 & 1 - 2 n_m n_e \end{array} \right)
\ee
Note that $M_\infty$ is not of this form.

\noindent
The global consistency conditions on how to patch together the local data
is simply given by\footnote{In the Ising model the {\it gluing} of the 
different local data consists in the identification of a self-dual point 
$K=K^*$, where $K=J/k_B T << 1$ is the coupling at high temperature $T$ and 
$K^* >> 1$ is the coupling at low temperature given by 
$\sinh (2 K^*) = (\sinh (2 K))^{-1}$, $J$ is the strength of the interaction
between nearest neighbors and $k_B$ the Boltzman constant. This determines 
exactly the critical temperature of the phase transition $T_c$ given by
$\sinh (2 J / k_B T_c) = 1$ \cite{ket1}.} 
\be
M_{+\Lambda^2}  \cdot  M_{-\Lambda^2} = M_\infty
\ee
which follows from the fact that the loops around $\pm \Lambda^2$ can be 
smoothly pull around the Riemann sphere to give the loop at infinity. 
By using the expression (\ref{genmon}) we can obtain the solution of this 
equation given by
\be
M_{+\Lambda^2} = 
M_{(1,0)} = \left( \begin{array}{cc} 1 & 0  \\ -2 & 1 \end{array} \right)
\quad  M_{-\Lambda^2} = 
M_{(1,1)} = \left( \begin{array}{cc} -1 & 2  \\ -2 & 3 \end{array} \right)
\ee
and we see that the particles becoming massless are indeed monopoles and 
dyons as conjectured.

\noindent
The monodromy matrices generate the subgroup $\Gamma_2$ of the full duality 
group $SL(2, \Z)$ given in (\ref{gamma2}). 

\bigskip

\noindent
We have now all the ingredients and we can write down the announced Fuchsian 
equation\footnote{This is a second order differential equation therefore it 
is equivalent to a Fuchsian system (\ref{fuchs}) with $p=2$. Note also that
the poles of $A(z)$ become second order.} 
\be\label{fuchs2}
\frac{d^2 f(z)}{d z^2} = A(z) f(z)
\ee
where \cite{bil}
\be
A(z) = -\frac{1}{4} {\Big [}\frac{1- \lambda_1^2}{(z+1)^2}
+ \frac{1- \lambda_2^2}{(z-1)^2} - 
\frac{1- \lambda_1^2 - \lambda_2^2 + \lambda_3^2}{(z+1)(z-1)}{\Big ]}
\ee
$z \equiv u/\Lambda^2$ and $A(z)$ exhibits the described singularities
at $z=\pm1$ and $z=\infty$. Seiberg and Witten have found that the 
coefficients are $\lambda_1 = \lambda_2 = 1$ and $\lambda_3 = 0$, thus 
\be\label{az}
A(z) = -\frac{1}{4} \frac{1}{(z+1)(z-1)}
\ee
The two solutions of (\ref{fuchs2}) with $A(z)$ in (\ref{az}), are given in 
terms of hypergeometric functions. By using their integral representation 
one finally obtains \cite{sw}, \cite{bil}
\bea
f_1(z) \equiv a_D (z) &=& 
\frac{\sqrt2}{\pi} \int_1^z dx \frac{\sqrt{x - z}}{\sqrt{x^2 -1}} \\
&& \nonumber \\
f_2(z) \equiv a (z) &=& 
\frac{\sqrt2}{\pi} \int_{-1}^1 dx \frac{\sqrt{x - z}}{\sqrt{x^2 -1}}
\eea
We can invert the second equation to obtain $z(a)$ then substitute this into
$a_D(z)$ to obtain $a_D(a) = \partial_a {\cal F}(a)$. Integrating with respect
to $a$ yields to ${\cal F} (a)$. Thus the theory is solved!

\noindent
As noted above this expression of ${\cal F} (a)$ is not globally valid on 
${\cal M}_q$, but only near $u=\infty$. For the other two regions one has
\cite{ler} ${\cal F}_D (a_D)$ near $u=+ \Lambda^2$ and 
${\cal F}_D (a -2a_D)$ near $u=- \Lambda^2$. The unicity of this solution 
was proved in \cite{lor9}.

\bigskip

\noindent
Let us conclude this quick {\it tour de force} on SW model by saying that
this theory is surely an exciting laboratory to study the behavior of gauge 
theories at the quantum core. Nevertheless it is strongly based on N=2 
Susy and, at the present status of the experiments, Nature does not even show 
any clear evidence for N=1 Susy\footnote{There is an intense search for N=1 
superparticles in the accelerators. For instance, the next generation of 
linear colliders will run at ranges of final energy 2 TeV \cite{kal}, where 
signals of N=1 Susy are expected.}!

\section{The computation of the effective $Z$}

\noindent
As we hope is clear from the previous Sections, the mass formula 
\be
M = |Z| = \sqrt2 |n_e a + n_m a_D|
\ee
plays a major role in SW model. Let us stress here again that the knowledge
of the central charge $Z$ amounts to the knowledge of the mass formula.

\noindent
In a nutshell the important features of $Z$ are:
\begin{itemize}
\item{It allows for SSB of the gauge symmetry within the supersymmetric 
theory.}
\item{It gives the complete and exact mass spectrum. Namely it fixes the 
masses for the elementary particles as well as the collective excitations.}
\item{It exhibits an explicit SL(2, $\Z$) duality symmetry whereas this 
symmetry is not a symmetry of the theory in the Noether sense.}
\item{In the quantum theory it is the most important global piece of 
information at our disposal on ${\cal M}_q$. Therefore it is vital for the 
exact solution of the model.}
\end{itemize}

\noindent
It is then not surprising that, following the paper of Seiberg and Witten, 
there has been a big interest in the computation of the mass formula in the 
quantum case. As a matter of fact, in their paper there is no direct proof of 
this formula but only a check that the bosonic terms of the SU(2) high 
energy effective Hamiltonian for a magnetic monopole admit a BPS lower bound
given by $\sqrt2 |{\cal F}'(a)|$ \cite{sw}.

\noindent
A similar type of BPS computation, only slightly more general, has been 
performed in \cite{roc}. There the authors considered again the SU(2) 
high energy effective Hamiltonian but this time for a dyon, namely also 
the electric field contribution was considered. By Legendre transforming 
the Lagrangian given in (\ref{asu2}), one sees that the bosonic terms of the 
Hamiltonian, in the gauge ${\cal D}_0 A^a = {\cal D}_0 A^{\dag a} = 0$ and 
for vanishing Higgs potential, are given by
\be
H = \frac{1}{8\pi} {\rm Im} \int d^3x {\cal F}^{a b}
(E^a_i E^b_i + B^a_i B^b_i + 2 {\cal D}_i A^a {\cal D}_i A^{\dag b}_i)
\ee
where the electric and magnetic fields are defined as $E^a_i = v^a_{0 i}$ and 
$B^a_i = \frac{1}{2} \eps_{0ijk} v^{jk}$, respectively, $a,b=1,2,3$ are the 
SU(2) indices and $i,j,k=1,2,3$ are the spatial Minkowski indices, 
${\cal F}$ is a function of the scalar fields only. 
By using the Bogomolnyi {\it trick} to complete the square one can write this
part of the Hamiltonian as the sum of two contributions, one dynamical and one
topological: $H = H_0 + H_{\rm top}$. Explicitly we have
\bea
H_0 &=& \frac{1}{8\pi} {\rm Im} \int d^3x {\cal F}^{a b}
(B^a_i + i E^a_i + \sqrt2 {\cal D}_i A^a)
(B^a_i - i E^a_i + \sqrt2 {\cal D}_i A^{\dag a}) \nonumber \\
&& \\
H_{\rm top} &=& - \frac{\sqrt2}{8\pi} {\rm Im} \int d^3x {\cal F}^{a b}
{\Big (} (B^a_i - i E^a_i) {\cal D}_i A^a +
(B^a_i + i E^a_i) {\cal D}_i A^{\dag a}{\Big )} \label{Htop}
\eea
Of course the topological term (\ref{Htop}) is the lower bound for $H$. The 
inequality $H \ge H_{\rm top}$ is saturated when the configurations of the 
fields satisfy the BPS equations \cite{bps} (BPS configurations)
\be
B^a_i + i E^a_i + \sqrt2 {\cal D}_i A^a = 0
\ee
Note that these equations hold in the classical case with no changes.

\noindent
The authors in \cite{roc} found that 
\be
H_{\rm top} = \sqrt2 |n_e a + n_m {\cal F}'(a)|
\ee
therefore they identified the r.h.s. of this equation with the modulus 
of the central charge $|Z|$.

\noindent
This computation is rather unsatisfactory since it only considers the bosonic
contributions to $|Z|$ and, due to Susy, one has to expect fermionic terms to 
play a role. Furthermore it is too an indirect computation of the central 
charge. The complete and direct computation has to involve the Noether 
supercharges constructed from the Lagrangian (\ref{asu2}). As discussed 
earlier, Witten and Olive \cite{wo} have done that in the classical case. But 
for the effective case a direct and complete derivation is in order. We shall 
dedicate most of our attention to this point in the rest of this thesis. 

\noindent
Firstly we shall concentrate on the U(1) low energy sector of the 
theory, since the U(1) massless fields are supposed to be the only ones 
contributing to the central charge. As a warming-up we shall re-obtain the 
classical results of Witten and Olive \cite{wo}. Then we shall move to the 
U(1) effective case to compare this case with the classical one and give 
the first direct and complete derivation of the mass formula \cite{io}.

\noindent
The SU(2) high energy sector is analyzed in the last Chapter. The main 
interest there is to check the role of the massive fields in SU(2)/U(1) with 
respect to the central charge.

\noindent
The other interest, not less important, is the application of the Noether 
procedure to find effective supercurrents and charges, as explained in some 
details in the first Chapter.

\bigskip

\noindent
The kind of computation we are considering also seems to follow from a 
geometric analysis of the N=2 vector multiplet in \cite{van}, where, however, 
the authors' aim there is completely different, the fermionic contribution is 
not present and there is no mention of Noether charges. On the other hand 
an independent complete computation \cite{syl} was performed while we were 
working on the SU(2) sector. We shall present our independent results for 
the SU(2) sector referring to this computation as a double-checking of our 
formulae.


\chapter{SW U(1) Low Energy Sector}

\noindent
In this Chapter we shall construct the Noether Susy currents and charges for 
the SW U(1) low energy Action. The second Section is dedicated to the 
classical case, where we shall set up a canonical formalism, necessary for the 
implementation of the Noether procedure for constructing the currents and the 
charges, as explained in Chapter 1. In this Section the classical central 
charge of the N=2 Susy is re-obtained. The result is in agreement with 
\cite{wo}. In the third Section we shall deal with the highly non trivial 
case of the quantum corrected theory. We shall show that the canonical setting 
still survives, but many delicate issues have to be handled with care. We shall
compute the non trivial contributions to the full currents, which we 
christened $V_\mu$ in Chapter 1. Then, after having tested these results 
by obtaining the Susy transformations from the Susy charges, we shall compute
the effective central charge $Z$. This computation is the first complete
and direct proof of the correctness of SW mass formula and it appears in 
\cite{io}.

\section{Introduction}

\noindent
There exist two massless N=2 Susy multiplets with maximal helicity 1 or 
less: the vector multiplet and the scalar multiplet \cite{west}, \cite{1001}. 
We are interested in the vector multiplet $\Psi$, also referred to as the N=2 
Yang-Mills multiplet, for the moment in its Abelian formulation. Its spin 
content is $(1,\frac{1}{2},\frac{1}{2},0,0)$ and, in terms of physical fields,
it can accommodate 1 vector field $v_\mu$, 2 Weyl fermions $\psi$ and 
$\lambda$, one complex scalar $A$.
The N=2 vector multiplet can be arranged into two N=1 multiplets, the vector 
(or Yang-Mills) multiplet $W$ and the scalar multiplet $\Phi$, related by 
$R$-symmetry: $\psi \lra -\lambda$, $E^\dag \lra E$ and 
$v^\mu \to -v^\mu$ (charge conjugation). In terms of component fields the 
N=1 multiplets are given by
\be
W = (\lambda_\alpha, \; v_\mu,  \; D)
\quad {\rm and} \quad 
\Phi = (A, \; \psi_\alpha, \; E)
\ee
where $E$ and $D$ are the (bosonic) dummy 
fields\footnote{We use the same symbol $E$ for the electric field and 
for the dummy field. Its meaning will be clear from the context.}. 
Note that $W$ is a real 
multiplet and $\Phi$ is complex. This means that $v_\mu$ and $D$ are real, and
$W$ contains also $\blambda$, as can be seen by the Susy transformations given 
below. Of course the complex conjugate of $\Phi$ is given by  
$\Phi^\dag = (E^\dag, \; \bpsi_{\dalpha}, \; A^\dag)$.

\noindent
The N=2 Susy transformations of these fields are well known \cite{west},
\cite{1001}, \cite{lik}. In our notation they are given by \cite{dorey}

\centerline{\it first supersymmetry, parameter $\eps_1$}

\bea
\delta_1 A &=& {\sqrt 2} \eps_1 \psi \nonumber \\
\delta_1 \psi^\alpha &=& {\sqrt 2}\epsilon_1^\alpha E \label{trns1}\\
\delta_1 E &=& 0 \nonumber 
\eea
\bea
\delta_1 E^\dag &=& i {\sqrt 2} \epsilon_1 \not\!\partial \bar\psi \nonumber \\
\delta_1 {\bar \psi}_{\dot \alpha} &=& - i {\sqrt 2} \epsilon_1^\alpha 
               \not\!\partial_{\alpha \dot \alpha} A^\dagger \\
\delta_1 A^\dagger &=& 0 \nonumber 
\eea
\bea
\delta_1 \lambda^\alpha &=& 
- \epsilon^\beta_1 ( \sigma^{\mu \nu \: \alpha}_{\: \beta} v_{\mu \nu}
    - i \delta^\alpha_\beta D ) \nonumber \\
\delta_1 v^\mu &=& i \epsilon_1 \sigma^\mu {\bar \lambda} \quad
\delta_1 D = - \epsilon_1 \not\!\partial \bar\lambda \label{trns2} \\
\delta_1 {\bar \lambda}_{\dot \alpha} &=& 0 \nonumber 
\eea

\centerline{\it second supersymmetry, parameter $\eps_2$}

\bea
\delta_2 A &=& -{\sqrt 2} \epsilon_2 \lambda \nonumber \\
\delta_2 \lambda^\alpha &=& -{\sqrt 2}\epsilon_2^\alpha E^\dag  \label{trns3}\\
\delta_2 E^\dag &=& 0 \nonumber 
\eea
\bea
\delta_2 E &=& - i {\sqrt 2} \epsilon_2 \not\!\partial 
                     \bar\lambda \nonumber \\
\delta_2 {\bar \lambda}_{\dot \alpha} &=&  i {\sqrt 2} \epsilon_2^\alpha 
               \not\!\partial_{\alpha \dot \alpha} A^\dagger \\
\delta_2 A^\dagger &=& 0 \nonumber 
\eea
\bea
\delta_2 \psi^\alpha &=& 
- \epsilon^\beta_2 ( \sigma^{\mu \nu \: \alpha}_{\: \beta} v_{\mu \nu}
    + i \delta^\alpha_\beta D ) \nonumber \\
\delta_2 v^\mu &=& i \epsilon_2 \sigma^\mu {\bar \psi} \quad
\delta_2 D =  \epsilon_2 \not\!\partial \bar\psi \label{trns4} \\
\delta_2 {\bar \psi}_{\dot \alpha} &=& 0 \nonumber 
\eea
We note here that by $R$-symmetry we can obtain the second set of 
transformations by the first one, by simply replacing $1 \to 2$, 
$\psi \lra -\lambda$, $v^\mu \to -v^\mu$ and $E^\dag \lra E$ in the first 
set.

\noindent
The N=2 Yang-Mills low-energy effective Lagrangian, up to second derivatives 
of the fields and four fermions is given by \cite{monto}
\bea
{\cal L} &=&
\frac{\rm Im}{4\pi} {\Big (} - {\cal F}'' (A)
   [\partial_\mu A^\dag \partial^\mu A
    + \frac{1}{4} v_{\mu \nu}{\hat v}^{\mu \nu}
    + i \psi  \not\!\partial {\bar \psi} 
    + i \lambda  \not\!\partial {\bar \lambda}
    - (E E^\dag + \frac{1}{2} D^2)] \nonumber \\
&+& {\cal F}'''(A) 
    [\frac{1}{\sqrt 2} \lambda\sigma^{\mu\nu}\psi v_{\mu\nu}
      -\frac{1}{2}(E^\dag \psi^2 + E \lambda^2)
      + \frac{i}{\sqrt 2} D \psi \lambda ] 
     + {\cal F}'''' (A)[\frac{1}{4} \psi^2\lambda^2] {\Big )} \nonumber \\
&& \label{lu1eff}
\eea
where ${\cal F}(A)$ is the prepotential discussed in the last Chapter, the 
prime indicates derivative with respect to the scalar field; the fields 
appearing are the ones remaining massless after SSB, they describe the whole 
effective dynamics; 
$v_{\mu \nu} = \partial_\mu v_\nu - \partial_\nu v_\mu$
is the Abelian vector field strength, 
$v^*_{\mu \nu}= \eps_{\mu \nu \rho \sigma} v^{\rho \sigma}$ is its dual, 
$\hat{v}_{\mu \nu} = v_{\mu \nu} + \frac{i}{2} v^*_{\mu \nu}$ is its self-dual
projection and 
$\hat{v}^\dag_{\mu \nu} = v_{\mu \nu} - \frac{i}{2} v^*_{\mu \nu}$ its 
anti-self-dual projection. Note that if we define the electric and magnetic 
fields as usual, $E^i=v^{0i}$ and $B^i=\frac{1}{2} \eps^{0ijk}v_{jk}$, 
respectively, we have $\hat{v}^{0i} = E^i + i B^i$ and 
$\hat{v}^{\dag 0i} = E^i - i B^i$. Susy constraints all the fields to 
be in the same representation of the gauge group as the vector field, 
namely the adjoint representation. In the U(1) case this representation is
trivial, and the derivatives are standard rather than covariant. We notice 
here that $v_0$ plays the role of a Lagrange multiplier, and the associate 
constraint is the Gauss law. Thus, by taking the derivative of $\cal L$ with 
respect to $v_0$ we obtain the quantum modified Gauss law for this theory, 
namely
\be\label{gauss1}
0 = \frac{\partial {\cal L}}{\partial v_0} = \partial_i \Pi^i
\ee
where $\Pi^i = \partial {\cal L} / \partial(\partial_0 v_i)$ is the conjugate
momentum of $v_i$, given by
\be
\Pi^i = -({\cal I} E^i - {\cal R} B^i)
+ \frac{1}{i \sqrt2} ({\cal F}''' \lambda\sigma^{0i}\psi 
- {{\cal F}^\dag}''' \blambda \bsigma^{0i} \bpsi) 
\ee
and ${\cal F} = {\cal R} + i {\cal I}$.

\section{The classical case}

\noindent
We shall study, for the moment, the classical limit of this Lagrangian. At 
this end it is sufficient to recall that in the classical case there is no 
running of the coupling constant, therefore there is only one global 
description at any scale of the energy. Thus we can use the expression 
(\ref{fu1}) to write the classical limit as
\be\label{clim}
{\cal F}(A) \to \frac{1}{2} \tau A^2
\ee
where $\tau$ is the complex coupling constant already introduced
\be
\tau = \tau_R + i \tau_I = \frac{\theta}{2\pi} + i \frac{4\pi}{g^2}
\ee
In this limit the second line of (\ref{lu1eff}) vanishes and the first line 
becomes
\be
{\cal L} = 
\frac{\rm Im}{4 \pi} \Big{(} - \tau [\partial_\mu A^\dag \partial^\mu A
+ \frac{1}{4} v_{\mu \nu}{\hat v}^{\mu \nu}
+ i \psi  \not\!\partial {\bpsi} + i \lambda  \not\!\partial {\blambda}
-  (E E^\dag + \frac{1}{2} D^2)] \Big{)}
\ee
By using ${\rm Im} (z w) = z_I w_R + z_R w_I = \frac{1}{2i} (zw - z^*w^*)$ 
we can write explicitly this Lagrangian as
\bea
{\cal L} & = & 
- \frac{1}{g^2}  \Big{(} \frac{1}{4} v_{\mu \nu} v^{\mu \nu}
+ \partial_\mu A^\dagger \partial^\mu A
- (E E^\dag + \frac{1}{2} D^2) \Big{)}
- \frac{\theta}{64 \pi^2}v_{\mu \nu} v^{* \mu \nu} \nonumber \\
&& -\frac{1}{4\pi}  \Big{(}
\frac{\tau}{2} \psi  \not\!\partial {\bpsi}
- \frac{\tau^*}{2} {\bpsi} {\not\!\bar\partial} \psi
+ \frac{\tau}{2} \lambda  \not\!\partial {\blambda}
- \frac{\tau^*}{2} {\blambda} {\not\!\bar\partial} \lambda \Big{)} 
\label{lu1class}
\eea
The non canonical momenta from the Lagrangian (\ref{lu1class}) are given by
\bea
\pi^\mu_A &\equiv& \frac{\partial \cal L}{\partial \; \partial_\mu A} 
= - \frac{1}{g^2} \partial^\mu A^\dagger \\ 
\pi^\mu_{A^\dag} &\equiv& 
\frac{\partial \cal L}{\partial \; \partial_\mu A^\dag} 
= - \frac{1}{g^2} \partial^\mu A = (\pi^\mu_A)^\dagger \\
\Pi^{\mu \nu} &\equiv& \frac{\partial \cal L}{\partial \; \partial_\mu v_\nu} 
= - \frac{1}{g^2} v^{\mu \nu} - \frac{\theta}{16 \pi^2} v^{* \mu \nu}
= -\frac{1}{8 \pi i} (\tau \hat{v}^{\mu \nu} - \tau^* \hat{v}^{\dag \mu \nu})
\eea
and
\bea
4 \pi (\pi^\mu_{\bar \psi})_{\dot\alpha} =
 \frac{\tau}{2} \psi^\alpha \sigma^\mu_{\alpha \dot\alpha} \quad
4 \pi (\pi^\mu_{\psi})^{\alpha} =
 -\frac{\tau^*}{2} \bar\psi_{\dot\alpha} \bar\sigma^{\mu \dot\alpha \alpha} \\
4 \pi (\pi^\mu_{\bar \lambda})_{\dot \alpha} =
 \frac{\tau}{2} \lambda^\alpha \sigma^\mu_{\alpha \dot \alpha} \quad
4 \pi (\pi^\mu_{\lambda})^{\alpha} =
 -\frac{\tau^*}{2} \bar\lambda_{\dot\alpha} \bar\sigma^{\mu \dot\alpha \alpha}
\eea
where 
$\pi^\mu_{\chi} \equiv \frac{\partial \cal L}{\partial \; \partial_\mu \chi}$,
$\chi = \psi, \lambda, \bpsi, \blambda$.

\noindent
Now we want to compute the Susy $\heartsuit$ Noether currents for this theory.
In fact we have only to compute the first Susy current, since by R-symmetry,
charge conjugation and complex conjugation we can obtain the other currents. 
As explained in the first Chapter, it is matter to compute the $V_\mu$ part 
of the current. In the classical case this is an easy matter. Thus, by taking 
the variation {\it off-shell} of (\ref{lu1class}) under the first Susy 
transformations in (\ref{trns1})-(\ref{trns2}),
$
\delta_1 {\cal L} = \partial_\mu V_1^\mu
$,
we obtain 
\bea
V_1^\mu &=& \pi_A^\mu \delta_1A + \Pi^{\mu \nu}\delta_1 v_\nu \nonumber \\
 && + \frac{\tau^*}{\tau} \delta_1 {\bar \psi} \pi_{\bar\psi}^\mu
+ \delta_1 \psi \pi_\psi^\mu
+ \delta_1^D \lambda \pi_{\lambda}^\mu
+ \frac{\tau }{\tau^*} \delta_1^v \lambda \pi_{\lambda}^\mu \label{V1class}
\eea
where again $\delta^X Y$ stands for the term in the variation of $Y$ that 
contains $X$, for instance 
$\delta_1^D \lambda^\alpha \equiv i \eps_1^\alpha D$.
The total current $J_1^\mu$ is then given by\footnote{We choose to explicitly
keep the Susy parameters $\eps_L^\alpha$, $L=1,2$. This simplifies some 
computations involving spinors. Therefore $J_1^\mu$ stands for 
$\eps^1 J_1^\mu$ and also for $\eps_1 J^{1 \mu}$. In the following we shall
not keep track of the position of these indices, they will be treated as 
labels.}
\bea
J_1^\mu &=& N^\mu_1 - V_1^\mu \nonumber \\
        &=& \pi_A^\mu \delta_1A + \Pi^{\mu \nu}\delta_1 v_\nu 
          + \delta_1 \psi \pi_{\psi}^\mu 
          + \delta_1 \lambda \pi_{\lambda}^\mu
          + \delta_1 \bar\psi \pi_{\bar\psi}^\mu - V_1^\mu \nonumber \\
	&=& \frac{2i\tau_I}{\tau} \delta_1 \bar\psi \pi_{\bar\psi}^\mu
 - \frac{2i\tau_I}{\tau^*} \delta_1^{\rm on} \lambda \pi_{\lambda}^\mu 
\eea
where $N^\mu_1$ is the rigid current, $\delta_1 {\blambda} = 0$ and 
$\delta_1^{\rm on} \lambda$ stand for the variation of $\lambda$ with dummy 
fields on-shell (there are no dummy fields in the variation of $\bpsi$).
In this case this means $E=D=0$ and one could also wonder if they are simply 
canceled in the total current. But, in agreement with our recipe, we shall see
later that indeed the dummy fields, and {\it only} them, have been 
automatically projected on-shell.

\noindent
If we set $\theta = 0$ in this non-canonical setting, we recover the same type
of expression, $J^\mu = 2 N^\mu_{\rm fermi}$, for the total 
current obtained in the massive WZ model, namely
\be
J_1^\mu|_{\theta = 0} =  2 (\delta_1^{\rm on} \lambda \pi_{\lambda}^\mu 
	  + \delta_1 \bpsi \pi_{\bpsi}^\mu)
\ee
Once again we see that the double counting of the fermionic degrees of freedom 
provides a very compact formula for the currents. All the informations are 
contained in the fermionic sector, since the variations of the fermions 
contain the bosonic momenta. Unfortunately this does not seem to be the case
for the effective theory, as we shall see in the next Section.

\noindent
We have now to integrate by parts in the fermionic sector of the 
Lagrangian (\ref{lu1class}) to obtain a proper phase space. Everything 
proceeds along the same lines as for the WZ model. The fermionic sector 
becomes 
\be
{\cal L}^I_{\rm fermi} = - i \frac{1}{g^2} 
(\psi \not\!\partial {\bar \psi} + \lambda  \not\!\partial {\blambda})
\ee
where with $I$ we indicate one of the two possible choices ($\bpsi$ and 
$\blambda$ are the fields). Thus the canonical fermionic momenta are 
\be
(\pi^{I \mu}_{\bpsi})_{\dalpha} =
  \frac{i}{g^2} \psi^\alpha \sigma^\mu_{\alpha \dalpha} 
\quad
(\pi^{I \mu}_{\blambda})_{\dalpha} =
  \frac{i}{g^2} \lambda^\alpha \sigma^\mu_{\alpha \dalpha}
\ee
and $\pi^{I \mu}_\psi = \pi^{I \mu}_\lambda =0$.
In this case there is no effect of the partial integration on the bosonic
momenta since $\partial_\mu \tau = 0$, we shall see that this is not longer 
the case for the effective theory. Also, the partial integration changes 
$V^\mu$, but, of course, also $N^\mu$ changes accordingly and they still 
combine to give the same total current $J^\mu$. Namely 
\bea
N^{I \mu}_1 &=&  \pi^\mu_A \delta_1 A  + 
\Pi^{\mu \nu} \delta_1 v_\nu + \delta_1 \bpsi \pi^\mu_{\bpsi} \\
V_1^{I \mu} &=&   \pi^\mu_A \delta_1 A
+ \frac{1}{8 \pi i} \tau^* \eps_1 \sigma_\nu \blambda v^{* \mu \nu} 
\eea
and 
\bea
J_1^{I \mu} &=& N^{I \mu}_1 - V^{I \mu}_1 \nonumber \\ 
&=& \delta_1 \bpsi \pi_{\bpsi}^{I \mu} 
-\frac{i}{g^2}\eps_1 \sigma_\nu \blambda {\hat v}^{\mu \nu}   
\label{jfclass} \\
&=& -\sqrt2 \eps_1 \sigma^\nu \bsigma^\mu \pi_{\nu A}
+ \Pi^{\mu \nu} \delta_1 v_\nu
- \frac{1}{8 \pi i} \tau^* \eps_1 \sigma_\nu \blambda v^{* \mu \nu}
\label{jbclass}
\eea
where we used the identities
\bea
\delta_1 \bpsi \pi_{\bpsi}^{I \mu} &=& 
-\sqrt2 \eps_1 \sigma^\nu \bsigma^\mu \pi_{\nu A} \\
-\frac{i}{g^2}\eps_1 \sigma_\nu \blambda {\hat v}^{\mu \nu} &=&
\Pi^{\mu \nu} \delta_1 v_\nu
- \frac{1}{8 \pi i} \tau^* \eps_1 \sigma_\nu \blambda v^{* \mu \nu} 
\eea
The point we make here is that the current is once and for all given by
\be
J_1^\mu = \frac{1}{g^2}
(\sqrt2 \eps_1 \sigma^\nu \bsigma^\mu \partial_\nu A^\dag 
-i \eps_1 \sigma_\nu \blambda {\hat v}^{\mu \nu})
\ee
but its content in terms of canonical variables changes according to partial 
integration. Furthermore one has to conveniently re-express the current 
obtained via Noether procedure to obtain the expression (\ref{jfclass}) or
(\ref{jbclass}) in terms of bosonic or fermionic momenta and transformations,
respectively. Note also that $\theta$ does not appear in the explicit formula,
as could be expected.

\noindent
The next step is to choose a gauge for the vector field and define the 
conjugate momenta (remember that the metric is given by 
$\eta^{\mu \nu} = {\rm diag}(-1, 1, 1, 1)$). We shall work in the temporal 
gauge for the vector field $v^0 = 0$, the conjugate momenta are then given by
\be
\pi_A \equiv \pi^0_A  = - \frac{1}{g^2} \partial^0 A^\dagger \quad
\pi_{A^\dagger} \equiv \pi^0_{A^\dag} = - \frac{1}{g^2} \partial^0 A  
\ee
\be
\Pi^i \equiv \Pi^{0 i} = -\frac{1}{8 \pi i} 
(\tau \hat{v}^{0 i} - \tau^* \hat{v}^{\dag 0 i})
\ee
and\footnote{See Appendix \ref{poisson} on the conventions for these momenta.}
\be
\pi^I_{\bpsi} \equiv \frac{i}{g^2} \psi \sigma^0 \quad
\pi^I_{\blambda} \equiv \frac{i}{g^2} \lambda \sigma^0
\ee
With this choice the first Susy charge is given by
\bea
\eps_1 Q^I_1 &\equiv& \int d^3 x J_1^{I 0} = 
\frac{1}{g^2} \int d^3 x 
(\sqrt2 \eps_1 \sigma^\nu \bsigma^0 \psi \partial_\nu A^\dagger 
-i \eps_1 \sigma_i \bar\lambda {\hat v}^{0 i}) \label{q1class} \\
&=& \int d^3 x (\delta_1 \bpsi \pi^I_{\bpsi} - 
\frac{i}{g^2} \eps_1 \sigma_i \blambda {\hat v}^{0 i}) \label{q1fclass} \\
&=& \int d^3 x (\delta_1 A \pi_A 
+ \sqrt2 \eps_1\sigma^i\bsigma^0\psi \partial_i A^\dagger  
+ \delta_1 v_i \Pi^i
+ \frac{i \tau^*}{8 \pi} \eps_1 \sigma_i \blambda v^{* 0 i}) \nonumber \\
\label{q1bclass}        
\eea
The other charges are obtained by R-symmetry, charge conjugation and 
complex conjugation. They are given by 
\bea
\eps_2 Q^I_2 &\equiv& \int d^3 x J^{I 0}_2 = \frac{1}{g^2} \int d^3 x 
(-\sqrt2 \eps_2 \sigma^\nu \bsigma^0 \lambda \partial_\nu A^\dag
- i \eps_2 \sigma_i \bpsi {\hat v}^{0 i}) \label{q2class} \\
{\beps}_{1} {\bar Q}^I_{1} &\equiv& \int d^3 x J_1^{I \dag 0} =
\frac{1}{g^2} \int d^3 x 
(\sqrt2 \beps_1 \bsigma^\nu \sigma^0 \bpsi \partial_\nu A 
-i \beps_1 \bsigma_i \lambda {\hat v}^{\dag 0 i}) \label{bq1class}\\
{\beps}_{2} {\bar Q}^I_{2} &\equiv& \int d^3 x J_2^{I \dag 0} =
\frac{1}{g^2} \int d^3 x 
(- \sqrt2 \beps_2 \bsigma^\nu \sigma^0 \blambda \partial_\nu A 
-i \beps_2 \bsigma_i \psi {\hat v}^{\dag 0 i}) \label{bq2class}
\eea
Of course one needs to rearrange also these expressions in terms of 
conjugate momenta and fields transformations as we did in 
(\ref{q1fclass}) and (\ref{q1bclass}) for the first Susy charge.

\subsection{Transformations and Hamiltonian from the $Q_\alpha$'s}

\noindent  
We now want to test the correctness of these charges by commuting them to 
obtain the Susy transformations of the fields and the Hamiltonian. At this 
end we first introduce the basic non zero equal-time graded Poisson brackets, 
given by (see also Appendix \ref{poisson})
\bea
\{A(x), \pi_A (y) \}_{-} &=& \{A^\dagger (x), \pi^\dagger_A (y) \}_{-} =
\delta^{(3)} (\vec{x} - \vec{y}) \\ 
\{v_i (x), \Pi^j (y) \}_{-} &=& \delta_i^j \delta^{(3)} (\vec{x} - \vec{y})
\eea
and
\be
\{ \bpsi_{\dalpha} (x), \pi_{\bpsi}^{\dbeta} (y) \}_{+} = 
\{ \blambda_{\dalpha} (x), \pi_{\blambda}^{\dbeta} (y) \}_{+} =
\delta_{\dalpha}^{\dbeta} \delta^{(3)} (\vec{x} - \vec{y}) 
\ee
Due to the conventions used for the graded Poisson brackets, for the fermions 
we act with the charge from the left while for the bosons we act from the 
right. We shall call $\Delta_1$ the transformation induced by our charge 
$\eps_1 Q^I_1$ in (\ref{q1class}).
For the bosonic transformations we use the expression (\ref{q1bclass}),
whereas for the fermions we use the expression (\ref{q1fclass}). Thus we 
obtain 
\bea
\Delta_1 A (x) &\equiv& \{ A (x) \; , \; \eps_1 Q_1^I \}_{-} 
= \int d^3 y 
\{ A (x) \; , \; \sqrt2 \eps_1 \psi (y) \pi_A (y) + {\rm irr.} \}_{-}
\nonumber \\
&=& \sqrt2 \eps_1 \psi (x) = \delta_1 A (x) \\
\Delta_1 A^\dagger (x) &\equiv&
\{ A^\dagger (x) \; , \; \eps_1 Q_1^I \}_{-} = 0 = \delta_1 A^\dagger (x) \\
\Delta_1 v_i (x) &\equiv& \{ v_i (x) \; , \; \eps_1 Q_1^I \}_{-} \nonumber \\
&=& \int d^3 y 
\{ v_i (x) \; , \;  \Pi^j (y) \}_{-} \delta_1 v_j (y) = \delta_1 v_i (x) \\
\Delta_1 \bar\psi_{\dot\alpha} (x) &\equiv& 
\{\eps_1 Q_1^I \; , \;  \bar\psi_{\dot\alpha} (x) \}_{-} \nonumber \\
&=& \int d^3 y  \delta_1 {\bar\psi}_{\dot\beta} (y) 
\{ \pi^{I \dot\beta}_{\bar\psi} (y) \;,\; \bar\psi_{\dot\alpha} (x) \}_{+}
= \delta_1 \bar{\psi}_{\dot\alpha} (x) \\
\Delta_1 \bar\lambda_{\dot\alpha} (x) &\equiv&
\{\eps_1 Q_1^I  \; , \;  {\bar\lambda}_{\dot\alpha} (x) \}_{-} = 0 =
\delta_1 {\bar\lambda}_{\dot\alpha} (x)
\eea
where ``irr.'' stands for terms irrelevant for the Poisson brackets.
The transformations for $\psi$ and $\lambda$ have to be obtained by acting
with the charge on the conjugate momenta of $\bpsi$ and $\blambda$, 
respectively
\be
\Delta_1 \pi^I_{\bpsi \dalpha} (x) \equiv 
\{\eps_1 Q_1^I  \; , \;  \pi^I_{\bar\psi \dot\alpha} (x) \}_{-} 
=  0
\ee
since 
$\pi^I_{\bpsi \dalpha} = \frac{i}{g^2} \sigma^0_{\alpha \dalpha} \psi^\alpha$
we have $\Delta_1 \psi^\alpha = 0 = \delta_1^{\rm on} \psi^\alpha$.
For $\lambda$ we have
\bea
\Delta_1 \pi^I_{\blambda \dalpha} (x) &\equiv& 
\{\eps_1 Q_1^I  \; , \;  \pi^I_{\blambda \dalpha} (x) \}_{-} 
= \int d^3 y (-\frac{i}{g^2} \eps^\alpha_1 \sigma_{i \alpha \dbeta}
\{ \blambda^{\dbeta} (y) \;,\; \pi_{\blambda \dalpha} (x) \}_{+} \nonumber \\
&=& - \frac{i}{g^2} \eps^\alpha_1 \sigma_{i \alpha \dalpha} \hat{v}^{0i} (x) 
\eea
by multiplying both sides by $\bsigma^{0 \dalpha \beta}$ and using 
the identities given in Appendix \ref{not} we obtain 
\be
\Delta_1 \lambda^\beta (x) = - (\eps_1 \sigma^{\mu \nu})^\beta v_{\mu \nu}
= \delta^{\rm on} \lambda^\beta (x)
\ee
Note that, due to the gauge chosen for the vector field, we only reproduce
the transformations up to $v^0$. 

\noindent
Thus $\Delta_1 \equiv \delta_1^{\rm on}$. By a similar computation, that we 
shall not write down here, we see that the same happens for $\bar\Delta_1$, 
$\Delta_2$ and $\bar\Delta_2$. Therefore our charges are the correct ones, 
in the spirit described in Chapter 1. Note that we did not need to improve 
the current (therefore the charge) in order to produce the right 
transformations.

\noindent
One could also check that the Hamiltonian obtained by Legendre transforming
the Lagrangian agrees with the one obtained from our charges. 

\noindent
The Susy algebra for N=2, introduced in Chapter 2, in terms of Poisson 
brackets is given by\footnote{Note that we keep all the indices $L,M$ in the
lower position. This reflects our choice to work with $\eps_L Q_L$ as 
explained earlier.}
\bea
\{ Q_{L \alpha} \; , \; \bar{Q}_{M \dalpha} \}_{+} &=& 
2 i \; \sigma^{\mu}_{\alpha \dalpha} P_\mu \delta_{L M}\\
\{ Q_{L \alpha} \; , \; Q_{M \beta} \}_{+} &=&
2 i \; Z \eps_{\alpha \beta}  \eps_{LM}   \\
\{ \bar{Q}_{L \dalpha} \; , \; \bar{Q}_{M \dbeta} \}_{+} &=&
2 i \; Z^* \eps_{\dalpha \dbeta} \eps_{LM}
\eea
where $L,M = 1,2$.
The Hamiltonian is then simply obtained as
\be
H = - \frac{i}{4} \bsigma^{0 \dalpha \alpha}
\{ Q_{L \alpha} \; , \; \bar{Q}_{L \dalpha} \}_{+}
\ee
where $L=1$ or $L=2$, and we define $H = P^0 = -P_0$.

\noindent
We shall not write down the details of this easy computation here, since in 
the next Chapter we shall spend some time on the effective Hamiltonian of the 
SU(2) effective theory. The result of the classical computation, performed for 
$\theta = 0$, is 
\be
H = \int d^3 x {\big [} g^2 {\big (} \pi_A \pi_A^\dag 
+ \frac{1}{2} \vec{E}^2 {\big )}
+ \frac{1}{g^2} {\big (} \vec{\nabla} A \cdot\vec{\nabla} A^\dag
+  \frac{1}{2} \vec{B}^2 - i \psi {\not\!\nabla} \bar\psi
- i \lambda {\not\!\nabla} \bar\lambda {\big )} {\big ]}
\ee
where $E^i = v^{0i}$ and $B^i = \frac{1}{2} \epsilon^{0 i j k} v_{j k}$.
The same result is obtained by Legendre transforming the Lagrangian
(\ref{lu1class}) for $\theta =0$. 

\subsection{The central charge}

\noindent
We can now compute the central charge $Z$ for the classical theory from the
algebra above given. Let us start by computing the six terms (three pairs)
contributing to the centre
\bea
\{ \epsilon_1 Q_1 , \epsilon_2 Q_2 \}_{-} &=& \int d^3 x d^3 y 
{\Big (} \frac{i}{8 \pi} \{ \Pi^i ,   v^{* 0 j} \}_{-} \delta_1 v_i \tau^* 
\eps_2 \sigma_j {\bpsi} 
\nonumber \\
&+& \frac{i}{8 \pi} \{v^{* 0 i} , \Pi^j \}_{-} 
\delta_2 v_j \tau^* \eps_1 \sigma_i {\blambda} 
\label{a1} \\
&+& \Pi^i \delta_2 \blambda_{\dalpha} 
\{ \delta_1 v_i , \pi^{\dalpha}_{\blambda} \}_{-}  
\nonumber \\
&+& \Pi^j \delta_1 \bpsi_{\dalpha} 
\{\pi^{\dalpha}_{\bpsi} , \delta_2 v_j \}_{-}
\label{b1} \\
&+& \frac{i}{8 \pi} \delta_1 {\bpsi}_{\dalpha}
 \{ \pi^{\dalpha}_{\bpsi} , \eps_2 \sigma_j {\bpsi} \}_{-}  
\tau^*  v^{* 0 j} 
\nonumber \\
&+& \frac{i}{8 \pi} \delta_2 {\blambda}_{\dalpha}
 \{\eps_1 \sigma_i {\blambda} , 
\pi^{\dalpha}_{\blambda}\}_{-}  
\tau^* v^{* 0 i} 
{\Big )} \label{b2} 
\eea
On the one hand, terms (\ref{a1}) give 
\be
\frac{i}{4\pi} \tau^* \int d^3 x \partial_i  
(\eps_1 \sigma^0 \bpsi \eps_2 \sigma^i \blambda -
 \eps_2 \sigma^0 \blambda \eps_1 \sigma^i \bpsi)
\ee
On the other hand, terms (\ref{b1}) and (\ref{b2}) give 
\be
\sqrt2 \eps_1 \eps_2 
\int d^3 x {\Big (} \partial_i (2 \Pi^i A^\dag 
+ \frac{1}{4\pi} v^{* 0i} A^\dag_D) 
+  2 (\partial_i \Pi^i) A^\dag + 
\frac{1}{4\pi} (\partial_i v^{* 0i}) A^\dag_D {\Big )} 
\ee
where $A^\dag_D = \tau^* A^\dag$ is the classical analogue of the dual
of the scalar field, as discussed in the previous Chapter.

\noindent
Getting rid of $\eps_1$ and $\eps_2$ and summing over the spinor 
indices ( $\{ \epsilon_1 Q_1 , \epsilon_2 Q_2 \}_{-} 
= - \eps_1^\alpha \eps_2^\beta \{ Q_{1 \alpha} , Q_{2 \beta} \}_{+}$ and
$\eps_{\alpha \beta} \eps^{\alpha \beta} = - 2$) we can write the centre as
\be
Z = \frac{i}{4} \eps^{\alpha \beta} \{ Q_{1 \alpha} , Q_{2 \beta} \}_{+}
\ee
obtaining 
\bea
Z &=& \int d^3 x {\Big (} \partial_i [i \sqrt2 
(\Pi^i A^\dag + \frac{1}{4\pi} B^i A^\dag_D) 
- \frac{1}{4\pi} \tau^* \bpsi \bsigma^{i 0} \blambda ]\nonumber \\
&+& i \sqrt2 [(\partial_i \Pi^i) A^\dag + 
\frac{1}{4\pi} (\partial_i v^{* 0i}) A^\dag_D] {\Big )} 
\eea
By using the Bianchi identities, $\partial_i v^{* 0i} = 0$, and the classical 
limit of the Gauss law (\ref{gauss1}), we are left with a total divergence. 
The final expression for $Z$ is then given by
\be\label{zclass}
Z= i \sqrt2 \int d^2 \vec{\Sigma} \cdot 
(\vec\Pi A^\dag + \frac{1}{4\pi} \vec{B} A^\dag_D)
\ee
where $d^2\vec\Sigma$ is the measure on the sphere at infinity $S^2_\infty$, 
and we have made the usual assumption that $\bpsi$ and $\blambda$ fall off 
at least like $r^{-{3\over 2}}$. This is the classical result discussed in the 
previous Chapter. Note that we ended up with the anti-holomorphic centre.

\noindent
When we define the electric and magnetic charges {\it \`a la} Witten and Olive,
we can write 
\be
Z = i \sqrt2 (n_e a^* + n_m a^*_D)
\ee
where $<0|A^\dag|0> = a^*$ and $n_e$, $n_m$ are the electric and magnetic 
quantum numbers, respectively.

\section{The effective case}

\noindent
We now want to move to the interesting case of the effective theory described 
by the Lagrangian (\ref{lu1eff}). Let us write down this Lagrangian explicitly
\bea
{\cal L} & = &
\frac{1}{2 i} {\bigg [}  -{\cal F}'' (A)
   [ \partial_\mu A^\dagger \partial^\mu A
     + \frac{1}{4} v_{\mu \nu}{\hat v}^{\mu \nu}
     + i \psi  \not\!\partial {\bar \psi} 
     + i \lambda  \not\!\partial {\bar \lambda}
     - (E E^\dag + \frac{1}{2} D^2)] \nonumber \\
&  & + {\cal F}''' (A) 
    [ \frac{1}{\sqrt 2} \lambda \sigma^{\mu \nu} \psi v_{\mu \nu}
      -\frac{1}{2} (E^\dag \psi^2 + E \lambda^2)
      + \frac{i}{\sqrt 2} D \psi \lambda ] \nonumber \\
&  & + {\cal F}'''' (A) 
    [\frac{1}{4} \psi^2 \lambda^2]  \nonumber \\
& + & {{\cal F}^{\dag}}'' (A^\dagger)
   [ \partial_\mu A^\dagger \partial^\mu A
     + \frac{1}{4} v_{\mu \nu}{\hat v}^{\dagger \mu \nu}
     + i {\bar \psi} {\not\!\bar\partial} \psi 
     + i {\bar \lambda} {\not\!\bar\partial} \lambda
     - (E E^\dag + \frac{1}{2} D^2)]  \nonumber \\
&  & + {{\cal F}^{\dag}}''' (A^\dagger) 
    [\frac{1}{\sqrt 2} 
           {\bar \psi} {\bar \sigma}^{\mu \nu} {\bar \lambda} v_{\mu \nu}
      + \frac{1}{2} (E {\bar\psi}^2 + E^\dag {\bar\lambda}^2 )
      + \frac{i}{\sqrt 2} D \bar\psi \bar\lambda ] \nonumber \\
&  & - {{\cal F}^{\dag}}'''' (A^\dagger) 
        [\frac{1}{4} {\bar\psi}^2 {\bar\lambda}^2 ]  {\bigg ]} 
	\label{L1}
\eea
where, for the moment, we scale $\cal F$ by a factor of $4 \pi$.

\noindent
The non canonical momenta are given by
\be
\pi^{\mu}_A = - {\cal I} \partial^\mu A^{\dagger} \quad \quad
\pi^{\mu}_{A^\dagger} = (\pi^{\mu}_A)^\dagger
\ee
\be
\Pi^{\mu \nu} = -\frac{1}{2i} ({\cal F}'' {\hat v}^{\mu \nu} 
- {{\cal F}^\dag}'' {\hat v}^{\dag \mu \nu}) 
+ \frac{1}{i \sqrt2} ({\cal F}''' \lambda\sigma^{\mu \nu}\psi 
- {{\cal F}^\dag}''' \blambda \bsigma^{\mu \nu} \bpsi) \label{Pieff}
\ee
and
\bea
(\pi^\mu_{\bar \psi})_{\dot\alpha} =
  \frac{1}{2} {\cal F}^{\prime \prime}
  \psi^\alpha \sigma^\mu_{\alpha \dot\alpha} \quad
(\pi^\mu_{\psi})^{\alpha} =
  - \frac{1}{2} {\cal F}^{\dagger \prime \prime}
  \bar\psi_{\dot\alpha} \bar\sigma^{\mu \dot\alpha \alpha} \\
(\pi^\mu_{\bar \lambda})_{\dot \alpha} =
  \frac{1}{2} {\cal F}^{\prime \prime} 
  \lambda^\alpha \sigma^\mu_{\alpha \dot \alpha} \quad
(\pi^\mu_{\lambda})^{\alpha} =
  - \frac{1}{2} {\cal F}^{\dagger \prime \prime}
  \bar\lambda_{\dot\alpha} \bar\sigma^{\mu \dot\alpha \alpha}
\eea
where ${\cal F}'' = {\cal R} + i {\cal I}$.

\noindent
This time the dummy fields couple non trivially to the fermions. Their 
expression on-shell is given by 
\bea 
\label{D}    D & = & - \frac{1}{2 \sqrt 2}
                     ( f \psi \lambda + f^{\dagger}\bar \psi \bar \lambda) \\
\label{F}    E^\dag & = & - \frac{i}{4} (f \lambda^2 - f^{\dagger}\bar\psi^2)  \\
\label{Fdag} E & = & \frac{i}{4} 
(f^{\dagger} \bar\lambda^2 - f \psi^2)
\eea
where $f(A,A^{\dagger}) \equiv {\cal F}''' / {\cal I}$. 

\noindent
As in the classical case we can concentrate on the computation of the first 
Susy current $J_1^\mu$.The task, of course, is to find $V_1^\mu$. It turns 
out that its computation is by no means easy as shown in some details in 
Appendix \ref{compV}

\subsection{Computation of the effective $J_1^\mu$}

\noindent
To compute $V_1^\mu$ we first realize that, by varying $\cal L$ off-shell 
under $\delta_1$ given in (\ref{trns1})-(\ref{trns2}), there is no mixing of 
the ${\cal F}(A)$ terms with the ${\cal F}^\dag(A^\dag)$ terms. The structure 
of the Lagrangian (\ref{L1}) is 
\be
{\cal L} \equiv  \frac{1}{2 i} {\bigg [}  
\{ -{\cal F}'' \cdot [1] +
{\cal F}''' \cdot [2] +
{\cal F}'''' \cdot [3] \} - \{h.c.\}{\bigg ]}
\ee
The terms $[1]$ are bilinear in the fermions and in the bosons 
($[1] \sim [2F + 2B]$), the terms $[2]$ are products of terms bilinear in the 
fermions and linear in the bosons ($[2] \sim [2F \cdot 1B]$), finally the 
terms $[3]$ are quadrilinear in the fermions ($[3] \sim [4F]$). When we vary 
the $\cal F$ terms under $\delta_1$ we see that 
$(\delta_1 {\cal F}'''') [3] = 0$, whereas
\bea
{\cal F}'''' \delta_1 [3] \sim {\cal F}'''' (1B \cdot 3F) 
&{\rm combines} \; {\rm with}&
(\delta_1 {\cal F}''') [2] \sim {\cal F}'''' (1B \cdot 3F) \nonumber \\
{\cal F}''' \delta_1 [2] \sim {\cal F}''' (2B \cdot 1F + 3F) 
&{\rm combines} \; {\rm with}&
(\delta_1 {\cal F}'') [1] \sim {\cal F}''' (2B \cdot 1F + 3F) \nonumber
\eea
Similarly for the ${\cal F}^\dag$ terms. The aim is to write these variations 
as one single total divergence and express it in terms of momenta and 
variations of the fields.

\noindent
The $\eps_1$ computation, illustrated in Appendix \ref{compV}, gives
\bea
V_1^\mu &=& \delta_1 A \pi^\mu_A 
+  \frac{{{\cal F}^{\dagger}}''}{{\cal F}''} \delta_1 \bar{\psi} 
    \pi^\mu_{\bar\psi} 
+ \delta_1 \psi \pi^\mu_{\psi} 
+ \delta_1 \lambda \pi^\mu_{\lambda} \nonumber \\
&& + \frac{1}{2i} {{\cal F}^{\dagger}}'' \epsilon_1 \sigma_\nu \bar{\lambda} 
  v^{* \mu \nu}
+ \frac{1}{2\sqrt2} {{\cal F}^{\dagger}}''' \epsilon_1 \sigma^\mu \bar{\psi} 
 \bar{\lambda}^2 
\eea
This expression of $V_1^\mu$ is far from being a straightforward 
generalization of the classical one given in (\ref{V1class}). One could 
naively try to guess the effective $V_1^\mu$ by simply ``inverting the 
arrow'' of the classical limit (\ref{clim}),
${\cal F}'' \la \tau$, but this is not the case. In fact, 
there are many other substantial differences. The main three are: the dummy 
fields now have the quite complicated expressions in terms of functions of 
the scalar fields (the factors $f$) and fermionic bilinears given in 
(\ref{D})-(\ref{Fdag}); $\Pi^{\mu \nu}$ does not appear in $V_1^\mu$ and it 
will not be canceled in the total current, as in the classical case (it 
is now given by the expression (\ref{Pieff}), again with fermionic bilinears 
and functions of the scalar fields); the last term is an additional quantum 
factor which one could not have guessed. Of course the rigid current is 
formally identical to the classical one, namely
\be
N_1^\mu = \delta_1 A \pi^\mu_A 
+ \delta_1 v_\nu \Pi^{\mu \nu}
+ \delta_1 \bar{\psi} \pi^\mu_{\bar\psi} 
+ \delta_1 \psi \pi^\mu_{\psi} 
+ \delta_1 \lambda \pi^\mu_{\lambda}
\ee
Thus we can write down our total current as
\bea
J_1^\mu  &\equiv&  N_1^\mu - V_1^\mu \nonumber \\
        &=&  \frac{2 i {\cal I}}{{\cal F}''} \delta_1 \bar{\psi} 
             \pi^\mu_{\bar\psi}
          +  \Pi^{\mu \nu} \delta_1 v_\nu
          -  \frac{1}{2i} {{\cal F}^{\dagger}}'' \epsilon_1 \sigma_\nu 
              \bar{\lambda} v^{* \mu \nu}
          -  \frac{1}{2\sqrt2} {{\cal F}^{\dagger}}''' \epsilon_1 \sigma^\mu 
              \bar{\psi} \bar{\lambda}^2  \nonumber \\
&& \label{J1} \\
&=& \sqrt2 {\cal I} \eps_1 (\not\!\partial A^\dag) 
  \bsigma^\mu \psi
- \frac{1}{2i} {{\cal F}^\dag}'' \eps_1\sigma_\nu\blambda 
  v^{* \mu \nu}
- \frac{1}{2\sqrt2} {{\cal F}^\dag}''' \epsilon_1 \sigma^\mu \bpsi 
  \blambda^2 \nonumber \\
&& + [-\frac{1}{2} ({\cal F}'' {\hat v}^{\mu \nu} 
- {{\cal F}^\dag}'' {\hat v}^{\dag \mu \nu})   
+ \frac{1}{\sqrt2} ({\cal F}''' \lambda\sigma^{\mu \nu}\psi 
- {{\cal F}^\dag}''' \blambda \bsigma^{\mu \nu} \bpsi)]
\eps_1 \sigma_\nu \bar\lambda \nonumber \\
&& \label{Jexp}
\eea
The current (\ref{J1}) is not canonically expressed, due to partial integration
necessary in the fermionic sector of the kinetic terms in (\ref{L1}). 
Nevertheless if we explicitly write the current in terms of fields and their 
derivatives as in (\ref{Jexp}) this form will be insensitive to partial 
integration as we shall show in the next Subsection. We have seen that this is
true for simpler cases (the classical theory and the massive WZ toy model of
the first Chapter). In the effective case the matter is not trivial.

\noindent
A final remark is in order. If we conveniently rearrange the terms in 
(\ref{Jexp}) we can write 
\be
J_1^\mu =  \frac{2 i {\cal I}}{{\cal F}''} \delta_1 \bar{\psi} 
           \pi^\mu_{\bar\psi}
         -  \frac{2 i {\cal I}}{{\cal F}^{'' \dagger}} \delta_1^{\rm on} 
	    \lambda \pi^\mu_{\lambda}
         +  \frac{1}{2\sqrt2} {{\cal F}^{\dagger}}''' \epsilon_1 \sigma^\mu 
              \bar{\psi} \bar{\lambda}^2
         + \frac{1}{\sqrt2} {\cal F}''' \epsilon_1 \psi 
           \lambda \sigma^\mu \bar{\lambda} \label{j12}
\ee
and we see that the ``labour saving'' formula $J_\mu = 2 N^\mu_{\rm fermi}$, 
no longer holds.

\subsection{Canonicity}

\noindent
In this Subsection we digress for a moment to establish the canonicity in the
effective context. This is a delicate point since it affects not only the 
definition of the canonical momenta for the fermionic fields, as in the 
classical case, but even the definition of the canonical momenta for the 
scalar fields.

\noindent 
Let us extract the fermionic kinetic piece from (\ref{L1})
\be
{\cal L}_{\rm kin. fermi} = 
\frac{1}{2i}[ - {\cal F}'' i \psi  \not\!\partial {\bar \psi} 
+ {{\cal F}^\dagger}'' i \bar\psi  \not\!\bar\partial \psi 
+ (\psi \ra \lambda)]
\ee
If we call ${\cal L}^{I}$ the Lagrangian with $\bar\psi$ and $\bar\lambda$ as 
fields, it differs from $\cal L$ only in the fermionic kinetic piece and 
${{\cal F}^\dagger}'''$ type of terms
\be
{\cal L}^{I}_{\rm kin. fermi} = 
\frac{1}{2i}[ - {\cal F}'' i \psi \not\!\partial {\bar \psi} 
+ {{\cal F}^\dagger}'' i \psi \not\!\partial {\bar \psi}
- i {{\cal F}^\dagger}''' (\partial_\mu A^\dagger)
\bar\psi \bar\sigma^\mu \psi + (\psi \ra \lambda)]
\ee
Similarly for  ${\cal L}^{II}$ ($\psi$ and $\lambda$ as fields)
\be
{\cal L}^{II}_{\rm kin. fermi} = 
\frac{1}{2i}[ - {\cal F}'' i \bar\psi  \not\!\bar\partial \psi
+ {{\cal F}^\dagger}'' i \bar\psi  \not\!\bar\partial \psi 
+ i {\cal F}''' (\partial_\mu A)
\psi \sigma^\mu \bar\psi + (\psi \ra \lambda)]
\ee
The relation among the three Lagrangians is clearly
\bea
{\cal L} &=& {\cal L}^{I} +
\partial_\mu (\frac{1}{2} {{\cal F}^\dagger}'' (\bar\psi \bar\sigma^\mu \psi
+ \bar\lambda \bar\sigma^\mu \lambda)) \label{LI} \\
&=& {\cal L}^{II} +
\partial_\mu (- \frac{1}{2} {\cal F}'' (\psi \sigma^\mu \bar\psi
+ \lambda \sigma^\mu \bar\lambda)) \label{LII}
\eea
and the momenta change accordingly. 

\noindent
From ${\cal L}^{I}$:
\be
\pi^{I \mu}_A = \pi^{\mu}_A \quad \quad
\pi^{I \mu}_{A^\dagger} = - {\cal I} \partial^\mu A 
- \frac{1}{2} {{\cal F}^\dagger}''' (\bar\psi \bar\sigma^\mu \psi
+ \bar\lambda \bar\sigma^\mu \lambda) \label{pbI}
\ee
and 
\bea
(\pi^{I \mu}_{\bar \psi})_{\dot\alpha} =
  i{\cal I} \psi^\alpha \sigma^\mu_{\alpha \dot\alpha} \quad
(\pi^{I \mu}_{\psi})^{\alpha} = 0  \nonumber \\
(\pi^{I \mu}_{\bar \lambda})_{\dot \alpha} =
  i{\cal I} \lambda^\alpha \sigma^\mu_{\alpha \dot \alpha} \quad
(\pi^{I \mu}_{\lambda})^{\alpha} = 0 \label{pfI}
\eea

\noindent
From ${\cal L}^{II}$:
\be
\pi^{II \mu}_A = - {\cal I} \partial^\mu A^{\dagger} 
+ \frac{1}{2} {\cal F}''' (\psi \sigma^\mu \bar\psi
+ \lambda \sigma^\mu \bar\lambda)
\quad \quad
\pi^{II \mu}_{A^\dagger} = \pi^{\mu}_{A^\dagger} \label{pbII}
\ee
and
\bea
(\pi^{II \mu}_{\bar \psi})_{\dot\alpha} = 0  \quad
(\pi^{II \mu}_{\psi})^{\alpha} =
i{\cal I}\bar\psi_{\dot\alpha}\bar\sigma^{\mu \dot\alpha \alpha} \nonumber \\
(\pi^{II \mu}_{\bar \lambda})_{\dot \alpha} = 0 \quad
(\pi^{II \mu}_{\lambda})^{\alpha} =
   i{\cal I}\bar\lambda_{\dot\alpha}\bar\sigma^{\mu \dot\alpha \alpha}
\label{pfII}
\eea
Note that nothing changes for $\Pi^{\mu \nu}$, since 
\be
\Pi^{I \mu \nu} = \Pi^{II \mu \nu} = \Pi^{\mu \nu}
\ee
whereas in both cases $(\pi_A)^{\dagger} \ne \pi_{A^\dag}$.
From (\ref{LI}) follows that
\be
V_1^{\mu I} = V_1^\mu - 
\frac{1}{2} {{\cal F}^\dagger}''
\delta_1 (\bar\psi \bar\sigma^\mu \psi
+ \bar\lambda \bar\sigma^\mu \lambda) \label{V1I}
\ee
where $\delta_1 {\cal L}^{I} = \partial_\mu V_1^{\mu I}$ and 
$\delta_1 {\cal F}^\dagger = 0$.
Explicitly (\ref{V1I}) reads
\bea
V_1^{\mu I} &=& \delta_1 A \pi^\mu_A 
+  \frac{{\cal F}^{'' \dagger}}{{\cal F}''} \delta_1 \bar{\psi} 
    \pi^\mu_{\bar\psi} 
+ \delta_1 \psi \pi^\mu_{\psi} 
+ \delta_1 \lambda \pi^\mu_{\lambda} \nonumber \\
&& + \frac{1}{2i} {\cal F}^{'' \dagger} \epsilon_1 \sigma_\nu \bar{\lambda} 
  v^{* \mu \nu}
+ \frac{1}{2\sqrt2} {\cal F}^{''' \dagger} \epsilon_1 \sigma^\mu \bar{\psi} 
  \bar{\lambda}^2 \nonumber \\
&-& \frac{1}{2}{{\cal F}^\dagger}'' \delta_1 \bar\psi \bar\sigma^\mu \psi
- \frac{1}{2}{{\cal F}^\dagger}'' \bar\psi \bar\sigma^\mu \delta_1 \psi
- \frac{1}{2}{{\cal F}^\dagger}''\bar\lambda \bar\sigma^\mu \delta_1 \lambda 
\eea
the second, third and fourth terms in the first line cancel against 
the first, second and third terms in the third line respectively, 
therefore the fermionic momenta are absent from $V_1^{\mu I}$. But
also the rigid current changes to
\be
N_1^{\mu I} = \delta_1 A \pi^\mu_A 
+ \Pi^{\mu \nu}\delta_1 v_\nu 
+ \delta_1 \bar{\psi} \pi^{I \mu}_{\bar\psi}
\ee
thus, recalling that $J_1^{\mu I} = N_1^{\mu I} - V_1^{\mu I}$, we have
\be
J_1^{\mu I} =  \delta_1 \bar{\psi} 
             \pi^{I \mu}_{\bar\psi}
          +  \Pi^{\mu \nu} \delta_1 v_\nu
          -  \frac{1}{2i} {\cal F}^{'' \dagger} \epsilon_1 \sigma_\nu 
              \bar{\lambda} v^{* \mu \nu}
          -  \frac{1}{2\sqrt2} {\cal F}^{''' \dagger} \epsilon_1 \sigma^\mu 
              \bar{\psi} \bar{\lambda}^2 \label{JI}
\ee
which is identical to the one in (\ref{Jexp}) when we write explicitly
the transformations and the {\it new} momenta.

\noindent
At this point we have to check that the same thing happens with the 
other partial integration. Here things are slightly more complicated 
due to the fact that $\pi^{II \mu}_A$ is no longer equal to 
$\pi^{\mu}_A$ and $\delta_1 {\cal F}'' \ne 0$. As we shall see these 
two problems cancel each other.

\noindent
First let us look at $V_1^{II \mu}$. From (\ref{LII}) follows that
\bea
V_1^{II \mu} &=& V_1^{\mu} 
+ \frac{1}{2} {\cal F}''' \delta_1 A (\psi \sigma^\mu \bar\psi
+ \lambda \sigma^\mu \bar\lambda)
+ \frac{1}{2} {\cal F}'' \delta_1 (\psi \sigma^\mu \bar\psi
+ \lambda \sigma^\mu \bar\lambda) \nonumber \\
&=& \delta_1 A \pi^\mu_A 
+  \frac{{\cal F}^{'' \dagger}}{{\cal F}''} \delta_1 \bar{\psi} 
    \pi^\mu_{\bar\psi} 
+ \delta_1 \psi \pi^\mu_{\psi} 
+ \delta_1 \lambda \pi^\mu_{\lambda} \label{a} \\
&& + \frac{1}{2i} {\cal F}^{'' \dagger} \epsilon_1 \sigma_\nu \bar{\lambda} 
  v^{* \mu \nu}
+ \frac{1}{2\sqrt2} {\cal F}^{''' \dagger} \epsilon_1 \sigma^\mu \bar{\psi} 
  \bar{\lambda}^2 \nonumber \\
&-& \frac{1}{2}{\cal F}''
\delta_1 \bar\psi \bar\sigma^\mu \psi
- \frac{1}{2}{\cal F}''
 \bar\psi \bar\sigma^\mu \delta_1 \psi
- \frac{1}{2}{\cal F}''
\bar\lambda \bar\sigma^\mu \delta_1 \lambda \label{b} \\
&& + \delta_1 A \frac{1}{2}{\cal F}''' (\psi \sigma^\mu \bar\psi
+ \lambda \sigma^\mu \bar\lambda) \label{c}
\eea
The first term in (\ref{a}) combines with the last term in (\ref{c})
to give $\delta_1 A \pi_A^{II \mu}$; the second term in (\ref{a}) and the
the first in (\ref{b}) combine to 
$-i {\cal I} \delta_1 \bar\psi \bar\sigma^\mu \psi$; the third and fourth terms
in (\ref{a}) combine with the second and third terms in (\ref{b})
respectively. The final expression for $V_1^{II \mu}$ is then given by
\bea
V_1^{II \mu} &=& \delta_1 A \pi_A^{II \mu} 
-i {\cal I} \delta_1 \bar\psi \bar\sigma^\mu \psi
+  \delta_1 \psi \pi_{\psi}^{II \mu}
+ \delta_1 \lambda \pi_{\lambda}^{II \mu} \nonumber \\
&& + \frac{1}{2i} {{\cal F}^{\dagger}}'' \epsilon_1 \sigma_\nu \bar{\lambda} 
  v^{* \mu \nu}
+ \frac{1}{2\sqrt2} {{\cal F}^{\dagger}}''' \epsilon_1 \sigma^\mu \bar{\psi} 
  \bar{\lambda}^2
\eea
The rigid current is 
\be
N_1^{II \mu} = \delta_1 A \pi_A^{II \mu} 
+ \delta_1 v_\nu \Pi^{\mu \nu}
+ \delta_1 \psi \pi_{\psi}^{II \mu}
+ \delta_1 \lambda \pi_{\lambda}^{II \mu}
\ee
thus the total current is given by
\be
J_1^{II \mu} = i {\cal I} \delta_1 \bar\psi \bar\sigma^\mu \psi
- \frac{1}{2i} {\cal F}^{'' \dagger} \epsilon_1 \sigma_\nu \bar{\lambda} 
  v^{* \mu \nu}
- \frac{1}{2\sqrt2} {\cal F}^{''' \dagger} \epsilon_1 \sigma^\mu \bar{\psi} 
  \bar{\lambda}^2
+ \delta_1 v_\nu \Pi^{\mu \nu} \label{JII}
\ee
Again we see that writing explicitly the momenta and the transformations
we recover the expression (\ref{Jexp}).

\noindent
We conclude that the current is once and for all given by (\ref{Jexp}),
but in order to implement the canonical procedure we have to 
express that current {\it either} as in (\ref{JI}) {\it or} as in (\ref{JII}) 
and stick to it.

\bigskip

\noindent
We can now impose the temporal gauge for the vector field $v^0 = 0$ and 
introduce the canonical conjugate momenta of the fields. As in the classical
case we define $\pi_{\rm any \; field} \equiv \pi_{\rm any \; field}^0$. 
Thus it is simply matter to pick up the time component of (\ref{pbI}) and
(\ref{pfI}), for the Lagrangian ${\cal L}^I$, or  (\ref{pbII}) and
(\ref{pfII}), for the Lagrangian ${\cal L}^{II}$. Note that for 
$\Pi^i \equiv \Pi^{0i}$ there is no difference, it is always given by the 
time component of (\ref{Pieff}).

\noindent
The basic non zero equal time Poisson brackets are the same as in the 
classical case, namely
\bea
\{A(x), \pi_A (y) \}_{-} &=& \{A^\dag (x), \pi_{A^\dag} (y) \}_{-} =
\delta^{(3)} (\vec{x} - \vec{y}) \\ 
\{v_i (x), \Pi^j (y) \}_{-} &=& \delta_i^j \delta^{(3)} (\vec{x} - \vec{y})
\eea
and\footnote{See also Appendix \ref{not}. For instance, there we explain the 
conventions for 
$
\{ \bpsi^{\dalpha} \;,\; (\pi_{\bpsi})^{\dbeta}  \}_{+} = 
\{ (\pi_{\bpsi})^{\dbeta} \;,\;  \bpsi^{\dalpha} \}_{+} =
\eps^{\dalpha \dbeta} \delta^{(3)} (\vec{x} - \vec{y}) 
$.}
\bea 
\{ \bar\psi_{\dot\alpha} (x) \;,\; (\pi^{I}_{\bar \psi})^{\dot\beta} (y) \}_{+}
&=& \delta_{\dot\alpha}^{\dot\beta} \delta^{(3)} (\vec{x} - \vec{y}) 
\nonumber \\
\{ \bar\lambda_{\dot\alpha} (x) \;,\; 
(\pi^{I}_{\bar \lambda})^{\dot\beta} (y) \}_{+}
&=& \delta_{\dot\alpha}^{\dot\beta} \delta^{(3)} (\vec{x} - \vec{y}) 
\label{canfermiI}
\eea
or 
\bea
\{ \psi_{\alpha} (x) \;,\; (\pi^{II}_{\psi})^{\beta} (y) \}_{+}
&=& \delta_{\alpha}^{\beta} \delta^{(3)} (\vec{x} - \vec{y}) \nonumber \\
\{ \lambda_{\alpha} (x) \;,\; (\pi^{II}_{\lambda})^{\beta} (y) \}_{+}
&=& \delta_{\alpha}^{\beta} \delta^{(3)} (\vec{x} - \vec{y})
\label{canfermiII} 
\eea
but, many subtleties have to be handled with care. Classically there 
is no effect of the partial integration on the bosonic momenta. Effectively 
this is no longer the case, as we have seen, but their Poisson brackets do 
not depend on which Lagrangian one uses (${\cal L}^I$ or 
${\cal L}^{II}$), thus we did not write an index $I$ or $II$ on the momenta.

\noindent
On the other hand, the fermionic brackets, classically and effectively, 
do depend on the Lagrangian used. Nevertheless we could easily derive a 
formula which does not depend on the partial integration. At this end we have 
simply to notice that the expression of the conjugate momenta in the two 
settings, 
$
(\pi^{I}_{\bar \psi})_{\dot\alpha} =
  i{\cal I} \psi^\alpha \sigma^0_{\alpha \dot\alpha}
$
and 
$
(\pi^{II}_{\psi})^{\alpha} =
i{\cal I}\bar\psi_{\dot\alpha}\bar\sigma^{0 \dot\alpha \alpha}
$
(same for $\lambda$) implies that the canonical commutations
(\ref{canfermiI}) and (\ref{canfermiII}) are both equivalent to 
\be
\{\psi_{\alpha}(x) \;,\; \bar\psi_{\dot\alpha} (y)\}_{+} 
= - \frac{i}{\cal I} \sigma^0_{\alpha \dot\alpha} 
\delta^{(3)} (\vec{x} - \vec{y})  \label{fermi} 
\ee
(same for $\lambda$). Thus we can use either (\ref{fermi}) or one of the two 
canonical brackets (\ref{canfermiI}) and (\ref{canfermiII}).

\noindent
The other Poisson brackets are all zero. Note for instance that
\be
\{ \Pi^i \;,\; \chi \}_{-} = 0
\ee
where $\chi \equiv \psi, \lambda, \bpsi, \blambda$, even if the effective 
$\Pi^i$ has all the fermions. Furthermore, it is crucial to notice that the
usual assumption that the Poisson brackets of bosons and fermions are all 
zero no longer holds. Choosing for instance the first setting, we must have 
$\{ \pi_A \;,\; \bpsi \}_{-} = \{ \pi_{A^\dag} \;,\; \bpsi \}_{-} = 0$. If we 
take into account that 
\be
\{ \pi_A \;,\; {\cal I} \}_{-} = - \frac{1}{2i} {\cal F}''' \quad
\{ \pi_{A^\dag} \;,\; {\cal I} \}_{-} = \frac{1}{2i} {{\cal F}'''}^\dag
\ee
and the above mentioned definition of $\pi^I_{\bpsi}$ we also have
\bea
\{ \pi_A \;,\; \pi_{\bpsi} \}_{-} = 0 &\Rightarrow&
\{ \pi_A \;,\; \psi \}_{-} = -\frac{i}{2} f \psi \\
\{ \pi_{A^\dag} \;,\; \pi_{\bpsi} \}_{-} = 0 &\Rightarrow&
\{ \pi_{A^\dag} \;,\; \psi \}_{-} = + \frac{i}{2} f^\dag \psi
\eea
where $f = {\cal F}''' / {\cal I}$. 
Similar formulae hold for $\lambda$.

\subsection{Verification that the  $Q_\alpha$'s generate the Susy 
transformations}

\noindent
At this point really we have to verify that our charges produce the given Susy 
transformations. As explained in the first Chapter, this is a very delicate
point for Susy and, more generally, for any space-time symmetry. In the 
simple case of the classical theory we succeeded in doing that, but for
the highly non trivial effective theory we have to be more careful. For 
instance the charges $\hat{Q}_\alpha$ obtained by letting the Susy parameters 
become local (see Eq.(\ref{jsylv})) do not work in this sense, as can be 
seen in \cite{syl}.

\noindent
In the following Section we shall choose the setting $I$ to compute the 
centre $Z$. For the moment we want to show how in both cases our charges 
produce the Susy transformations. 

\noindent
If we choose the current (\ref{JI}), the charge is given by
\be
\eps_1 Q^I_1 = \int d^3 x {\big [} \delta_1 \bar{\psi} 
             \pi^{I}_{\bar\psi} +  \Pi^i \delta_1 v_i
          -  \frac{1}{2i} {{\cal F}^{\dag}}'' \epsilon_1 \sigma_i 
              \bar{\lambda} v^{* 0 i}
          -  \frac{1}{2\sqrt2} {{\cal F}^{\dag}}''' \epsilon_1 \sigma^0 
              \bar{\psi} \bar{\lambda}^2  {\big ]} \label{QI}
\ee
Using the same conventions as for the classical case, we shall
call $\Delta_1$ the transformation induced by this charge. Again we have to 
conveniently express it in terms of fermionic and bosonic variables. 

\noindent
We have: 
\bea
\Delta_1 A (x) \equiv \{ A (x) \; , \; \eps_1 Q_1^I \}_{-} &=& \int d^3 y 
\{ A (x) \; , \;  \delta_1 \bar{\psi} (y) 
\pi^{I}_{\bar\psi} (y) \}_{-} \nonumber \\
&=& \int d^3 y 
\{ A (x) \; , \; \sqrt2 \eps_1 \sigma^\nu \bar\sigma^0 \psi (y)
{\cal I} (y) \partial_\nu A^\dagger (y) \}_{-} \nonumber \\
&=& \int d^3 y
\{ A (x) \; , \; \sqrt2 \eps_1 \psi (y) \pi_A^I (y) + {\rm irr.} \}_{-}
\nonumber \\
&=& \sqrt2 \eps_1 \psi (x) = \delta_1 A (x) \label{d1A}
\eea
where ``irr.'' stands for terms irrelevant for the Poisson bracket.
\be
\Delta_1 A^\dagger (x) \equiv
\{ A^\dagger (x) \; , \; \eps_1 Q_1^I \}_{-} = 0 = \delta_1 A^\dagger (x)
\ee
\be
\Delta_1 v_i (x) \equiv \{ v_i (x) \; , \; \eps_1 Q_1^I \}_{-} = \int d^3 y 
\{ v_i (x) \; , \;  \Pi^j (y) \}_{-} \delta_1 v_j (y) 
= \delta_1 v_i (x)
\ee
\bea
\Delta_1 \bar\psi_{\dot\alpha} (x) \equiv 
\{\eps_1 Q_1^I \; , \;  \bar\psi_{\dot\alpha} (x) \}_{-} &=&
\int d^3 y  \delta_1 {\bar\psi}_{\dot\beta} (y) 
\{ \pi^{I \dot\beta}_{\bar\psi} (y) \;,\; \bar\psi_{\dot\alpha} (x) \}_{+}
\nonumber \\
&=& \delta_1 \bar{\psi}_{\dot\alpha} (x)
\eea
\be
\Delta_1 \bar\lambda_{\dot\alpha} (x) \equiv
\{\eps_1 Q_1^I  \; , \;  {\bar\lambda}_{\dot\alpha} (x) \}_{-} = 0 =
\delta_1 {\bar\lambda}_{\dot\alpha} (x)
\ee
For $\Delta_1 \pi^I_{\bar\psi \dot\alpha}$
some attention is due to the fact that $\pi^I_{\bar\psi \dot\alpha}$
is a product of a bosonic function $\cal I$ and of a fermion $\psi$. On the 
one hand
\bea
\Delta_1 \pi^I_{\bar\psi \dot\alpha} (x) \equiv 
\{\eps_1 Q_1^I  \; , \;  \pi^I_{\bar\psi \dot\alpha} (x) \}_{-} 
&=& \int d^3 y ( - \frac{1}{2\sqrt2} {\cal F}^{''' \dagger} 
\{ \epsilon_1 \sigma^0 \bar{\psi} \;,\; \pi^I_{\bar\psi \dot\alpha} (x) \}_{-}
\bar{\lambda}^2 ) \nonumber \\
&=& - \frac{1}{2\sqrt2} {\cal F}^{''' \dagger} 
\eps_1^\alpha \sigma^0_{\alpha \dot\alpha} \bar{\lambda}^2
\eea
on the other hand,  writing explicitly $\pi^I_{\bar\psi \dot\alpha}$ we 
have 
\be
\Delta_1 \pi^I_{\bar\psi \dot\alpha} (x) =
\frac{1}{\sqrt2} {\cal F}''' \eps_1 \psi \psi^\alpha 
\sigma^0_{\alpha \dot\alpha} 
+ i {\cal I} \sigma^0_{\alpha \dot\alpha} \Delta_1 \psi^\alpha
\ee
where we have used 
$
\Delta_1 {\cal I} = \frac{1}{2i} 
({\cal F}''' \Delta_1 A - {{\cal F}^\dagger}''' \Delta_1 A^\dagger )
$.
Thus, by comparing the two expressions for 
$\Delta_1 \pi^I_{\bar\psi \dot\alpha}$ we obtain
\be
\Delta_1 \psi^\alpha = \sqrt2 \eps_1^\alpha 
(-\frac{i}{4} f \psi^2 + \frac{i}{4} f^\dagger {\bar\lambda}^2) 
= \sqrt2 \eps_1^\alpha E_{\rm on} = \delta_1^{\rm on} \psi^\alpha
\ee
where we have used the expression (\ref{Fdag}) for $E$ on-shell and 
the Fierz identity 
$\psi_\alpha \psi^\beta = - \frac{1}{2} \delta_\alpha^\beta \psi^2$.

\noindent
More labour is needed to compute  $\Delta_1 \lambda$ from 
$\Delta_1 \pi^I_{\bar\lambda \dot\alpha}$
and we leave this to Appendix \ref{Atrns}. 
The result of that computation is the following
\be \label{dl}
\Delta_1 \lambda^\beta =
- \eps_1^\alpha (\sigma^{\mu \nu})_\alpha^\beta v_{\mu \nu}
+ i \eps_1^\beta (-\frac{1}{2\sqrt2}(f \psi \lambda 
+ f^\dagger \bar\psi \bar\lambda)) = \delta_1^{\rm on} \lambda^\beta
\ee
where we have used again the expression of the dummy fields on shell
(\ref{D}).

\noindent
We can conclude that in the effective case as well 
$\Delta_1 \equiv \delta^{\rm on}_1$. Thus our effective charge
$\eps_1 Q_1^I$ correctly generates
the first supersymmetry transformations (\ref{trns1})-(\ref{trns2}).
The second set of supersymmetry transformations (\ref{trns3})-(\ref{trns4})
is obtained by
first replacing the charge in (\ref{QI}) by its R-symmetric 
counterpart\footnote{Note that under R-symmetry $\Pi^i \ra - \Pi^i$ due to 
$v^i \ra - v^i$ and 
$\lambda \sigma^{0 i} \psi \ra + \psi \sigma^{0 i} \lambda 
= - \lambda \sigma^{0 i} \psi$.} 
\be
\eps_2 Q^I_2 = \int d^3 x {\big [} \delta_2 \bar{\lambda} 
             \pi^{I}_{\bar\lambda} +  \Pi^i \delta_2 v_i
          -  \frac{1}{2i} {{\cal F}^{\dag}}'' \epsilon_2 \sigma_i 
              \bar{\psi} v^{* 0 i}
          +  \frac{1}{2\sqrt2} {{\cal F}^{\dag}}''' \epsilon_2 \sigma^0 
              \bar{\lambda} \bar{\psi}^2  {\big ]} \label{Q2I}
\ee
and then performing for $\Delta_2$ the same kind of computations we have 
done so far for $\Delta_1$. We immediately see that this charge reproduces 
the correct transformations for $A$ (``R mirror'' computation of (\ref{d1A})), 
$A^\dagger$ ($\pi^I_{A^\dagger}$ is absent), $v_i$ (trivial), $\bar\lambda$ 
(trivial) and $\bar\psi$ ($\pi^I_{\bar\psi}$ absent). 
By a direct ``R mirror'' check we also reconstruct the transformations for 
$\lambda$ and $\psi$. We give in Appendix \ref{Atrns} the explicit 
computation for the tricky one, $\Delta_2 \psi$.

\noindent
We also leave to Appendix \ref{Atrns} the interesting check that 
$Q_1^{II}$ as well generates the transformations (\ref{trns1})-(\ref{trns2}). 
As we shall show there, in this case the delicate point is to handle the 
${\cal F}'''$ term in $\pi_A^{II}$. This problem is absent for 
$\eps_1 Q^I_1$ due to the fact that $\pi_A^I = \pi_A$ and there is no 
$\pi_{A^\dagger}^I$ in the charge, as we expect being 
$\delta_1 A^\dagger = 0$. Note that even if we use $\eps_1 Q^I_1$
the same problem will appear in handling ${\bar \eps}_1 {\bar Q}^I_1$ where 
we have $\pi_{A^\dagger}^I$. This means that we have to express 
the time derivative of the scalar field in terms of the correspondent 
canonical momentum and commute this expression with the fields and momenta.
From this follows that it is simpler to compute the central charge $Z$ with 
the $Q^I$'s, whereas for the Hamiltonian there is no such a computational 
advantage. Thus we shall choose the ``${\cal L}^I$ setting'' for our 
computations, being now sure that the results will be the same in the other 
setting.

\subsection{The central charge}

\noindent
Another interesting check is the computation of the Hamiltonian $H$. Since we 
shall perform this computation in the SU(2) sector we do not show it here.
The main point is the computation of the central charge. Now it is an easy
matter. 

\noindent
Let us first write the R-symmetric of (\ref{QI}) given by
\be
\eps_2 Q_2 = \int d^3 x {\Big(} \Pi^j \delta_2 v_j
+ \delta_2  \blambda \pi_{\blambda} 
+ \frac{i}{2}  {{\cal F}^{\dag}}'' 
\eps_2 \sigma_j {\bpsi} v^{* 0 j}
+ \frac{1}{2\sqrt2}  {{\cal F}^{\dag}}'''
\eps_2 \sigma^0 \blambda \bpsi^2 {\Big )}
\ee
We have simply to commute the two charges as we did in the classical case, 
paying due attention to the subtleties discussed earlier. The eight terms 
(four pairs) different from zero are 
\bea
\{ \epsilon_1 Q_1 , \epsilon_2 Q_2 \}_{-} &=& \int d^3 x d^3 y 
{\Big (} \{ \Pi^i ,  v^{* 0 j} \}_{-} 
\delta_1 v_i \frac{i}{2}  {{\cal F}^{\dag}}'' \eps_2 \sigma_j \bpsi 
\nonumber \\
&+& \{v^{* 0 i} , \Pi^j \}_{-} 
\delta_2 v_j \frac{i}{2}  {{\cal F}^{\dag}}'' \eps_1 \sigma_i \blambda 
\label{aa1} \\
&+& \delta_1 {\bpsi}_{\dalpha} 
\{ \pi^{\dalpha}_{\bpsi} , \bpsi^2 \}_{-}  
\frac{1}{2\sqrt2}  {{\cal F}^{\dag}}''' \eps_2 \sigma^0 \blambda 
\nonumber \\
&-& \delta_2 \blambda_{\dalpha}
\{\blambda^2 ,  \pi^{\dalpha}_{\blambda} \}_{-}  
\frac{1}{2\sqrt2}  {{\cal F}^{\dagger}}''' \eps_1 \sigma^0 {\bpsi}
 \label{aa2} \\
&+& \Pi^i \delta_2 \blambda_{\dalpha} 
\{ \delta_1 v_i , \pi^{\dalpha}_{\blambda} \}_{-}  
\nonumber \\
&+& \Pi^j \delta_1 \bpsi_{\dalpha} 
\{\pi^{\dalpha}_{\bpsi} , \delta_2 v_j \}_{-}
\label{bb1} \\
&+& \delta_1 {\bpsi}_{\dalpha} 
\{ \pi^{\dalpha}_{\bpsi} , \eps_2 \sigma_j {\bpsi} \}_{-}  
\frac{i}{2}  {{\cal F}^{\dag}}''  v^{* 0 j} 
\nonumber \\
&+& \delta_2 {\blambda}_{\dalpha}
\{\eps_1 \sigma_i {\blambda} , \pi^{\dalpha}_{\blambda} \}_{-}  
\frac{i}{2}  {{\cal F}^{\dag}}''  v^{* 0 i} 
{\Big )} \label{bb2} 
\eea
The terms (\ref{aa1}) combine to a term in any respect similar to the 
classical counterpart (\ref{a1}). It is of the form 
${{\cal F}^\dag}'' \partial (\bar\psi\bar\lambda)$. When we write explicitly 
$\delta_1 \bpsi = -i\sqrt2 \eps_1 \not\!\!\partial A^\dag$ and 
$\delta_2 \blambda = i\sqrt2 \eps_2 \not\!\!\partial A^\dag$ we see that the 
terms (\ref{aa2}) combine to a term of the form 
$(\partial {{\cal F}^\dag}'') \bpsi\blambda$. Thus from these terms we obtain 
the total divergence given by
\be
\int d^3 x \partial_i [i {{\cal F}^{\dag}}''
(\eps_1 \sigma^0 \bpsi \eps_2 \sigma^i \blambda -
\eps_2 \sigma^0 \blambda \eps_1 \sigma^i \bpsi)]
\ee
Again by explicitly writing $\delta_1 \bpsi$ and $\delta_2 \blambda$, we see 
tha the terms (\ref{bb1}) and  (\ref{bb2}) give a total divergence and two
additional terms, as in the classical case
\be
\int d^3 x {\Big (} \partial_i [2 \sqrt2 \Pi^i A^\dag 
+ \sqrt2 v^{* 0i} {{\cal F}'}^\dag] + 2\sqrt2 (\partial_i \Pi^i) A^\dag
+ \sqrt2 (\partial_i v^{* 0i}) {{\cal F}'}^\dag {\Big )} \eps_1 \eps_2
\ee
Imposing the Bianchi identities and the Gauss law, dropping the Susy parameters
$\eps_1$ and $\eps_2$, using the formula
$
Z=\frac{i}{4} \epsilon^{\alpha \beta} \{ Q_{1 \alpha} , Q_{2 \beta} \}_{+}
$, reintroducing the factor $4\pi$ and dropping the fermionic
term as in the classical case, we can write
\be\label{zeffu1}
Z = i \sqrt2 \int d^2 \vec{\Sigma} \cdot 
(\vec{\Pi} A^\dag + \frac{1}{4\pi} \vec{B} A^\dag_D)
\ee
where $d^2 \vec{\Sigma}$ is the measure on the sphere at infinity $S^2_\infty$,
$B^{i} = \frac{1}{2} \eps^{0ijk} v_{jk}$ as in the classical case,
and we introduced the SW dual of the scalar field $A^\dag$
\be
A^\dag_D \equiv {{\cal F}'}^\dag (A^\dag)
\ee
Surprisingly enough the expression (\ref{zeffu1}) is {\it formally} identical 
to the classical one given in (\ref{zclass}). 
We see that the topological nature of $Z$ is sufficient to protect its form at
the quantum level. All one has to do is to use a little dictionary and replace
classical quantities with their quantum counterparts.

\noindent
Thus we can apply exactly the same logic as in the classical case and define 
the electric and magnetic charges {\it \`a la} Witten and Olive. The final 
expression is  
\be
Z = i \sqrt2 (n_e a^* + n_m a^*_D)
\ee
where $<0|A^\dag|0> = a^*$, $<0|A^\dag_D|0> = a_D^*$ and $n_e$, $n_m$ are the 
electric and magnetic quantum numbers, respectively.

\noindent
Eventually we proved the SW mass formula. At this end we can simply use the 
BPS type of argument given in \cite{sw} or \cite{roc}, noticing that 
our direct computation includes fermions but they occur as a total divergence
which falls off fast enough to give contribution on $S^2_\infty$. Thus 
\be
M = |Z| = \sqrt2 |n_e a + n_m {\cal F}'(a)|
\ee

\noindent
A last remark is now in order. The U(1) low energy theory is invariant under
the linear shift ${\cal F} (A) \to {\cal F} (A) + c A$. This produces an 
ambiguity in the definition of $Z$. For this and other purposes we want also 
to analyse the SU(2) high energy theory in the next Chapter.

\chapter{SW SU(2) High Energy Sector}

\noindent
We want now to generalize the results obtained in the previous Chapter 
to the high energy sector taking into account all the effective SU(2) fields, 
massive and massless. 

\noindent
We intend to clarify the following points. First, the U(1) Lagrangian is 
invariant under the linear shift ${\cal F}(A) \to {\cal F}(A) + c A$, where 
$c$ is a c-number. In principle, this induces an ambiguity in the central 
charge due to the presence of ${\cal F}'(A)$. This ambiguity can be removed in
the full high energy theory, where the prepotential is a function of the SU(2)
Casimir $A^a A^a$, and such a linear shift is not allowed, since it would 
break the SU(2) gauge symmetry. Second, we want to see what is the role in the
mass formula of the heavy fields. Third, the SU(2) theory has non trivial 
features, absent in the low energy sector, as for instance, a non trivial 
Gauss law. We want to test our Susy Noether recipe on this more complicated 
ground as well, even if we do not expect any change with respect to the U(1)
case.

\noindent
In the first Section we shall construct a unique charge $Q_{1 \alpha}$ 
for the first Susy starting from its U(1) limit. In the second Section we 
shall commute this charge with its complex conjugate to obtain the Hamiltonian
and, by Legendre transforming it, the Lagrangian. The Chapter ends with the 
computation of the central charge $Z$ by commuting $Q_{1 \alpha}$ with its 
R-symmetric counterpart.

\section{The SU(2) Susy charges}

\noindent
To construct the SU(2) Susy charges\footnote{As in the Abelian case we can 
concentrate on the first Susy charge.} we shall write the most natural 
generalization of the U(1) Susy charges obtained in the last Chapter, impose 
canonicity and define the SU(2) fields and conjugate momenta and finally
fix them by requiring that they generate the given Susy transformations.

\noindent
Before starting our journey let us introduce the SU(2) notation we shall use
and make few remarks.

\noindent
A generic SU(2) vector is defined as $\vec{X} = \frac{1}{2} \sigma^a X^a$ with
$a=1,2,3$ and we follow the summation convention. The $\sigma^a$'s are
the standard Pauli matrices satisfying 
$[\sigma^a, \sigma^b] = 2 i \eps^{a b c} \sigma^c$, where $\eps^{a b c}$ are 
the structure constants of SU(2), and 
${\rm Tr} \sigma^a \sigma^b = 2 \delta^{a b}$. The covariant derivative 
and the vector field strength are given by
${\cal D}_\mu \vec{X} = \partial_\mu \vec{X} - i [\vec{v}_\mu , \vec{X}]$ 
and $\vec{v}_{\mu \nu}=\partial_\mu \vec{v}_\nu - \partial_\nu \vec{v}_\mu 
- i [\vec{v}_\mu , \vec{v}_\nu]$, respectively.

\noindent
We shall work in components thus it is convenient to write down these formulae 
explicitly
\be
{\cal D}_\mu X^a = \partial_\mu X^a +  \eps^{a b c} v^b_\mu X^c
\ee 
and 
\be
v^a_{\mu \nu}=\partial_\mu v^a_\nu - \partial_\nu v^a_\mu 
+  \eps^{abc} v^b_\mu v^c_\nu
\ee

\noindent
Some authors, \cite{ket}, \cite{dorey}, \cite{bil}, keep the renormalizable 
SU(2) gauge coupling $g$ even in the effective theory (for instance their 
covariant derivatives are defined as 
${\cal D}_\mu X^a = \partial_\mu X^a + g \eps^{a b c} v^b_\mu X^c$). 
This is somehow misleading since, as discussed earlier, in SW theory the 
effective coupling is once and for all given by $\tau(a) = {\cal F}''(a)$. Of 
course the microscopic theory is scale invariant before SSB\footnote{As a 
matter of fact, it is invariant under the full superconformal group.}, and a 
redefinition of the fields $g X \to X$ does no harm. The matter is less clear 
in the effective theory, where even the definition of what is a field poses 
some problems and scale invariance is lost after SSB. Therefore we prefer to 
follow the conventions of Seiberg and Witten \cite{sw}, where already at 
microscopic level the $g$ is absorbed in the definition of the fields and 
only appears in the overall factor $1 / g^2$ (see also our expression for 
the U(1) classical Lagrangian in (\ref{lu1class})).

\noindent
Nevertheless we can keep track of $g$ since by charge conjugation\footnote{In 
the Abelian case we implemented the R-symmetry as $\psi \lra -\lambda$, 
$E \lra E^\dag$ and $v_\mu \to - v_\mu$ (charge conjugation) when 
$\eps_1 \to \eps_2$. Noticing that  the doublet
$\left(\begin{array}{c} \psi \\ \lambda \end{array} \right)$ transforms
under
$-\left(\begin{array}{c c} 0 & 1 \\ 1 & 0 \end{array}\right) \in {\rm U(2)}$
we can say that there is room for the charge conjugation $v_\mu \to - v_\mu$ 
which becomes the $\Z_2$ discrete part of U(2).} 
$g \to -g$ (see for instance \cite{iz}), which in our notation becomes 
$\eps^{abc} \to -\eps^{abc}$.

\noindent
Finally, as we have seen in Chapter 2, the SU(2) prepotential $\cal F$ is a 
holomorphic function of the SU(2) gauge Casimir $A^a A^a$, $a = 1,2,3$. Our 
$\cal F$ corresponds to the function ${\cal H}$ in Seiberg and Witten 
conventions \cite{sw}. For some properties of this function see also 
Appendix \ref{su2comp}.

\bigskip

\noindent
Let us now write down the SU(2) generalization \cite{dorey} of the U(1) Susy 
transformations given in (\ref{trns1})-(\ref{trns4}) 

\centerline{\it first supersymmetry, parameter $\eps_1$}

\bea
\delta_1 \vec{A} &=& \sqrt2 \eps_1 \vec{\psi} \nonumber \\
\delta_1 \vec{\psi}^\alpha &=& \sqrt2 \eps_1^\alpha \vec{E}\\
\delta_1 \vec{E}  &=& 0 \nonumber 
\eea
\bea
\delta_1 \vec{E}^\dag &=& i \sqrt2 \eps_1 \not\!{\cal D} \vec{\bpsi}
+ 2 i [\vec{A}^\dag , \eps_1 \vec{\lambda}] 
\nonumber \\
\delta_1 \vec{\bpsi}_{\dalpha} &=& - i \sqrt2 \eps_1^\alpha 
\not\!{\cal D}_{\alpha \dalpha} \vec{A}^\dag \\
\delta_1 \vec{A}^\dag &=& 0 \nonumber 
\eea
\bea
\delta_1 \vec{\lambda}^\alpha &=& 
- \eps^\beta_1 ( \sigma^{\mu \nu \: \alpha}_{\: \beta} \vec{v}_{\mu \nu}
    - i \delta^\alpha_\beta \vec{D} ) \nonumber \\
\delta_1 \vec{v}^\mu &=& i \eps_1 \sigma^\mu \vec{\blambda} \quad
\delta_1 \vec{D} = - \eps_1 \not\!{\cal D} \vec{\blambda} \\
\delta_1 \vec{\blambda}_{\dalpha} &=& 0 \nonumber 
\eea

\centerline{\it second supersymmetry, parameter $\eps_2$}

\bea
\delta_2 \vec{A} &=& - {\sqrt 2} \epsilon_2 \vec{\lambda} \nonumber \\
\delta_2 \vec{\lambda}^\alpha &=& 
- {\sqrt 2}\epsilon_2^\alpha \vec{E}^\dag \\
\delta_2 \vec{E}^\dag  &=& 0 \nonumber 
\eea
\bea
\delta_2 \vec{E} &=& - i \sqrt2 \eps_2 \not\!{\cal D} \vec{\blambda}
+ 2 i  [\vec{A}^\dag , \eps_2 \vec{\psi}] 
\nonumber \\
\delta_2 \vec{\blambda}_{\dalpha} &=& i \sqrt2 \eps_2^\alpha 
\not\!{\cal D}_{\alpha \dalpha} \vec{A}^\dag \\
\delta_2 \vec{A}^\dag &=& 0 \nonumber 
\eea
\bea
\delta_2 \vec{\psi}^\alpha &=& 
- \eps^\beta_2 ( \sigma^{\mu \nu \: \alpha}_{\: \beta} \vec{v}_{\mu \nu}
    + i \delta^\alpha_\beta \vec{D} ) \nonumber \\
\delta_2 \vec{v}^\mu &=& i \eps_2 \sigma^\mu \vec{\bpsi} \quad
\delta_2 \vec{D} =  \eps_2 \not\!{\cal D} \vec{\bpsi} \\
\delta_2 \vec{\bpsi}_{\dalpha} &=& 0 \nonumber 
\eea
We want now to construct the SU(2) Susy charges that generate these 
transformations by generalizing the U(1) Susy charges obtained in the last
Chapter. At this end we simply write down the explicit expression of the first
U(1) charge as the spatial integral of the time component of the explicit 
current $J^\mu_1$ given in (\ref{Jexp})
\bea
\eps_1 Q_1^{\rm U(1)} &=& \int d^3x {\Big (} 
\sqrt2 {\cal I} \eps_1 (\not\!\partial A^\dag) \bsigma^0 \psi
- \frac{1}{2i} {{\cal F}^\dag}'' \eps_1\sigma_i\blambda v^{* 0 i}
- \frac{1}{2\sqrt2} {{\cal F}^\dag}''' \eps_1 \sigma^0 \bpsi 
  \blambda^2 \nonumber \\
&+ &[-\frac{1}{2}({\cal F}''{\hat v}^{0i} 
-{{\cal F}^\dag}'' {\hat v}^{\dag 0 i})   
+\frac{1}{\sqrt2}({\cal F}'''\lambda\sigma^{0 i}\psi 
-{{\cal F}^\dag}'''\blambda\bsigma^{0i}\bpsi)]
\eps_1\sigma_i\blambda {\Big )} \nonumber \\
\eea
where, for the moment, we scale $\cal F$ by a factor of $4\pi$.

\noindent
Then we generalize this expression in the most natural way, namely
\bea
\eps_1 Q_1^{\rm SU(2)} &\equiv& \int d^3x {\Big (} 
\sqrt2 {\cal I}^{ab} \eps_1 (\not\!\partial A^{a \dag}) \bsigma^0 \psi^b
- \frac{1}{2i} {\cal F}^{a b \dag} \eps_1\sigma_i\blambda^a v^{* b 0 i}
 \nonumber \\
&&- \frac{1}{2\sqrt2} {\cal F}^{a b c \dag} \eps_1 \sigma^0 \bpsi^a 
  \blambda^b \blambda^c + [-\frac{1}{2}({\cal F}^{a b}\hat{v}^{a0i} 
-{\cal F}^{ab\dag}\hat{v}^{\dag a0i}) \nonumber \\
&& +\frac{1}{\sqrt2}({\cal F}^{bcd}\lambda^c\sigma^{0i}\psi^d 
-{\cal F}^{bcd\dag}\blambda^c\bsigma^{0i}\bpsi^d)]
\eps_1\sigma_i\blambda^b {\Big )}  \label{Qexp} 
\eea
where ${\cal F}^{a_1 \cdots a_n} 
= \partial^n {\cal F} / \partial A^{a_1} \cdots \partial A^{a_n}$ (see 
Appendix \ref{su2comp}) and ${\cal F}^{ab}={\cal R}^{ab} + i {\cal I}^{ab}$.

\noindent
In order to impose a canonical form to this charge we have to define the
conjugate momenta of the fields and therefore produce the transformations
above given. The SU(2) {\it version} of the U(1) conjugate momenta is 
given by
\be
\Pi^{a i} = -\frac{1}{2i} 
({\cal F}^{a b} {\hat v}^{0 i b} - 
{\cal F}^{a b \dagger} {\hat v}^{\dagger 0 i b})
+ \frac{1}{i \sqrt2} {\cal F}^{a b c} \lambda^b \sigma^{0 i} \psi^c 
- \frac{1}{i \sqrt2} {\cal F}^{a b c \dagger} 
     \bar\lambda^b \bar\sigma^{0 i} \bar\psi^c \label{Pisu2}
\ee
\be
\pi^a_A = - {\cal I}^{a b} \partial^0 A^{\dagger b} \quad \quad
\pi^a_{A^\dagger} = - {\cal I}^{a b} \partial^0 A^b 
- \frac{1}{2} {{\cal F}^\dagger}^{a b c} (\bar\psi^b \bar\sigma^0 \psi^c
+ \bar\lambda^b \bar\sigma^0 \lambda^c) \label{pbsu2}
\ee
\be
\pi^a_{\bar \psi} = i {\cal I}^{a b} \psi^b \sigma^0 
\quad \quad
\pi^a_{\bar \lambda} = i {\cal I}^{a b} \lambda^b  \sigma^0 \label{pfsu2}
\ee
where we chose the {\it setting $I$} (see the correspondent U(1) expressions
given in the previous Chapter in (\ref{pbI}), (\ref{pfI}) and (\ref{Pieff})) 
and the temporal gauge for the vector field (thus ${\cal D}^0 = \partial^0$).

\noindent
With these definitions the charge (\ref{Qexp}) becomes
\be
\epsilon_1 Q_1^{\rm SU(2)} = 
 \int d^3 x {\Big(} \Pi^{a i} \delta_1 v^a_i
+ \delta_1  {\bar\psi}^a \pi^a_{\bar\psi} 
+ \frac{i}{2}  {\cal F}^{\dagger a b} 
\epsilon_1 \sigma_i {\bar\lambda}^a v^{* 0 i b}
- \frac{1}{2\sqrt2}  {\cal F}^{\dagger a b c}
\epsilon_1 \sigma^0 {\bar\psi}^a \bar\lambda^b \bar\lambda^c{\Big )} 
\label{Qfirst}
\ee
By using the same techniques as in the Abelian case, we see that this charge 
correctly generates the transformations that do not involve dummy fields,
namely
\be
\delta_1 v^a_i \quad 
\delta_1 A^a \quad \delta_1 A^{\dag a} \quad 
\delta_1 \bar\psi^a \quad 
\delta_1 \bar\lambda^a
\ee
To produce the other transformations one needs the explicit expressions of 
the dummy fields on shell. At this point we notice that one of the main 
differences between the U(1) and the SU(2) theories relies on the coupling of 
the dummy fields $D^a$ to the scalar fields to give the Higgs 
potential, after elimination \cite{west}. This potential (and the Yukawa 
potential) can never be reproduced in the Hamiltonian by commuting the charge 
(\ref{Qfirst}) with its complex conjugate. Therefore we see that this first 
generalization needs to be improved.

\noindent
We can obtain the missing terms by considering the classical (microscopic)
SU(2) Lagrangian ${\cal L}_{\rm class}^{\rm SU(2)}$ and solving the 
Euler-Lagrange equations for $D^a$. The result is given by 
\be
(D^a)^{\rm on}_{\rm class} = i \eps^{a b c} A^b A^{c \dag}
\ee
where we used the standard expression for ${\cal L}_{\rm class}^{\rm SU(2)}$
(see, for instance, \cite{bil} and \cite{dorey}). Note that 
$(E^a)_{\rm class}^{\rm on} = 0$.
From the recipe given in the first Chapter and extensively applied in the 
U(1) case, we know that the charge has to produce the transformations 
with the dummy fields on shell. $D^a$ appears in the transformation of 
$\lambda^a$ therefore we want to produce $\delta^{D}_1 \lambda^a$ 
from $\{ \epsilon_1 Q^{\rm SU(2)}_1, \pi^a_{\blambda} \}$, where 
$\pi^a_{\blambda} = i \tau_I \lambda^a  \sigma^0$ is the classical limit of 
the SU(2) effective conjugate momentum of $\blambda^a$ given in (\ref{pfsu2}).
Thus we conclude that a missing term in the classical charge is given by
\be\label{dclassical}
i  \tau_I \epsilon_1 \sigma^0 \bar\lambda^a 
\epsilon^{a c d} A^c A^{d \dagger}
\ee
Furthermore this term is the only missing term, because once it is
added then we obtain all the correct Susy transformations.
The term in the effective charge that will produce (\ref{dclassical}) in the 
classical limit is evidently
\be
i  {\cal I}^{a b} \epsilon_1 \sigma^0 \bar\lambda^b 
\epsilon^{a c d} A^c A^{d \dagger}
\ee
and it is therefore clear that we must add such a term to the charge in 
(\ref{Qfirst}). 

\noindent
We shall see in the next Section that, as in the classical theory, this term 
is the {\it only} new term that is required. It produces the Higgs and Yukawa 
potential in the Hamiltonian and it is responsible for most of the new terms 
in the centre.

\noindent
Thus finally, the first Susy SU(2) effective charge is given by 
\bea
\eps_1 Q_1 &=& 
 \int d^3 x {\Big(} \Pi^{a i} \delta_1 v^a_i
+ \delta_1  {\bpsi}^a \pi^a_{\bpsi} 
+ \frac{i}{2}  {\cal F}^{\dag a b} 
\eps_1 \sigma_i {\blambda}^a v^{* 0 i b}  \nonumber \\
&& - \frac{1}{2\sqrt2}  {\cal F}^{\dag a b c}
\eps_1 \sigma^0 {\bpsi}^a \blambda^b \blambda^c
+ i{\cal I}^{a b} \eps_1 \sigma^0 \blambda^b 
\eps^{a c d} A^c A^{d \dag} {\Big )} \label{Qsu21}
\eea
where we have dropped the label ``SU(2)''.

\noindent
We conclude this Section by writing the R-symmetric counterpart of 
(\ref{Qsu21}), given by
\bea
\eps_2 Q_2 &=& \int d^3 x {\Big(} \Pi^{b j} \delta_2 v^b_j
+ \delta_2  {\blambda}^b \pi^b_{\blambda} 
+ \frac{i}{2}  {\cal F}^{\dag c d} 
\eps_2 \sigma_j {\bpsi}^c v^{* 0 j d}  \nonumber \\
&& + \frac{1}{2\sqrt2}  {\cal F}^{\dag d e f}
\eps_2 \sigma^0 {\blambda}^d \bpsi^e \bpsi^f 
+ i{\cal I}^{e f} \eps_2 \sigma^0 \bpsi^f 
\eps^{e g h} A^g A^{h \dag} {\Big )} \label{Qsu22}
\eea
which generates the second set of Susy transformations above given, and 
the complex conjugate of (\ref{Qsu21}), given by
\bea
\bar\epsilon_1 \bar{Q}_1 &=& 
\int d^3 x {\Big(} \Pi^{a i} \bar\delta_1 v^a_i
+ \sqrt2 {\cal I}^{a b} \beps_1 \not\!\bar{\cal D} A^a \sigma^0 \bpsi^b 
+ \frac{i}{2}  {\cal F}^{a b} \beps_1 \sigma_i \lambda^a v^{* 0 i b} 
\nonumber \\
&& + \frac{1}{2\sqrt2}  {\cal F}^{a b c}
\beps_1 \bsigma^0 \psi^a \lambda^b \lambda^c 
- \beps_1 \pi^a_{\blambda} \eps^{a c d} A^c A^{d \dagger} 
{\Big )} \label{bQsu21}
\eea
where we introduced the conjugate momentum $\pi^a_{\blambda}$.

\section{Hamiltonian and Lagrangian}

\noindent
We are now in the position to compute the Hamiltonian $H$. The main points
here are: first to compare the SU(2) Lagrangian obtained by Legendre 
transforming $H$ with the U(1) Lagrangian and with the Lagrangian obtained by 
superfield expansion; second to obtain the non trivial Gauss law 
for the SU(2) theory. The last point is vital for our purpose, since our main 
interest is the computation of the central charge where the Gauss law is 
expected to play an important role.

\noindent
We computed $H$ by taking the Poisson brackets of $\eps_1 Q_1$ given in 
(\ref{Qsu21}) with $\beps_1 \bar{Q}_1$ given in (\ref{bQsu21}), then getting 
rid of $\eps_1$ and $\beps_1$ ($ \{ \eps_1 Q_1 \; , \; \beps_1 \bar{Q}_1 \}_{-}
= \eps_1^\alpha \beps_1^{\dalpha} 
\{ Q_{1 \alpha} \; , \; \bar{Q}_{1 \dalpha} \}_{+} $)
and finally taking the trace with $\bsigma^{0 \dalpha \alpha}$. 
The final formula for $H$ is 
\be
H = - \frac{i}{4} \bsigma^{0 \dalpha \alpha}
\{ Q_{1 \alpha} \; , \; \bar{Q}_{1 \dalpha} \}_{+}
\ee
where we defined $H = P^0 = -P_0$.
This lengthy computation is illustrated in some details in Appendix 
\ref{su2comp}. Its final result is given by
\bea
H &=& \int d^3x {\Big (} -\frac{1}{2}({\cal I}^{ab})^{-1} \Pi^{ai} \Pi^{bi}
- ({\cal I}^{ab})^{-1} {\cal R}^{bc} \Pi^{ai} B^{i c} 
- \frac{1}{2}({\cal I}^{ab})^{-1}{\cal F}^{\dag ab}{\cal F}^{ef}
B^{i b}B^{i f} \nonumber \\
&&-({\cal I}^{ab})^{-1} \pi^a_A (\pi^b_A)^\dag 
+ {\cal I}^{ab} ({\cal D}^i A^a)({\cal D}^i A^{\dag b}) \nonumber \\
&& + i {\cal I}^{ab} \psi^a\sigma^i{\cal D}_i\bpsi^b 
+ i {\cal I}^{ab} \lambda^a\sigma^i{\cal D}_i\blambda^b 
+ \frac{1}{2} (\partial_i {\cal F}^{\dag ab}) \blambda^a \sigma^i \lambda^b 
\nonumber \\
&& - \frac{1}{\sqrt2}{\cal I}^{ad}\eps^{abc}  
(A^c \bpsi^d \blambda^b + A^{\dag c} \psi^d \lambda^b) 
+ \frac{1}{2}{\cal I}^{ab}\eps^{acd} \eps^{bfg}
A^c A^{\dag d} A^f A^{\dag g} \nonumber \\
&& +\frac{i}{\sqrt2} ({\cal I}^{af})^{-1} {\cal F}^{feg} 
\psi^e\sigma_{i0}\lambda^g (\Pi^{ia}+ {\cal F}^{\dag ab} B^{ib}) \nonumber \\
&& -\frac{i}{\sqrt2} ({\cal I}^{ec})^{-1}{\cal F}^{\dag abc} 
\bpsi^a\bsigma_{i0}\blambda^b (\Pi^{ie}+ {\cal F}^{ed} B^{id}) \nonumber \\
&& + \frac{1}{16} {\cal F}^{\dag efg} {\cal F}^{cad} ({\cal I}^{gc})^{-1}
\bpsi^e\bpsi^f \psi^a\psi^c 
+\frac{1}{16} {\cal F}^{\dag abc} {\cal F}^{efg} ({\cal I}^{ae})^{-1}
\blambda^b\blambda^c \lambda^f\lambda^g \nonumber \\
&& + \frac{3}{16} {\cal F}^{\dag bec} {\cal F}^{\dag efg} ({\cal I}^{ab})^{-1}
\bpsi^a\bpsi^c \blambda^f\blambda^g 
+ \frac{3}{16} {\cal F}^{bec} {\cal F}^{efg} ({\cal I}^{ab})^{-1}
\psi^a\psi^c \lambda^f\lambda^g \nonumber \\ 
&& -\frac{1}{2i} 
(\frac{1}{4}{\cal F}^{abcd} \psi^a\psi^b \lambda^c\lambda^d
- \frac{1}{4}{\cal F}^{\dag abcd} \bpsi^a\bpsi^b \blambda^c\blambda^d) 
{\Big)} \nonumber \\
&& + \int d^3x \partial_i 
(\frac{1}{2} {\cal F}^{\dag ab} \blambda^a \bsigma^i \lambda^b
- \frac{i}{2} {\cal I}^{ab} \bpsi^a \bsigma^i \psi^b) \label{Hsu2}
\eea
where $E^{a i} = v^{a 0 i}$ and $B^{a i} = \frac{1}{2} \eps^{0ijk} v^a_{jk}$ 
are the SU(2) generalization of the electric and magnetic fields, respectively.
 
\noindent
If we call ``classical'' the terms whose factors are at most second 
derivatives of $\cal F$, we see that the first four lines contain only
``classical'' terms, modulo the two fermions contribution to $\Pi^{ai}$ 
(see (\ref{Pisu2})) and the term due to partial integration. The first line 
are the e.m. terms, and if we write 
\be
\Pi^{ai} = -({\cal I}^{a b} E^{bi} + {\cal R}^{a b} B^{bi}) + \Pi^{ai}_{\rm F}
\label{Pisu22}
\ee
where $\Pi^{ai}_{\rm F}$ is the purely quantum two fermions piece, it is easy 
to see that the ``classical'' terms combine to give 
$-{\cal I}^{ab}(E^{ai}E^{bi} + B^{ai}B^{bi})$. The second and third lines 
are the standard terms one would expect for the complex scalars and the 
fermions, modulo the last term in the third line on which we shall comment in 
a moment. As promised we reproduced the Yukawa and Higgs potentials, given by 
the first and second term in the fourth line, respectively. The other terms 
are purely quantum terms. There we have the two fermions terms coupled to the
e.m. fields and momenta and the four fermions term. To check the correctness
of these terms we have to consider the correspondent terms in the Lagrangian 
and compare them with their U(1) limit.

\noindent
We notice here that, in the last line, we kept a total divergence to 
explicitly show that we partially integrated the fermionic kinetic terms, in 
order to fix the phase space $(\bpsi, \blambda ; \pi_{\bpsi} , \pi_{\blambda})$
we started with. It turns out that this total divergence is not symmetric with
respect to $\psi$ and $\lambda$ and this is reflected in the last term of the 
third line where only $\lambda$-terms appear. This means that the Lagrangian 
we shall obtain by Legendre transforming $H$ will be slightly different 
from the one expected. Nevertheless the difference will not affect the 
conjugate momenta (\ref{Pisu2})-(\ref{pfsu2}), therefore the charges 
(\ref{Qsu21})-(\ref{bQsu21}) above constructed are not affected by this 
asymmetry. Furthermore this problem is entirely due to the non trivial partial
integration in the effective case. As explained in the previous Chapter this 
does not affect the explicit expression of the currents and charges. 

\noindent
A last remark on the computation of this Hamiltonian, is that we extensively 
exploited the SU(2) generalization of the Poisson brackets defined in the last
Chapter. In particular we made use of the ``setting independent'' formula for 
the fermions
\be
\{ \bpsi^a_{\dalpha} \; , \; \psi^b_\alpha \}_{+}
= \{ \blambda^a_{\dalpha} \; , \; \lambda^b_\alpha \}_{+} 
= -i ({\cal I}^{ab})^{-1} \sigma^0_{\alpha \dalpha}
\ee
and the non trivial brackets between bosons and fermions
\bea
\{ \pi^a_A \; , \; \psi^b_\alpha \}_{-} 
&=& -\frac{i}{2} ({\cal I}^{bc})^{-1} {\cal F}^{cad} \psi^d_\alpha \\
\{ \pi^a_{A^\dag} \; , \; \psi^b_\alpha \}_{-} 
&=& +\frac{i}{2} ({\cal I}^{bc})^{-1} {\cal F}^{\dag cad} \psi^d_\alpha
\eea
similarly for $\lambda$.

\noindent
We want now to Legendre transform the Hamiltonian (\ref{Hsu2}) to obtain the 
correspondent Lagrangian. Recalling that our metric is 
$\eta^{\mu \nu} = {\rm diag}(-1,1,1,1)$ we have 
\be
{\cal L} = 
\partial_0 v^{i a} \Pi^{i a} + \partial_0 A^a \pi_A^a
+ \partial_0 A^{\dag a} \pi_{A^\dag}^a
+ \partial_0 \bpsi^a \pi_{\bpsi}^a
+ \partial_0 \blambda^a \pi_{\blambda}^a - {\cal H}
\ee
where $H = \int d^3x {\cal H}$ and we discarded the total derivative in the 
last line of (\ref{Hsu2}).

\noindent
Since for the potential terms (Yukawa, Higgs and four fermions terms) we have
simply to reverse the sign, let us concentrate on the other terms.

\noindent
In the e.m. sector we obtain
\bea
{\cal L}_{\rm e.m.} &=& 
- \frac{1}{2} {\cal I}^{ab} (E^{i a} E^{i b} - B^{i a} B^{i b})
- {\cal R}^{ab} E^{i a} B^{i b} \nonumber \\
&&-\frac{1}{2i} 
{\Big (} \Pi_{i {\rm F}}^{a \psi \lambda} (E^{i a} + i B^{i a})
+ \Pi_{i {\rm F}}^{a \bpsi \blambda} (E^{i a} - i B^{i a}) {\Big )} 
\nonumber \\
&& + \frac{3}{2} ({\cal I}^{ab})^{-1} \Pi^{i a}_{\rm F} \Pi^{i b}_{\rm F}
\eea
where the second line corresponds to the fifth and sixth lines of the 
Hamiltonian (\ref{Hsu2}), we used the expression (\ref{Pisu22}) with 
$\Pi^{a}_{i {\rm F}} \equiv \Pi_{i {\rm F}}^{a \psi \lambda} 
+ \Pi_{i {\rm F}}^{a \bpsi \blambda}$ and the last line is the 
four fermions term that has to be combined with the other four fermions terms.

\noindent
On the other hand, in the sector with the kinetic terms for the fermions and 
the scalars we obtain
\bea
{\cal L}_{\rm scalar-fermi} &=& 
- {\cal I}^{ab}( \partial_0 A^a \partial^0 A^{\dag b} +
{\cal D}_i A^a {\cal D}^i A^{\dag b}) \nonumber \\
&& - i {\cal I}^{ab} 
(\psi^a \sigma^0 \partial_0 \bpsi^b +  \psi^a \sigma^i {\cal D}_i \bpsi^b
+  \lambda^a \sigma^0 \partial_0 \blambda^b  
+  \lambda^a \sigma^i {\cal D}_i \blambda^b ) \nonumber \\
&& - \frac{1}{2} (\partial_\mu {\cal F}^{\dag ab}) 
\blambda^a \bsigma^\mu \lambda^b 
- \frac{1}{2} (\partial_0 {\cal F}^{\dag ab}) 
\bpsi^a \bsigma^0 \psi^b \nonumber 
\eea
where we used the $\pi_{A^\dag} \ne (\pi_A)^\dag$ given in (\ref{pfsu2}). 
Note again here that $\psi$ and $\lambda$ in the last line, do not have the 
same factor, as explained above.

\noindent
If we reintroduce $v^0$ in these expressions, our Lagrangian density is given 
by
\be
{\cal L} \equiv {\cal L}_1 + {\cal L}_2
\ee
where 
\bea
{\cal L}_1 &=&  \frac{1}{2i} {\Big [} 
- \frac{1}{4} {\cal F}^{ab} v^{a \mu \nu} \hat{v}^b_{\mu \nu}
+ \frac{1}{4} {\cal F}^{\dag ab} v^{a \mu \nu} \hat{v}^{\dag b}_{\mu \nu}
{\Big ]} \nonumber \\
&& - {\cal I}^{ab} {\Big [} {\cal D}_\mu A^a {\cal D}^\mu A^{\dag b}
+ i \psi^a \not \!\!{\cal D} \bpsi^b  
+ i \lambda^a \not\!\!{\cal D} \blambda^b \nonumber \\
&& - \frac{1}{\sqrt2} \eps^{adc}  
(A^c \bpsi^b \blambda^d + A^{\dag c} \psi^b \lambda^d) 
+ \frac{1}{2} \eps^{acd}  \eps^{bfg}
A^c A^{\dag d} A^f A^{\dag g} {\Big ]} \nonumber \\
&& - \frac{1}{2} (\partial_\mu {\cal F}^{\dag ab}) 
\blambda^a \bsigma^\mu \lambda^b 
- \frac{1}{2} (\partial_0 {\cal F}^{\dag ab}) 
\bpsi^a \bsigma^0 \psi^b 
\eea
contains the ``classical'' terms, and
\bea
{\cal L}_2 &=& \frac{1}{2i} {\Big [} \frac{1}{\sqrt2} {\cal F}^{abc} 
\lambda^a \sigma^{\mu \nu} \psi^b v^c_{\mu \nu}
- \frac{1}{\sqrt2} {\cal F}^{\dag abc} 
\blambda^a \bsigma^{\mu \nu} \bpsi^b v^c_{\mu \nu} {\Big ]} \nonumber \\
&& +\frac{3}{16} ({\cal I}^{ab})^{-1} {\Big [}
{\cal F}^{acd} {\cal F}^{bef} 
(\psi^d\psi^f \lambda^c\lambda^e - \psi^d\lambda^e \lambda^c\psi^f)
- {\cal F}^{gbc} {\cal F}^{gef} \psi^a\psi^c \lambda^e\lambda^f \nonumber \\ 
&& + {\cal F}^{\dag acd} {\cal F}^{\dag bef}
(\bpsi^d\bpsi^f \blambda^c\blambda^e - \bpsi^d\blambda^e \blambda^c\bpsi^f)
- {\cal F}^{\dag gbc} {\cal F}^{\dag gef}
\bpsi^a\bpsi^c \blambda^e\blambda^f \nonumber \\
&& + {\cal F}^{\dag acd} {\cal F}^{\dag bef}
(\lambda^c\sigma^0\bpsi^f \psi^d\sigma^0\blambda^e 
- \lambda^c\sigma^0\blambda^e \psi^d\sigma^0\bpsi^f ) {\Big ]} \nonumber \\
&& - \frac{1}{16} ({\cal I}^{ab})^{-1} {\cal F}^{\dag acd} {\cal F}^{bef} 
(\bpsi^c\bpsi^d \psi^e\psi^f +  
\blambda^c\blambda^d \lambda^e\lambda^f) \nonumber \\
&& + \frac{1}{2i} 
(\frac{1}{4}{\cal F}^{abcd} \psi^a\psi^b \lambda^c\lambda^d
- \frac{1}{4}{\cal F}^{\dag abcd} \bpsi^a\bpsi^b \blambda^c\blambda^d) 
\label{l4fermi}
\eea
contains the purely quantum terms.
The second and third lines of (\ref{l4fermi}) come from the combination of 
$\frac{3}{2} \Pi_{\rm F}$ and the four fermions in the Hamiltonian, whereas the
fourth line comes from $\frac{3}{2} \Pi_{\rm F}$ alone. 
In Appendix \ref{su2comp} we shall show that the factors are in agreement 
with the U(1) correspondent ones.

\noindent
Note also that the above given expression for the SW SU(2) high-energy 
effective Lagrangian is in agreement with the one obtained directly by
superfield expansion in \cite{syl}.

\noindent
What is left is to produce the Gauss law descending from this Lagrangian.
At this end we have only to consider the terms in the Lagrangian that contain
the Lagrange multiplier $v^{g 0}$ and define the associated Gauss constraint 
as we did in the U(1) sector (see (\ref{gauss1})). We obtain 
\bea
0 = \frac{\partial {\cal L}}{\partial v^{g 0}} &=& 
\frac{1}{2i} [\partial_i (-{\cal F}^{g b} {\hat v}_0^{b i}
+ \sqrt2 {\cal F}^{gbc} \lambda^b \sigma^i_0 \psi^c) \nonumber \\
&& - \eps^{gad} v^{i d} {\cal F}^{ab} \hat{v}^b_{i 0}
+ \eps^{gcd} \sqrt2 {\cal F}^{abc} \lambda^a \sigma^i_0 \psi^b v^d_i
- h.c.] \nonumber \\
&+& \eps^{gac} {\cal I}^{ab}
(A^c {\cal D}_0 A^{\dag b} + A^{\dag c} {\cal D}_0 A^b 
+ i \psi^b \sigma_0 \bpsi^c + i \lambda^b \sigma_0 \blambda^c) 
\eea
Recalling the definition of the conjugate momentum $\Pi^{g i}$ of $v^g_i$ 
given in (\ref{Pisu2}) and the definition of the covariant derivative, 
${\cal D}_\mu X^a = \partial_\mu X^a +  \eps^{a b c} v^b_\mu X^c$, we see that
the first two lines give ${\cal D}_i \Pi^{g i}$. Thus we have
\be \label{gauss}
{\cal D}_i \Pi^{i g} = - \eps^{gac} {\cal I}^{ab}
(A^c {\cal D}^0 A^{\dag b} + A^{\dag c} {\cal D}^0 A^b 
+ i \psi^b \sigma^0 \bpsi^c + i \lambda^b \sigma^0 \blambda^c)
\ee
which is the required Gauss law.

\section{Computation of the central charge}

\noindent
We can now compute the central charge for the SU(2) theory. As we did in the
U(1) sector, we first compute the Poisson brackets of $\eps_1 Q_1$ and
$\eps_2 Q_2$ given in (\ref{Qsu21}) and (\ref{Qsu22}), respectively. 
The non zero contributions are given by
\bea
\{ \epsilon_1 Q_1 , \epsilon_2 Q_2 \}_{-} &=& \int d^3 x d^3 y 
{\Big (} \{ \Pi^{a i} ,   v^{* 0 j d} \}_{-} 
\delta_1 v^a_i \frac{i}{2}  {\cal F}^{\dagger c d} 
\epsilon_2 \sigma_j {\bar\psi}^c 
\nonumber \\
&+& \{v^{* 0 i b} , \Pi^{c j}\}_{-} 
\delta_2 v^c_j \frac{i}{2}  {\cal F}^{\dagger a b} 
\epsilon_1 \sigma_i {\bar\lambda}^a 
 \label{A1} \\
&+& \delta_1 {\bar\psi}_{\dot\alpha}^a 
\{ \pi^{\dot\alpha a}_{\bar\psi} , \bar\psi^e \bar\psi^f \}_{-}  
 \frac{1}{2\sqrt2}  {\cal F}^{\dagger d e f}
\epsilon_2 \sigma^0 {\bar\lambda}^d 
 \nonumber \\
&+& \delta_2 {\bar\lambda}_{\dot\alpha}^d 
\{\pi^{\dot\alpha d}_{\bar\lambda} , \bar\lambda^b \bar\lambda^c\}_{-}  
\frac{1}{2\sqrt2}  {\cal F}^{\dagger a b c}
\epsilon_1 \sigma^0 {\bar\psi}^a 
 \label{A2} \\
&+& \Pi^{a i} \delta_2 \bar\lambda_{\dot\alpha}^b 
\{ \delta_1 v^a_i , \pi^{\dot\alpha b}_{\bar\lambda} \}_{-}  
 \nonumber \\
&+& \Pi^{b j} \delta_1 \bar\psi_{\dot\alpha}^b 
\{\pi^{\dot\alpha a}_{\bar\psi} , \delta_2 v^b_j \}_{-}
 \label{B1} \\
&+& \delta_1 {\bar\psi}_{\dot\alpha}^a 
\{ \pi^{\dot\alpha a}_{\bar\psi} , \epsilon_2 \sigma_j {\bar\psi}^c \}_{-}  
 \frac{i}{2}  {\cal F}^{\dagger c d}  v^{* 0 j d} 
 \nonumber \\
&+& \delta_2 {\bar\lambda}_{\dot\alpha}^b
\{\epsilon_1 \sigma_i {\bar\lambda}^a , 
\pi^{\dot\alpha b}_{\bar\lambda}\}_{-}  
 \frac{i}{2}  {\cal F}^{\dagger a b}  v^{* 0 i b} 
\label{B2} \\
&+& \{ \Pi^{a i} , \delta_2 {\bar\lambda}_{\dot\alpha}^b \}_{-}
\delta_1 v^a_i \pi^{\dot\alpha b}_{\bar\lambda} 
\nonumber \\
&+& \{ \delta_1 {\bar\psi}_{\dot\alpha}^a , \Pi^{b j} \}_{-}
\delta_2 v^b_j  \pi^{\dot\alpha a}_{\bar\psi}  
\label{N1}\\
&+& i {\cal I}^{e f} \delta_1 {\bar\psi}_{\dot\alpha}^a 
\{\pi^{\dot\alpha a}_{\bar\psi} ,  \epsilon_2 \sigma^0 \bar\psi^f\}_{-} 
 \epsilon^{e g h} A^g A^{h \dagger}
\nonumber \\
&+& i {\cal I}^{a b} \delta_2 {\bar\lambda}_{\dot\alpha}^c 
\{ \epsilon_1 \sigma^0 \bar\lambda^b , \pi^{\dot\alpha c}_{\bar\lambda}\}_{-} 
\epsilon^{a d e} A^d A^{e \dagger}  
\label{N2} \\
&+& \{ \pi^b_A , {\cal I}^{ef} \} \delta_1 A^b i  
\epsilon_2 \sigma^0 \bar\psi^f \epsilon^{egh} A^g A^{\dagger h} 
\nonumber \\
&+& \{{\cal I}^{ab} ,  \pi^e_A \} \delta_2 A^e i  
\epsilon_1 \sigma^0 \bar\lambda^b \epsilon^{acd} A^c A^{\dagger d} 
\label {N3} \\
&+& \{ \pi^b_A , A^g \} \delta_1 A^b i {\cal I}^{ef}
\epsilon_2 \sigma^0 \bar\psi^f \epsilon^{egh} A^{\dagger h} 
\nonumber \\
&+& \{ A^c , \pi^e_A \} \delta_2 A^e i {\cal I}^{ab} 
\epsilon_1 \sigma^0 \bar\lambda^b \epsilon^{acd} A^{\dagger d} {\Big )}
\label {N4} 
\eea
The eight terms (four pairs) (\ref{A1}) - (\ref{B2}) are simply the SU(2)
{\it version} of the Abelian computation. 
On the one hand, terms (\ref{A1}) and  (\ref{A2}) give
\be
\int d^3 x \partial_i [i {\cal F}^{a b \dag}
(\epsilon_1 \sigma^0 \bar\psi^a \epsilon_2 \sigma^i \bar\lambda^b -
 \epsilon_2 \sigma^0 \bar\lambda^a \epsilon_1 \sigma^i \bar\psi^b)]
\ee
where the terms (\ref{A1}) give 
${\cal F}^{\dag ab} \partial (\bar\psi^a\bar\lambda^b)$ type of term 
and the terms (\ref{A2}) give 
$(\partial {\cal F}^{\dag ab}) \bar\psi^a\bar\lambda^b$ type of term.
Note that there is no SU(2) contribution coming from the covariant 
derivative ${\cal D} A^\dag$.

\noindent
On the other hand, terms (\ref{B1}) and  (\ref{B2}) give 
\be
\int d^3 x {\Big (} \partial_i [2 \sqrt2 \Pi^{a i} A^{\dag a} 
+ \sqrt2 v^{* 0i a} A^{\dag a}_D] 
+ 2\sqrt2 ({\cal D}_i \Pi^{a i}) A^{a \dag}
+ \sqrt2 ({\cal D}_i v^{* 0i a}) A^{\dag a}_D {\Big )} \eps_1 \eps_2
\ee
where again $B^{a i} = \frac{1}{2} \eps^{0ijk} v^a_{jk}$ and we introduced 
the SW dual of $A^{a \dag}$, $A^{a \dag}_D \equiv {\cal F}^{a \dag}$.
The Bianchi identities ${\cal D}_i v^{* 0i a} = 0$ can be applied in this
case as well, thus we expect the eight new terms (four pairs) 
(\ref{N1})-(\ref{N4}) to contribute to the Gauss law only. Let us look at 
them one by one.

\noindent
The terms (\ref{N1}) give
\be
\int d^3 x  \; i \sqrt2 {\cal I}^{a e} \epsilon^{a c d} A^{\dagger d}
(\epsilon_1 \sigma^i \bar\sigma^0 \psi^e \epsilon_2 \sigma_i \bar\psi^c + 
\epsilon_2 \sigma^i \bar\sigma^0 \lambda^e \epsilon_1 \sigma_i \bar\lambda^c)
\ee
the terms (\ref{N2}) give
\bea
&& - \int d^3 x \;
 \sqrt2 {\cal I}^{a b} ({\cal D}_\mu A^{\dagger b}) \epsilon^{a d e}
A^d A^{\dagger e} (\epsilon_1 \sigma^0 \bar\sigma^\mu \epsilon_2 +
\epsilon_1 \sigma^\mu \bar\sigma^0 \epsilon_2) \nonumber \\
&=& - \int d^3 x  2 \sqrt2  \pi^a_A
\epsilon^{a d e} A^d A^{\dagger e} 
\epsilon_1 \epsilon_2 
\eea
where we used $\pi^a_A = - {\cal I}^{a b} \partial^0 A^{\dagger b}$ and 
$(\sigma^0 \bar\sigma^\mu + \sigma^\mu \bar\sigma^0)_\alpha^\beta = 
- 2 \eta^{0 \mu} \delta_\alpha^\beta$.

\noindent
The terms (\ref{N3}) give
\be
\int d^3 x \; -\frac{1}{\sqrt2} {\cal F}^{efb} \epsilon^{egh} A^g A^{\dagger h}
(\epsilon_1 \psi^b \epsilon_2 \sigma^0 \bar\psi^f +
\epsilon_2 \lambda^b \epsilon_1 \sigma^0 \bar\lambda^f)
\ee
Finally terms (\ref{N4}) give
\be
\int d^3 x \; - i \sqrt2 {\cal I}^{ef} \epsilon^{egh} A^{\dagger h}
(\epsilon_1 \psi^g \epsilon_2 \sigma^0 \bar\psi^f +
\epsilon_2 \lambda^g \epsilon_1 \sigma^0 \bar\lambda^f)
\ee
As usual we now get rid of $\eps_1$ and $\eps_2$, sum over the spinor 
indices ($ \{ \eps_1 Q_1 , \eps_2 Q_2 \}_{-} 
=-\eps_1^\alpha \eps_2^\beta \{ Q_{1 \alpha} , Q_{2 \beta} \}_{+}$
and $\eps_{\alpha \beta} \eps^{\alpha \beta}=-2$) and write the centre as 
$Z=\frac{i}{4}\epsilon^{\alpha \beta} \{ Q_{1 \alpha} , Q_{2 \beta} \}_{+}$.
Collecting all the contributions we obtain
\bea
Z &=& \int d^3 x {\Big (} \partial_i  
[i \sqrt2 (\Pi^{a i} A^{\dag a} + B^{a i} A^{\dag a}_D)
- {\cal F}^{\dag ab} \bar\psi^a \bar\sigma^{i 0} \bar\lambda^b] 
\nonumber \\
&+& i \sqrt2 ({\cal D}_i \Pi^{ai}) A^{\dagger a} \nonumber \\
&-& \sqrt2 {\cal I}^{be} \epsilon^{bcd} A^{\dagger d} 
(\psi^e \sigma^0 \bar\psi^c + \lambda^e \sigma^0 \bar\lambda^c) \nonumber \\
&-&\frac{1}{2\sqrt2}({\cal I}^{be} \epsilon^{bcd} 
+ {\cal I}^{bc} \epsilon^{bed})
A^{\dagger d} (\psi^e \sigma^0 \bar\psi^c + \lambda^e \sigma^0 \bar\lambda^c)
\nonumber \\
&-& i \sqrt2  \epsilon^{abc} A^b A^{\dagger c} \pi^a_A \nonumber \\
&+& \frac{i}{4 \sqrt2} {\cal F}^{abc} \epsilon^{ade} A^d A^{\dagger e}
(\psi^c \sigma^0 \bar\psi^b +
\lambda^c \sigma^0 \bar\lambda^b) {\Big )} \label{app}
\eea
As shown in Appendix \ref{su2comp} we can recast the terms in the last line
into ${\cal F}^{ab}$ type of terms and combine them with the similar terms 
(remember that $2i {\cal I}^{ab} = ({\cal F}^{ab} - {\cal F}^{\dag ab})$).
The result of this recombination is given by 
\bea
Z &=& \int d^3 x {\Big (} \partial_i  
[i \sqrt2 (\Pi^{a i} A^{\dag a} + B^{a i} A^{\dag a}_D)
- {\cal F}^{\dag ab} \bar\psi^a \bar\sigma^{i 0} \bar\lambda^b] \nonumber \\
&+& i\sqrt2 [(D_i \Pi^{ai}) A^{\dagger a}  
+ i {\cal I}^{be} \epsilon^{bcd} A^{\dagger d} 
(\psi^e \sigma^0 \bar\psi^c + \lambda^e \sigma^0 \bar\lambda^c) \nonumber \\
&& -  \epsilon^{abc} A^b A^{\dagger c} \pi^a_A]  {\Big )}
\eea
We see from here that the terms which are not a total divergence, given 
in the second and third lines above, simply cancel due to the Gauss law 
(\ref{gauss}) obtained in the previous Section
\be
D_i \Pi^{ai} = - \epsilon^{abc} {\cal I}^{bd}
(A^c \partial^0 A^{\dagger d} + A^{\dagger c} \partial^0 A^d 
+ i (\psi^d \sigma^0 \bar\psi^c + \lambda^d \sigma^0 \bar\lambda^c))
\nonumber 
\ee
Eventually we are left with the surface terms that vanish when the SU(2) 
gauge symmetry is not broken down to U(1). If we break the symmetry along a 
flat direction of the Higgs potential, say $a=3$, we recover the same result
we found in the U(1) sector. In other words we see that on the sphere at
infinity
\bea
Z &=& \int d^2 \vec{\Sigma} \cdot 
[i \sqrt2 (\vec{\Pi^a} A^{\dag a} + \frac{1}{4\pi} \vec{B}^a A^{\dag a}_D) 
- \frac{1}{4\pi} {\cal F}^{\dag ab} \bpsi^a\vec{\bsigma}\blambda^b)] 
\nonumber \\
&\to& i \sqrt2 \int d^2 \vec{\Sigma} \cdot 
(\vec{\Pi^3} A^{\dagger 3} + \frac{1}{4\pi} \vec{B}^3 A^{\dagger 3}_D)
\eea
where $\vec{\bsigma} \equiv (\bsigma^{01}, \bsigma^{02}, \bsigma^{03})$ and 
we reintroduced the factor $4\pi$. We made the usual assumption that the 
bosonic massive fields in the SU(2)/U(1) sector ($a=1,2$) and all the 
fermionic fields fall off faster than $r^3$, whereas the scalar massless 
field ($a=3$) and its dual tend to their Higgs v.e.v.'s $a^*$ and $a^*_D$, 
respectively.

\noindent
We conclude that the fields in the massive sector, have no effect on the mass 
formula.

\appendix

\chapter{Proof of Noether Theorem}\label{mercaldo}

\noindent
The following proof is based on Ref.s \cite{lop} and \cite{merc}. 

\noindent
Let us consider the Action 
\be
{\cal A}_\Omega = \int_\Omega d^4 x {\cal L} (\Phi_i , \partial \Phi_i)
\ee
where $\Omega$ is the space-time volume of integration. The infinitesimal 
transformations of the coordinates, of the fields and of the derivatives of 
the fields are given respectively by 
\bea
x_\mu &\to& x'_\mu = x_\mu + \delta x_\mu \\
\Phi_i (x) &\to& \Phi'_i (x') = \Phi_i (x) + \delta \Phi_i (x) \\
\partial_\mu \Phi_i (x) &\to& \partial'_\mu \Phi'_i (x') 
= \partial_\mu \Phi_i (x) + \delta \partial_\mu \Phi_i (x)
\eea
note that $\delta$ does not commute with the derivatives. 

\noindent
When we act with this transformation the Action changes to
\be
{\cal A}'_{\Omega'} = \int_{\Omega'} d^4 x' {\cal L} 
(\Phi'_i , \partial' \Phi'_i)
\ee
If the transformation is a symmetry we have 
${\cal A}'_{\Omega'} - {\cal A}_{\Omega} = 0$, therefore at the first order we 
obtain
\bea
0 &=& {\cal A}'_{\Omega'} - {\cal A}_{\Omega} \nonumber \\
&=& \int_\Omega d^4 x {\Big [} (1 + \partial_\rho \delta x^\rho)
{\Big (} {\cal L} (\Phi_i , \partial \Phi_i) 
+ \frac{\partial {\cal L}}{\partial \Phi_i} \delta \Phi_i 
+ \frac{\partial {\cal L}}{\partial (\partial_\mu \Phi_i)} 
\delta \partial_\mu \Phi_i{\Big )} - {\cal L}{\Big ]} \nonumber \\
&=& \int_\Omega d^4 x 
{\Big (}\frac{\partial {\cal L}}{\partial \Phi_i} \delta \Phi_i 
+ \frac{\partial {\cal L}}{\partial (\partial_\mu \Phi_i)} 
\delta \partial_\mu \Phi_i + {\cal L} \partial_\rho x^\rho {\Big )} \label{AA}
\eea
where $(1 + \partial_\rho \delta x^\rho)$ is the 
Jacobian of the  change of coordinates from $x'$ to $x$ at the first order.

\noindent
Let us now introduce another variation $\delta^*$ that commutes
with the derivatives. If we do so we can write 
\be
\delta \Phi_i = \partial_\mu \Phi_i (x) \delta x^\mu 
+ \delta^* \Phi_i \quad{\rm and}\quad 
\delta (\partial_\mu \Phi_i) = \partial_\mu \partial_\nu \Phi_i (x) 
\delta x^\nu + \delta^* \partial_\mu \Phi_i
\ee
Substituting these back in (\ref{AA}) we obtain
\bea 
&& \int_\Omega d^4 x {\Big [} \partial_\mu 
{\Big (} \frac{\partial {\cal L}}{\partial (\partial_\mu \Phi_i)} 
\delta^* \Phi_i {\Big )} 
+ {\Big (} \frac{\partial {\cal L}}{\partial \Phi_i} \partial_\mu \Phi_i 
+ \frac{\partial {\cal L}}{\partial (\partial_\nu \Phi_i)} 
\partial_\mu \partial_\nu \Phi_i {\Big )} \delta x^\mu 
+ {\cal L} \partial_\mu \delta x^\mu {\Big ]} \nonumber \\
&=& \int_\Omega d^4 x  { \Big [} \partial_\mu 
{\Big (} \frac{\partial {\cal L}}{\partial (\partial_\mu \Phi_i)} 
\delta^* \Phi_i {\Big )} 
+ {\Big (} \frac{\partial {\cal L}}{\partial x^\mu} \delta x^\mu 
+ {\cal L} \partial_\mu \delta x^\mu{ \Big )} {\Big ]}   \\
&=& \int_\Omega d^4 x ({\rm E.L.})_i \delta^* \Phi^i \nonumber
\eea
which finally gives the wanted conservation law on-shell 
$\partial_\mu J^\mu = 0$ where
\be
J^\mu = \Pi_i^\mu \delta^* \Phi^i + {\cal L} \delta x^\mu
\ee
This leads to the identification $V^\mu = - {\cal L} \delta x^\mu$
introduced in Section \ref{1.1}. If we write back the 
space-time dependent variations $\delta \Phi^i$ we obtain
\bea 
J^\mu &=& \Pi_i^\mu \delta \Phi^i -
(\Pi_i^\mu \partial^\nu \Phi_i - \eta^{\mu \nu} {\cal L}) \delta x_\nu 
\nonumber \\
&=& \Pi_i^\mu \delta \Phi^i - T^{\mu \nu} \delta x_\nu \label{epsx}
\eea
that leads to the definition (\ref{t}) of the energy-momentum 
tensor $T^{\mu \nu}$.

\chapter{Notation and Spinor Algebra}\label{not}

\noindent
Let us say here that in Susy conventions and notations are not a trivial matter
at all. We follow the conventions of Wess and Bagger \cite{wb} with no changes.
Fortunately these conventions are becoming more and more popular and this is 
one of the reasons why we chose them. Rather than filling pages with
well known formulae we refer to the Appendices A and B in \cite{wb}.
Here we shall comment on some of those conventions and show the formulae more 
relevant for our computations.

\section{Crucial conventions}

\noindent
The spinors are Weyl two components in Van der Waerden notation.
Spinors with undotted indices transform under the representation
$(\frac{1}{2},0)$ of $SL(2,\C)$ and spinors with dotted indices
transform under the conjugate representation $(0,\frac{1}{2})$.
The relations between Dirac, Majorana and Weyl spinors are given by
\be
\Psi_{\rm Dirac} = 
\left( \begin{array}{c} \psi_\alpha \\  \blambda^{\dalpha} \end{array} \right)
\quad
\Psi_{\rm Majorana} = 
\left( \begin{array}{c} \psi_\alpha \\  \bpsi^{\dalpha} \end{array} \right)
\ee
The sigma matrices are standard Pauli matrices
\be
\sigma^0 = \left( \begin{array}{cc} -1 & 0 \\ 0 & -1 \end{array} \right)
\quad
\sigma^1 = \left( \begin{array}{cc} 0 & 1 \\ 1 & 0 \end{array} \right)
\ee
\be
\sigma^2 = \left( \begin{array}{cc} 0 & -i \\ i & 0 \end{array} \right)
\quad
\sigma^3 = \left( \begin{array}{cc} 1 & 0 \\ 0 & -1 \end{array} \right)
\ee
The relation with the gamma matrices is given by
\be
\gamma^\mu = 
\left( \begin{array}{cc} 0 & \sigma^\mu \\ \bsigma^\mu & 0 \end{array} \right)
\ee 
The metric is $\eta_{\mu \nu} = {\rm diag}(-1,1,1,1)$. 
To rise and lower the spinor indices 
we use $\eps_{\alpha \beta}$ and $\eps^{\alpha \beta}$, where 
$\eps_{2 1} = \eps^{1 2} = - \eps_{1 2} = - \eps^{2 1} = 1$. Also
$\eps_{0 1 2 3} = -1$.

\noindent
The position of the spinor indices is not negotiable
and is given once and for all by
\be 
\sigma^\mu_{\: \alpha \dot \alpha}
\quad
{\bar \sigma}^{\mu \: {\dot \alpha} \alpha}
\quad
\sigma^{\mu \nu \: \beta}_{\: \alpha}
\quad
\bsigma^{\mu \nu \: \dot \alpha}_{\: \quad \dot \beta}
\ee
where
\bea
\sigma^{\mu \nu \: \beta}_{\: \alpha} &=&
\frac{1}{4} (\sigma_{\alpha \dalpha}^\mu \bsigma^{{\dalpha} \beta \: \nu}
- \sigma_{\alpha \dalpha}^\nu \bsigma^{{\dalpha} \beta \: \mu}) 
\label{sigmn}\\
\bsigma^{\mu \nu \: \dalpha}_{\: \quad \dot \beta} &=&
\frac{1}{4} (\bsigma^{{\dalpha} \alpha \: \mu} \sigma^\nu_{\alpha \dot\beta}
- \bsigma^{{\dalpha} \alpha \: \nu} \sigma^\mu_{\alpha \dot\beta})
\label{bsigmn}
\eea
From $\sigma$ to $\bar \sigma$ and {\it vice versa}:
\bea
\sigma^\mu_{\: \alpha \dot \alpha} = 
\epsilon_{\alpha \beta} \epsilon_{\dot \alpha \dot \beta}
{\bar \sigma}^{\mu \: {\dot \beta} \beta}
&
{\bar \sigma}^{\mu \: {\dot \alpha} \alpha} =
\epsilon^{\dot \alpha \dot \beta} \epsilon^{\alpha \beta}
\sigma^\mu_{\: \beta \dot \beta} \\
\epsilon_{\alpha \beta} {\bar \sigma}_\mu^{\dot \alpha \beta} =
\epsilon^{\dot \alpha \dot \beta} \sigma_{\mu \alpha \dot \beta}
&
\epsilon_{\dot\alpha \dot\beta} {\bar \sigma}_\mu^{\dot \alpha \beta} =
\epsilon^{\alpha \beta} \sigma_{\mu \alpha \dot \beta}
\eea
To raise and lower spinor indices use A(9) in \cite{wb} always matching
the indices from left to right as follows:
\be
\psi^\alpha = \epsilon^{\alpha \beta} \psi_\beta \quad
\psi_\alpha = \epsilon_{\alpha \beta} \psi^\beta
\ee
of course
\be
\psi^\beta \epsilon_{\beta \alpha} = \epsilon_{\beta \alpha} \psi^\beta =
- \epsilon_{\alpha \beta} \psi^\beta = - \psi_\alpha
\ee
As explained in Section \ref{grass} momenta are on a different footing and 
the convention to raise and lower their indices is the opposite to the 
standard one. Namely
\be
\pi^\alpha = \epsilon^{\beta \alpha} \pi_\beta \quad
\pi_\alpha = \epsilon_{\beta \alpha} \pi^\beta
\ee
Quantities with one spinor index 
are grassmanian variables thus anti-commute:
\be
\psi_\alpha \chi_\beta = - \chi_\beta \psi_\alpha , \quad
\bar\psi_{\dot\alpha} \bar\chi_{\dot\beta} 
= - \bar\chi_{\dot\beta} \bar\psi_{\dot\alpha} , \quad
\bar\psi_{\dot\alpha} \chi_{\beta} = - \chi_{\beta} \bar\psi_{\dot\alpha}
\ee
But some care is needed due to the (subtle) convention
\be
\psi \chi \equiv \psi^\alpha \chi_\alpha = 
- \psi_\alpha \chi^\alpha
\ee
and 
\be
\bar\psi \bar\chi \equiv \bar\psi_{\dot\alpha} \bar\chi^{\dot\alpha} = 
- \bar\psi^{\dot\alpha} \bar\chi_{\dot\alpha}
\ee
that leads to 
\be
(\psi\chi)^\dag = \bar\psi \bar\chi 
\ee
with no minus sign.
Note that $\psi \chi = \chi \psi$ ($\bar\psi \bar\chi = \bar\chi \bar\psi$)
but $\pi \chi = - \chi \pi$ ($\bar\pi \bar\chi = - \bar\chi \bar\pi$) where
$\pi$ is a momentum. Explicitly writing the indices that means: 
$\pi^\alpha \chi_\alpha = \pi_\alpha \chi^\alpha$
and $\bar\pi_{\dot \alpha} \bar\chi^{\dot \alpha} = 
\bar\pi^{\dot \alpha} \bar\chi_{\dot \alpha}$.

\noindent
Quantities with two spinor indices are c-number matrices
\be
\epsilon_{\alpha \beta}, \quad \epsilon_{\dot \alpha \dot \beta}, \quad
\sigma^\mu_{\alpha \dot \beta}, \quad 
{\bar \sigma}^{\mu {\dot \alpha  \beta}}, \quad
(\sigma^{\mu \nu})_\alpha^{\quad \beta}, \quad
({\bar \sigma}^{\mu \nu})^{\dot \alpha}_{\quad \dot \beta},
\ee
For instance the (anti)commutator of $\sigma^\mu$ and 
${\bar \sigma}^\nu$ is with respect to the Minkowski indices $\mu , \nu$.

\noindent
Other formulae:
\be
\not\!\partial_{\alpha \dalpha} {\not\!\bar\partial}^{\dalpha \beta} =
- \delta_\alpha^\beta \Box \quad
{\not\!\bar\partial}^{\dot \alpha \alpha} \not\!\partial_{\alpha \dot \beta} =
- \delta^{\dot \alpha}_{\dot \beta} \Box
\ee
where
$
\not\!\partial_{\alpha \dot \alpha} \equiv 
\sigma^\mu_{\alpha \dot \alpha} \partial_\mu
$
and
$
{\not\!\bar\partial}^{\dot \alpha \alpha} \equiv 
\bar\sigma_\mu^{\dot \alpha  \alpha} \partial^\mu
$.

\noindent
Also
$
\psi \sigma^{\mu \nu} \chi = - \chi \sigma^{\mu \nu} \psi
$,
$
(\psi \sigma^{\mu \nu} \chi)^{\dagger} = 
- \bar\chi \bar\sigma^{\mu \nu} \bar\psi
$,
$
(\psi \not\!\partial \bar \psi)^\dagger =
- {\bar \psi} {\not\!\bar\partial} \psi
$.

\section{Useful algebra}

\noindent
Beside the Fierz identities given in (B.13) in \cite{wb} we also find 
\bea
\psi^\alpha \lambda^\beta - \psi^\beta \lambda^\alpha = 
- \epsilon^{\alpha \beta} \psi\lambda
&
\psi_\alpha \lambda_\beta - \psi_\beta \lambda_\alpha = 
\epsilon_{\alpha \beta} \psi\lambda \\
\psi_\alpha \lambda^\beta - \psi^\beta \lambda_\alpha =
- \delta^\beta_\alpha \psi\lambda
&
\psi^\alpha \lambda_\beta - \psi_\beta \lambda^\alpha = 
\delta_\beta^\alpha \psi\lambda \\
\bar\psi^{\dot\alpha} \bar\lambda^{\dot\beta} - 
\bar\psi^{\dot\beta} \bar\lambda^{\dot\alpha} =
\epsilon^{\dot\alpha \dot\beta} \bar\psi \bar\lambda
&
\bar\psi_{\dot\alpha} \bar\lambda_{\dot\beta} - 
\bar\psi_{\dot\beta} \bar\lambda_{\dot\alpha} = 
- \epsilon_{\dot\alpha \dot\beta} \bar\psi \bar\lambda \\
\bar\psi_{\dot\alpha} \bar\lambda^{\dot\beta} - 
\bar\psi^{\dot\beta} \bar\lambda_{\dot\alpha} = 
\delta^{\dot\beta}_{\dot\alpha} \bar\psi \bar\lambda
&
\bar\psi^{\dot\alpha} \bar\lambda_{\dot\beta} - 
\bar\psi_{\dot\beta} \bar\lambda^{\dot\alpha} = 
- \delta_{\dot\beta}^{\dot\alpha} \bar\psi \bar\lambda
\eea
Using the definitions and the properties given in (A.11), (A.14) and (A.15)
in \cite{wb} we find :
\bea
\sigma^{\mu \nu} \sigma^{\lambda} &=& \frac{1}{2}
(-\eta^{\lambda \nu} \sigma^{\mu} + \eta^{\lambda \mu} \sigma^{\nu}
+ i \epsilon^{\lambda \mu \nu \kappa} \sigma_{\kappa}) \\
\sigma^{\mu} \bar\sigma^{\nu \lambda} &=& \frac{1}{2}
(\eta^{\mu \lambda} \sigma^{\nu} - \eta^{\mu \nu} \sigma^{\lambda}
+ i \epsilon^{\mu \nu \lambda \kappa} \sigma_{\kappa}) \\
\bar\sigma^{\mu \nu} \bar\sigma^{\lambda} &=& \frac{1}{2}
(-\eta^{\lambda \nu} \bar\sigma^{\mu} + \eta^{\lambda \mu} \bar\sigma^{\nu}
- i \epsilon^{\lambda \mu \nu \kappa} \bar\sigma_{\kappa}) \\
\bar\sigma^{\mu} \sigma^{\nu \lambda} &=& \frac{1}{2}
(\eta^{\mu \lambda} \bar\sigma^{\nu} - \eta^{\mu \nu} \bar\sigma^{\lambda}
- i \epsilon^{\mu \nu \lambda \kappa} \bar\sigma_{\kappa})
\eea
which imply
\be
\sigma^{\mu \nu} \sigma_{\nu} = 
\sigma_\nu \bar\sigma^{\nu \mu} =
- \frac{3}{2} \sigma^\mu
\quad \quad
\bar\sigma^{\mu \nu} \bar\sigma_{\nu} = 
\bar\sigma_\nu \sigma^{\nu \mu} = 
- \frac{3}{2} \bar\sigma^\mu
\ee
Very useful is the version of the previous identities with free spinor indices
\bea
\sigma^{\mu \nu \: \beta}_{\: \alpha} \sigma_{\nu \: \gamma \dot\gamma} &=&
\frac{1}{2} (\sigma^\mu_{\: \delta \dot\gamma}
\epsilon_{\gamma \alpha} \epsilon^{\beta \delta}  -
\sigma^\mu_{\: \alpha \dot\gamma} \delta^\beta_\gamma) \label{sigsig1} \\
\sigma^{\mu \nu \: \beta}_{\: \alpha} \bar\sigma^{\dot\alpha \gamma}_{\nu} &=& 
\frac{1}{2} (\bar\sigma^{\mu \: \dot\alpha \delta} 
\epsilon_{\alpha \delta} \epsilon^{\beta \gamma} +
\bar\sigma^{\mu \: \dot\alpha \beta} \delta_\alpha^\gamma ) \label{sigsig2} \\
\bar\sigma^{\mu \nu \: \dot\alpha}_{\: \quad \dot\beta}
\bar\sigma^{\dot\gamma \gamma}_{\nu} &=& 
\frac{1}{2} (\bar\sigma^{\mu \: \dot\delta \gamma}
\epsilon^{\dot\alpha \dot\gamma}\epsilon_{\dot\delta \dot\beta} -
\bar\sigma^{\mu \: \dot\alpha \gamma} 
\delta^{\dot\gamma}_{\dot\beta}) \label{sigsig3} \\
\bar\sigma^{\mu \nu \: \dot\alpha}_{\: \quad \dot\beta}
\sigma_{\nu \: \alpha \dot\gamma} &=& \frac{1}{2}
(\sigma^\mu_{\: \alpha \dot\delta}
\epsilon^{\dot\alpha \dot\delta}\epsilon_{\dot\beta \dot\gamma} +
\sigma^\mu_{\: \alpha \dot\beta} \delta^{\dot\alpha}_{\dot\gamma}) 
\label{sigsig4}
\eea
We also have 
\bea
\sigma^{\mu \nu \beta}_\alpha \bsigma_{\mu \nu \dbeta}^{\dalpha}
&=& - \delta^{\dalpha}_{\dbeta} \delta_\alpha^\beta \\
\sigma^{0 \mu \beta}_\alpha \sigma_{\gamma 0 \mu}^\delta
&=& - \frac{1}{4} (\eps_{\alpha \gamma} \eps^{\beta \delta} 
+ \delta^\delta_\alpha \delta^\beta_\gamma) \\
\bsigma^{0 \mu \dalpha}_{\dot\beta} \bsigma^{\dot\gamma}_{0 \mu \dot\delta}
&=& - \frac{1}{4} (\eps^{\dalpha \dot\gamma} \eps_{\dot\beta \dot\delta} 
+ \delta_{\dot\delta}^{\dot\alpha} \delta_{\dbeta}^{\dot\gamma})
\eea
Also useful are the following identities:
\be
(\sigma^{\rho \sigma} \sigma^{\mu \nu})^{\alpha}_{\beta}
v_{\rho \sigma}v_{\mu \nu} = 
- \frac{1}{2} \delta^{\alpha}_{\beta} v_{\mu \nu} {\hat v}^{\mu \nu}
\quad \quad 
(\bar\sigma^{\rho \sigma} \bar\sigma^{\mu \nu})^{\dot\alpha}_{\dot\beta}
v_{\rho \sigma}v_{\mu \nu} = 
- \frac{1}{2} \delta^{\dot\alpha}_{\dot\beta} v_{\mu \nu} 
{\hat v}^{\dagger \mu \nu}
\ee
\bea
\sigma^{\rho \sigma}\sigma^{\mu}v_{\rho \sigma} &=& 
{\hat v}^{\mu \nu} \sigma_\nu  \nonumber \\
\bar\sigma^{\mu} \sigma^{\rho \sigma}v_{\rho \sigma} &=& 
- {\hat v}^{\mu \nu} \bar\sigma_\nu  \\
\sigma^{\mu} \bar\sigma^{\rho \sigma} v_{\rho \sigma} &=& 
- {\hat v}^{\dagger \mu \nu} \sigma_\nu \nonumber \\
\bar\sigma^{\rho \sigma} \bar\sigma^{\mu} v_{\rho \sigma} &=& 
{\hat v}^{\dagger \mu \nu} \bar\sigma_\nu \nonumber 
\eea
where 
\be
{\hat v}^{\mu \nu} = v^{\mu \nu} + \frac{i}{2} v^{* \mu \nu}
\quad \quad
{\hat v}^{\dagger \mu \nu} = v^{\mu \nu} - \frac{i}{2} v^{* \mu \nu}
\ee
and $v^{\mu \nu} = - v^{\nu \mu}$.

\section{A typical calculation}

\noindent
We present here an example of a typical calculation encountered during the 
lengthy computations we dealt with. 

\noindent
Often we have to reduce expressions of the form
\be
\lambda\sigma^{0 i}\psi \: \chi\sigma_{0 i}\varphi
\ee
In order to do that first we have to write in the spinor indices, then 
extract the matrices being careful about the position of the spinors involved.
Thus the expression above becomes
\be\label{lspcsf}
\lambda^\alpha \psi_\beta \chi^\gamma \varphi_\delta \: \:  
\sigma^{0 i \beta}_\alpha \sigma_{0 i \gamma}^\delta
\ee
Then we use the definition (\ref{sigmn}) to write the product of the matrices 
as 
\be
\frac{1}{4} 
(\sigma^0_{\alpha \dalpha} \bsigma^{i \dalpha \beta}
- \sigma^i_{\alpha \dalpha} \bsigma^{0 \dalpha \beta}) 
\sigma_{0 i \gamma}^\delta
\ee
using the properties (\ref{sigsig1}) and (\ref{sigsig2}) this becomes
\be
\frac{1}{8}
{\Big (} 
(\sigma^0 \bsigma_0)_\alpha^\eps \eps_{\gamma \eps} \eps^{\delta \beta}
+ (\sigma^0 \bsigma_0)_\alpha^\delta \delta_\gamma^\beta
- (\sigma_0 \bsigma^0)_\eps^\beta \eps_{\alpha \gamma} \eps^{\delta \eps}
+ (\sigma_0 \bsigma^0)_\gamma^\beta \delta_\alpha^\delta {\Big )}
\ee
using $(\sigma^0 \bsigma_0)_\alpha^\beta = - \delta_\alpha^\beta$ we obtain
\be
\frac{1}{8}
{\Big (} -\eps_{\gamma \alpha} \eps^{\delta \beta}
- \delta_\alpha^\delta \delta_\gamma^\beta
+ \eps_{\alpha \gamma} \eps^{\delta \beta}
- \delta_\gamma^\beta \delta_\alpha^\delta {\Big )} = - \frac{1}{4}
{\Big (} \eps_{\gamma \alpha} \eps^{\delta \beta}
+ \delta_\alpha^\delta \delta_\gamma^\beta {\Big )}
\ee
When we substitute this back in (\ref{lspcsf}), pay attention to the summation
conventions and commute the spinors we end up with
\be
-\frac{1}{4} (\psi\varphi \: \lambda\chi - \psi\chi \: \lambda\varphi)
\ee
In the case where $\varphi \equiv \lambda$ we can reduce the expression even
more using the Fierz identities given in (B.13) in \cite{wb}. In fact we can 
write
\be
\psi^\alpha \lambda_\alpha \: \lambda^\beta \chi_\beta =
-\frac{1}{2} \delta_\alpha^\beta \psi^\alpha \chi_\beta \: \lambda^2
= -\frac{1}{2}  \psi\chi \: \lambda^2
\ee
Thus for $\varphi \equiv \lambda$ the expression (\ref{lspcsf}) can be reduced 
to 
\be
\frac{3}{8} \psi\chi \: \lambda^2
\ee

\section{Derivation with respect to a grassmanian variable}\label{grass}

\noindent
The derivative $\frac{\delta}{\delta \psi}$ is a grassmanian variable
therefore anti commutes. From the general rule
$
\partial_\mu = \eta_{\nu \mu} \partial^\nu
$
it follows that the indices have to be raised and lowered with the 
opposite metric tensor with respect to the standard convention
\be
\frac{\delta}{\delta \psi^\alpha} = 
\epsilon_{\beta \alpha} \frac{\delta}{\delta \psi_\beta} 
\ee
This is crucial to get the signs right, for instance:
\be
\frac{\delta}{\delta \psi^\gamma} (\psi \psi) = 
\frac{\delta}{\delta \psi^\gamma} (\psi^{\alpha} \psi_{\alpha}) =
\epsilon_{\alpha \beta} 
\frac{\delta}{\delta \psi^\gamma} (\psi^{\alpha} \psi^{\beta}) =
\epsilon_{\alpha \beta} (\delta^\alpha_\gamma \psi^\beta -
\psi^\alpha \delta^\beta_\gamma) = + 2 \psi_\gamma
\ee
and
\be
\frac{\delta}{\delta \psi_\gamma} (\psi \psi) = 
\epsilon^{\beta \gamma} 
\frac{\delta}{\delta \psi^\beta} (\psi^{\alpha} \psi_{\alpha}) =
\epsilon^{\beta \gamma} (2 \psi_\beta) = - 2 \psi^\gamma
\ee
Similarly for dotted indices
\be
\frac{\delta}{\delta \bar\psi_{\dot\gamma}} (\bar\psi \bar\psi) = 
\frac{\delta}{\delta \bar\psi_{\dot\gamma}} 
(\bar\psi_{\dot\alpha} \bar\psi^{\dot\alpha}) = 
+ 2 \bar\psi^{\dot\gamma} \quad
\frac{\delta}{\delta \bar\psi^{\dot\gamma}} 
(\bar\psi_{\dot\alpha} \bar\psi^{\dot\alpha}) = 
- 2 \bar\psi_{\dot\gamma}
\ee
Form here it is clear why the momenta have to be treated with the opposite 
convention.

\chapter{Graded Poisson brackets}\label{poisson}

\noindent
We deal with c-number valued fields, i.e. non operator, in the classical 
as well as effective case. Therefore the Susy algebra has to be implemented 
via graded Poisson brackets, namely with Poisson brackets $\{ , \}_{-}$ and 
anti-brackets $\{ , \}_{+}$. We define the following equal time Poisson (anti) 
brackets\footnote{Following a nice argument given by Dirac \cite{dir}, this 
definition leads to the quantum {\it anti}-commutator for two fermions. The 
original argument relates classical Poisson brackets to the commutator 
$[B_1(x) , B_2(y)]_{-} \to i \hbar \{ B_1(x) , B_2(y)\}_{-}$, where the $B$'s
stand for bosonic variables. The generalization to fermions is naturally 
given by $ [F_1(x) , F_2(y)]_{+} \to i \hbar \{ F_1(x) , F_2(y)\}_{+}$
(we shall use the natural units $\hbar = c = 1$), where the $F$'s are 
fermionic variables.}
\bea
\{ B_1(x) , B_2(y)\}_{-}&\equiv& \int d^3 z 
 {\bigg (} \frac{\delta B_1(x)}{\delta \Phi(z)} 
          \frac{\delta B_2(y)}{\delta \Pi(z)} - 
	  \frac{\delta B_2(y)}{\delta \Phi(z)}
	  \frac{\delta B_1(x)}{\delta \Pi(z)} {\bigg )} \\
\{ B(x) , F(y)\}_{-} &\equiv& \int d^3 z 
 {\bigg (} \frac{\delta B(x)}{\delta \Phi(z)} 
          \frac{\delta F(y)}{\delta \Pi(z)} - 
	  \frac{\delta F(y)}{\delta \Phi(z)}
	  \frac{\delta B(x)}{\delta \Pi(z)} {\bigg )} \\
\{ F_1(x) , F_2(y)\}_{+} &\equiv& \int d^3 z 
 {\bigg (} \frac{\delta F_1(x)}{\delta \Phi(z)} 
          \frac{\delta F_2(y)}{\delta \Pi(z)} + 
	  \frac{\delta F_2(y)}{\delta \Phi(z)}
	  \frac{\delta F_1(x)}{\delta \Pi(z)} {\bigg )} 
\eea
where the $B$'s are bosonic and the $F$'s fermionic variables and $\Phi$
and $\Pi$ span the whole phase space.

\noindent
Form this definition it follows that the properties of the graded Poisson
brackets are the same as for standard commutators and anti-commutators
\be
\{ B_1 , B_2 \}_{-} = - \{ B_2 , B_1 \}_{-} \quad
\{ B , F \}_{-} = - \{ F , B \}_{-} \quad
\{ F_1 , F_2 \}_{+} = + \{ F_2 , F_1 \}_{+}
\ee

\noindent
Let us notice that only a formal algebraic meaning can be associated to the 
Poisson anti-bracket of two fermions, since there is no physical meaning for 
a {\it classical} fermion.

\noindent
The Susy algebra (\ref{susy1})-(\ref{susy3}) is modified by a factor $i$
due to the relation\footnote{See previous Note.} 
$[ , ]_{\pm} \to i \{ , \}_{\pm}$.

\noindent
The canonical equal-time Poisson brackets for a Lagrangian with $\phi$ and 
$\psi$ as boson and fermionic fields respectively are given by the usual 
relations
\be
\{ \phi (\vec{x} , t) \; , \; \pi_\phi (\vec{y} , t) \}_{-}
= \delta^{(3)}(\vec{x} - \vec{y}) \quad
\{ \psi^\alpha (\vec{x} , t) \; , \; \pi_{\psi \beta} (\vec{y} , t) \}_{+} 
= \delta^\alpha_\beta \delta^{(3)}(\vec{x} - \vec{y})
\ee
and
\be
\{ \phi \; , \; \phi \}_{-} = \{ \psi \; , \psi \; \}_{+} =
\{ \pi_\phi \; , \; \pi_\phi \}_{-} = \{ \pi_\psi \; , \pi_\psi \; \}_{+} 
= \{ \phi \; , \; \pi_\psi \}_{-} = \{ \psi \; , \; \pi_\phi \}_{-} = 0
\ee
The same structure survives at the effective level even if a great deal 
of care is required.

\noindent 
Note that  
\be
\{ \psi_\alpha \; , \; \pi_\psi^\beta \}_{+} = \delta_\alpha^\beta
\quad {\rm and} \quad 
\{ \psi^\alpha \; , \; \pi_{\psi \beta} \}_{+} = \delta^\alpha_\beta
\ee
are compatible iff $\pi_\alpha = - \eps_{\alpha \beta} \pi^\beta$ which 
is the convention explained in Appendix \ref{not}.
Note also that we impose 
\be
\{ \psi_\alpha \; , \; \pi_{\psi \beta} \}_{+} = \eps_{\alpha \beta}
=  \{ \pi_{\psi \alpha} \; , \;  \psi_\beta \}_{+} 
\ee
We use the graded Poisson brackets in the same spirit of derivatives, 
thus even if some of the variables involved are not dynamical we have to 
commute them. For instance 
$
\{ \epsilon \psi , \pi_\psi \}_{-} = \epsilon \{ \psi , \pi_\psi \}_{+}
- \{ \epsilon , \pi_\psi \}_{+} \psi = \epsilon \{ \psi , \pi_\psi \}_{+}
$
where $\psi$ ,  $\pi_\psi$ are dynamical and $\epsilon$ is just a grassmanian
parameter.

\noindent
Useful identities:
\bea
\{ B_1 \; , \; B_2 B_3 \}_{-} &=& \{ B_1 , B_2 \}_{-} B_3 
+ B_2 \{ B_1 , B_3 \}_{-} \\
\{ B_1  B_2 \; , \; B_3 \}_{-} &=& B_1 \{ B_2 , B_3 \}_{-}
+ \{ B_1 , B_3 \}_{-} B_2 
\eea
\bea
\{ F_1 \; , \; F_2 F_3 \}_{-} &=& \{ F_1 , F_2 \}_{+} F_3 
- F_2 \{ F_1 , F_3 \}_{+} \\
\{ F_1  F_2 \; , \; F_3 \}_{-} &=& F_1 \{ F_2 , F_3 \}_{+}
- \{ F_1 , F_3 \}_{+} F_2 
\eea
\bea
\{ F_1 F_2 \; , \; B \}_{-} &=& F_1 \{ F_2 , B \}_{-} 
+  \{ F_1 , B \}_{-} F_2 \\
\{ B \; , \; F_3 F_4 \}_{-} &=&  \{ B , F_3 \}_{-} F_4
+ F_3 \{ B , F_4 \}_{-} 
\eea
\bea
\{ F_1 F_2 \; , \; F_3 F_4 \}_{-} &=& F_1 \{ F_2 , F_3 \}_{+} F_4 
- F_1 \{ F_2 , F_4 \}_{+} F_3 \\
&& - F_2 \{ F_1 , F_3 \}_{+} F_4  
+ F_2 \{ F_1 , F_4 \}_{+} F_3 
\eea
where the $B$'s and the $F$'s are bosonic and fermionic variables respectively.

\chapter{Computation of the effective $V_\mu$} \label{compV}

\noindent
We first notice that, by varying off-shell the Lagrangian (\ref{L1}) 
(the one not integrated by parts) under the Susy transformations given in 
(\ref{trns1})-(\ref{trns4}), there is no mixing of the $\cal F$ terms with 
the ${\cal F}^\dag$ terms. As explained in Chapter 3, the structure of the 
Lagrangian is 
\bea
2 i {\cal L} &=&  - {\cal F}'' [2B+2F] + {\cal F}''' [1B 2F] 
+ {\cal F}'''' [4F] \\
&+& {{\cal F}^\dag}'' [2B+2F]^\dag - {{\cal F}^\dag}''' [1B 2F]^\dag
- {{\cal F}^\dag}'''' [4F]^\dag 
\eea
where $B$ and $F$ stand for bosonic and fermionic variables, respectively.

\noindent
For instance, if we vary the $\cal F$ terms under $\delta_1$ we have  
$(\delta_1 {\cal F}'''') [3] = 0$, whereas the other terms combine as follows
\bea
{\cal F}'''' \delta_1 [4F] \sim {\cal F}'''' (1B 3F) 
&{\rm with}&
(\delta_1 {\cal F}''') [1B 2F] \sim {\cal F}'''' (1B 3F) \nonumber \\
{\cal F}''' \delta_1 [1B 2F] \sim {\cal F}''' (2B 1F + 3F) 
&{\rm with}&
(\delta_1 {\cal F}'') [2B+2F] \sim {\cal F}''' (2B 1F + 3F) \nonumber
\eea
Finally there are terms ${\cal F}'' \delta_1 [2B + 2F]$, the naive 
generalization of the classical $V_1^\mu$. The aim is to write these 
quantities as one single total divergence and express it in terms of momenta 
and variations of the fields, that we write down again here
\be
\pi^{\mu}_A = - {\cal I} \partial^\mu A^{\dagger} \quad \quad
\pi^{\mu}_{A^\dagger} = (\pi^{\mu}_A)^\dagger 
\ee
\be
\Pi^{\mu \nu} = -\frac{1}{2i} ({\cal F}'' {\hat v}^{\mu \nu} 
- {{\cal F}^{\dagger}}'' {\hat v}^{\dagger \mu \nu}) 
+ \frac{1}{i \sqrt2} ({\cal F}''' \lambda\sigma^{\mu \nu}\psi 
- {\cal F}^{^\dagger \prime \prime \prime} 
     \bar\lambda \bar\sigma^{\mu \nu} \bar\psi) 
\ee
\bea
(\pi^\mu_{\bar \psi})_{\dot\alpha} =
  \frac{1}{2} {\cal F}^{\prime \prime}
  \psi^\alpha \sigma^\mu_{\alpha \dot\alpha} \quad
(\pi^\mu_{\psi})^{\alpha} =
  - \frac{1}{2} {\cal F}^{\dagger \prime \prime}
  \bar\psi_{\dot\alpha} \bar\sigma^{\mu \dot\alpha \alpha} \\
(\pi^\mu_{\bar \lambda})_{\dot \alpha} =
  \frac{1}{2} {\cal F}^{\prime \prime} 
  \lambda^\alpha \sigma^\mu_{\alpha \dot \alpha} \quad
(\pi^\mu_{\lambda})^{\alpha} =
  - \frac{1}{2} {\cal F}^{\dagger \prime \prime}
  \bar\lambda_{\dot\alpha} \bar\sigma^{\mu \dot\alpha \alpha} 
\eea
\bea
\delta_1 A &=& {\sqrt 2} \eps_1 \psi \nonumber \\
\delta_1 \psi^\alpha &=& {\sqrt 2}\epsilon_1^\alpha E \\
\delta_1 E &=& 0 \nonumber 
\eea
\bea
\delta_1 E^\dag &=& i {\sqrt 2} \epsilon_1 \not\!\partial \bar\psi \nonumber \\
\delta_1 {\bar \psi}_{\dot \alpha} &=& - i {\sqrt 2} \epsilon_1^\alpha 
               \not\!\partial_{\alpha \dot \alpha} A^\dagger \\
\delta_1 A^\dagger &=& 0 \nonumber 
\eea
\bea
\delta_1 \lambda^\alpha &=& 
- \epsilon^\beta_1 ( \sigma^{\mu \nu \: \alpha}_{\: \beta} v_{\mu \nu}
    - i \delta^\alpha_\beta D ) \nonumber \\
\delta_1 v^\mu &=& i \epsilon_1 \sigma^\mu {\bar \lambda} \quad
\delta_1 D = - \epsilon_1 \not\!\partial \bar\lambda \\
\delta_1 {\bar \lambda}_{\dot \alpha} &=& 0 \nonumber 
\eea
This computation is by no means easy. It is matter of
\begin{itemize}
\item{identifying similar terms and compare them}
\item{use partial integration cleverly: never throw away surface terms!}
\item{use extensively Fierz identities and spinor algebra}
\end{itemize}
We have found by direct computation $V_1^\mu$ and $\bar{V}_1^\mu$. Of course 
the first one has been the most difficult to find, since if one understands
how to proceed in the first case, the other cases become only lengthy checks.
We do not have the space here to explicitly show all the details. 
What we want to show explicitly in this Appendix, is only the simplest part 
of the computation of $V^\mu_1$, namely the contribution coming from the 
$\cal F$ terms. 

\noindent
Let us apply the scheme discussed above. First we consider the 
${\cal F}''''$ type of terms. If we find contributions from these terms we 
know that they cannot be canceled by terms coming from the rigid current 
$N_\mu$ and there is no hope to rearrange them in the form of on-shell 
dummy fields (they only contain ${\cal F}'''$ type of terms). This would then 
be a signal that by commuting the charges we could have contributions that
would spoil the SW mass formula. What we find is that the terms
\bea
{\cal F}'''' \delta_1 [4F] &=&
{\cal F}'''' \frac{1}{2}[(\delta_1\psi)\psi \lambda^2  
+ \psi^2 (\delta_1\lambda)\lambda] \nonumber \\
&=& {\cal F}'''' \frac{1}{2}[\sqrt2 \eps_1\psi \lambda^2 E  
-\eps_1 \sigma^{\mu \nu}\lambda v_{\mu \nu} \psi^2
+ i \eps_1\lambda \psi^2 D]
\eea
summed to the terms
\be
(\delta_1 {\cal F}''') [1B 2F] = 
{\cal F}'''' [\eps_1\psi \lambda\sigma^{\mu \nu}\psi v_{\mu \nu}
-\frac{1}{\sqrt2} E \eps_1\psi \lambda^2 
+i D \eps_1\psi \psi\lambda]
\ee
fortunately give zero.

\noindent
Let us then move to the next level, the ${\cal F}'''$ terms. In principle 
these terms can be present, since they appear in the expression of the 
on-shell dummy fields. We find that the terms
\bea
{\cal F}'''  [(\delta_1 1B) 2F] &=& {\cal F}'''
[\frac{1}{\sqrt 2} \lambda \sigma^{\mu \nu} \psi \delta_1 v_{\mu \nu}
-\frac{1}{2} (\delta_1 E^\dag) \psi\psi 
+\frac{i}{\sqrt 2} (\delta_1 D) \psi\lambda] \nonumber \\
&=& {\cal F}'''
[\frac{1}{\sqrt 2} \lambda \sigma^{\mu\nu} \psi 
2 i \eps_1 \sigma_\nu \partial_\mu \blambda
-\frac{i}{\sqrt2} \eps_1\not\!\partial\bpsi \psi\psi 
-\frac{i}{\sqrt 2} \eps_1\not\!\partial\blambda  \psi\lambda] \nonumber \\
&=& \frac{i}{\sqrt2} {\cal F}''' 
(2 \lambda\not\!\partial\blambda  \eps_1\psi
- \eps_1\not\!\partial\bpsi \psi\psi)
\eea
summed to the terms
\be
-(\delta_1 {\cal F}'') [2F] = -i \sqrt2 {\cal F}''' 
\lambda\not\!\partial\blambda  \eps_1\psi
-i \sqrt2 {\cal F}''' \psi\not\!\partial\bpsi \eps_1\psi)
\ee
again give zero. 

\noindent
What is left are the other ${\cal F}'''$ terms and the ${\cal F}''$ terms. 
There we find
\bea
-{\cal F}'' \delta_1 [2B+2F] &=& 
-{\cal F}'' [ (\partial_\mu A^\dagger) \partial^\mu (\delta_1 A)
+ \frac{1}{2} (\delta_1 v_{\mu \nu}) v^{\mu \nu}
+ \frac{i}{4} (\delta_1 v_{\mu \nu}) v^{* \mu \nu} \nonumber \\
&& + i (\delta_1 \psi)  \not\!\partial {\bar \psi} 
+ i \psi \not\!\partial (\delta_1 {\bar \psi}) 
+ i (\delta_1 \lambda)  \not\!\partial {\bar \lambda}
- E \delta_1 E^\dag - D \delta_1 D] \nonumber \\
&=& -{\cal F}'' [ (\partial_\mu A^\dagger) \partial^\mu (\sqrt2 \eps_1\psi)
+ (i \eps_1 \sigma_\nu (\partial_\mu \blambda)) v^{\mu \nu}
\nonumber \\
&& - \frac{1}{2} (i \eps_1 \sigma_\nu (\partial_\mu \blambda)) v^{* \mu \nu} 
+ i \sqrt2 \eps_1 \not\!\partial {\bar \psi} E 
- \sqrt2 \psi^\alpha (\not\!\partial_{\alpha \dalpha}
\not\!\partial^{\dalpha \beta} A^\dag) \eps_{1 \beta} 
\nonumber \\
&& - i \eps_1^\beta (\sigma^{\mu \nu \: \alpha}_{\: \beta} v_{\mu \nu}
- i\delta^\alpha_\beta D )
\not\!\partial_{\alpha \dalpha}{\blambda}^{\dalpha} \nonumber \\
&& - i \sqrt2 \eps_1 \not\!\partial {\bpsi}E  
- D \eps_1\not\!\partial {\blambda}] \nonumber \\
&=& -{\cal F}'' [ \sqrt2 \eps_1(\partial^\mu \psi) \partial_\mu A^\dag
+ \sqrt2 \eps_1\psi \Box A^\dag] \nonumber \\
&=& -{\cal F}'' \partial^\mu [\sqrt2 \eps_1\psi \partial_\mu A^\dag]
\label{1}
\eea
Thus we find the first non zero contribution. Let us note that this term 
would already be a total divergence if we impose the classical limit 
${\cal F}'' \to \tau$. Thus we can guess that the ${\cal F}'''$ terms 
have to combine to give the quantum piece missing in order to built up 
a total divergence when summed to the terms (\ref{1}). We find that
\bea
-\delta_1 {\cal F}'' [2B] &=& {\cal F}''' \sqrt2 \eps_1 \psi
[- \partial_\mu A^\dagger \partial^\mu A
- \frac{1}{4} v_{\mu \nu}{\hat v}^{\mu \nu} + E E^\dag + \frac{1}{2} D^2]
\nonumber
\eea
summed to
\bea
{\cal F}'''  [1B (\delta_1 2F)] &=& {\cal F}'''
[\frac{1}{\sqrt 2} (\delta_1\lambda) \sigma^{\mu \nu} \psi v_{\mu \nu}
+ \frac{1}{\sqrt 2} \lambda \sigma^{\mu \nu} (\delta_1 \psi) v_{\mu \nu} 
\nonumber \\
&& -\frac{1}{2} (E^\dag \psi\delta_1\psi + E \lambda\delta_1\lambda)
+\frac{i}{\sqrt 2} D (\delta_1 \psi) \lambda 
+\frac{i}{\sqrt 2} D \psi \delta_1 \lambda] \nonumber \\
&=& {\cal F}'''
[- \frac{1}{\sqrt 2} \eps_1 \sigma^{\mu \nu} \sigma^{\rho \sigma}\psi 
v_{\mu \nu} v_{\rho \sigma} 
+ \frac{i}{\sqrt 2} \eps_1 \sigma^{\mu \nu}\psi v_{\mu \nu} D \nonumber \\
&& + \lambda \sigma^{\mu \nu}\eps_1  v_{\mu \nu} E 
- E^\dag E \sqrt2 \eps_1\psi 
+ E \eps_1 \sigma^{\mu \nu}\lambda v_{\mu \nu})
\nonumber \\ 
&& -i \eps_1\lambda DE + i D E \eps_1\lambda 
- \frac{i}{\sqrt 2} D \eps_1\sigma^{\mu \nu} v_{\mu \nu}\psi  
- \frac{1}{\sqrt 2} D^2 \eps_1\psi ] \nonumber \\
&=& - {\cal F}''' 
[\frac{1}{\sqrt 2} \eps_1\sigma^{\mu \nu}\sigma^{\rho \sigma}
v_{\mu \nu}v_{\rho \sigma}\psi
+ E^\dag E \sqrt2 \eps_1\psi + \frac{1}{\sqrt 2} D^2 \eps_1\psi] \nonumber 
\eea
give
\be
( \partial^\mu {\cal F}'') [-\sqrt2 \eps_1\psi \partial_\mu A^\dag]
\label{2}
\ee
Collecting the two contributions (\ref{1}) and (\ref{2}) we end up with 
the wanted total divergence
\bea
&& {\cal F}'' \partial^\mu [-\sqrt2 \eps_1\psi \partial_\mu A^\dag] +
(\partial^\mu {\cal F}'') [-\sqrt2 \eps_1\psi \partial_\mu A^\dag] 
\nonumber \\
&=& \partial^\mu (-{\cal F}'' \sqrt2 \eps_1\psi \partial_\mu A^\dag) 
\nonumber \\
&=& \partial^\mu (\frac{{\cal F}''}{\cal I} \delta_1A \pi^\mu_A)
\label{3}
\eea
where the definitions of momenta and the Susy transformations were used.

\noindent
More labour is needed for the ${\cal F}^\dag$ terms. We only give the result 
of that computation here. We have 
\bea
\partial_\mu [{{\cal F}''}^\dag \sqrt2 \eps_1\psi \partial^\mu A^\dag
+ i {{\cal F}''}^\dag \eps_1 \sigma_\nu \blambda \hat{v}^{\mu \nu \dag}
+ {{\cal F}''}^\dag \eps_1 \sigma^\mu \blambda D \nonumber \\
+ {{\cal F}''}^\dag \sqrt2 \eps_1\sigma^\nu\sigma^\mu\psi \partial_\nu A^\dag
+ i \sqrt2 {{\cal F}''}^\dag \bpsi \bsigma^\mu \eps_1 E
+ \frac{i}{\sqrt2}{{\cal F}'''}^\dag \eps_1 \sigma^\mu\bpsi \lambda^2]
\nonumber
\eea
Using the definitions of the non canonical momenta and the Susy 
transformations of the fields, these terms can be recast into the following 
form
\bea
\partial_\mu [-\frac{{{\cal F}''}^\dag}{\cal I} \delta_1 A \pi^\mu_A
+ 2i \frac{{{\cal F}''}^\dag}{{\cal F}''} \delta_1 \bpsi \pi^\mu_{\bpsi}
+ 2 i \delta_1\lambda \pi_\lambda^\mu 
+ {{\cal F}^\dag}'' \eps_1\sigma_\nu\blambda v^{* \mu \nu} \nonumber \\
+ 2 i \delta_1\psi \pi_\psi^\mu
+ \frac{i}{\sqrt2}{{{\cal F}'''}^\dag} \eps_1\sigma^\mu\bpsi \lambda^2]
\label{4}
\eea
Summing up the terms (\ref{3}) and (\ref{4}) and dividing by $2i$ we obtain
the final expression
\bea
V_1^\mu &=& \delta_1 A \pi^\mu_A 
+  \frac{{{\cal F}^{\dagger}}''}{{\cal F}''} \delta_1 \bar{\psi} 
    \pi^\mu_{\bar\psi} 
+ \delta_1 \psi \pi^\mu_{\psi} 
+ \delta_1 \lambda \pi^\mu_{\lambda} \nonumber \\
&& + \frac{1}{2i} {{\cal F}^{\dagger}}'' \epsilon_1 \sigma_\nu \bar{\lambda} 
  v^{* \mu \nu}
+ \frac{1}{2\sqrt2} {{\cal F}^{\dagger}}''' \epsilon_1 \sigma^\mu \bar{\psi} 
 \bar{\lambda}^2 
\eea
As explained in Chapter 3 this form is not canonical and has to be modified 
according to the rules given there.

\chapter{Transformations from the U(1) effective charges} 
\label{Atrns}

\noindent
In this Appendix we shall complete the proof given in Chapter 3 that our U(1)
effective charges correctly generate the Susy transformations. 

\section{$\Delta_1 \lambda$ from $Q_1^{I}$}

\noindent
On the one hand
\bea
\Delta_1 \pi^I_{\bar\lambda \dot\alpha} (x) &\equiv&
\{\eps_1 Q_1  \; , \;  \pi^I_{\bar\lambda \dot\alpha} (x) \}_{-} \nonumber \\
&=& \int d^3 y (\Pi^i(y) \{ \delta_1 v_i(y) \;,\; 
\pi^I_{\bar\lambda \dot\alpha} (x) \}_{-} \nonumber \\
&& - \frac{1}{2i} {{\cal F}^\dagger}''(y) v^{* 0 i}(y) 
\{ \eps_1 \sigma_i \bar\lambda (y)
\;,\; \pi^I_{\bar\lambda \dot\alpha} (x) \}_{-} \nonumber \\
&& - \frac{1}{2\sqrt2} {{\cal F}^\dagger}'''(y) \eps_1 \sigma^0 \bar\psi (y) 
\{ {\bar\lambda}^2 (y) \;,\; \pi^I_{\bar\lambda \dot\alpha} (x) \}_{-} )
\eea
On the other hand
\be
\Delta_1 \pi^I_{\bar\lambda \dot\alpha} =
\frac{1}{\sqrt2} {\cal F}''' \eps_1 \psi \lambda^\alpha 
\sigma^0_{\alpha \dot\alpha} 
+ i {\cal I} \sigma^0_{\alpha \dot\alpha} \Delta_1 \lambda^\alpha
\ee
Equating the two expressions, writing explicitly $\Pi^i$ and collecting the 
terms according to the order of the derivative of $\cal F$ we have
\bea
i {\cal I} \sigma^0_{\alpha \dot\alpha} \Delta_1 \lambda^\alpha &=& 
-\frac{1}{2} ({\cal F}'' {\hat v}^{0 i} 
- {{\cal F}^{\dagger}}'' {\hat v}^{\dagger 0 i}) 
\eps_1^\alpha \sigma_{i \alpha \dot\alpha}
+ \frac{i}{2} {{\cal F}^\dagger}''  v^{* 0 i}
\eps_1^\alpha \sigma_{i \alpha \dot\alpha} \nonumber \\
&& + \frac{1}{\sqrt2} {\cal F}''' \lambda\sigma^{0 i}\psi
\eps_1^\alpha \sigma_{i \alpha \dot\alpha} 
- \frac{1}{\sqrt2} {\cal F}''' \eps_1 \psi \lambda^\alpha 
\sigma^0_{\alpha \dot\alpha} \nonumber \\
&& - \frac{1}{\sqrt2} {{\cal F}^\dagger}''' 
     \bar\lambda \bar\sigma^{0 i} \bar\psi
\eps_1^\alpha \sigma_{i \alpha \dot\alpha}
- \frac{1}{\sqrt2} {{\cal F}^\dagger}''' 
     \eps_1 \sigma^0 \bar\psi \bar\lambda_{\dot\alpha}
\eea
The terms are arranged such that in the first column there are terms from 
$\Pi^i$ and in the second the others. Now we notice that the terms in the 
first line should combine to give the term proportional to $v_{\mu \nu}$
and the other two lines should combine to give the term proportional 
to $D^{\rm on}$ in $\delta_1 \lambda$. 
First line:
\be \label{v}
-\frac{1}{2} ({\cal F}'' - {{\cal F}^{\dagger}}'') {\hat v}^{0 i}
\eps_1^\alpha \sigma_{i \alpha \dot\alpha} 
= -i {\cal I} (\eps_1 \sigma^{\mu \nu} \sigma^0)_{\dot\alpha} v_{\mu \nu}
\ee 
where the identity 
$
{\hat v}^{0 i} \sigma_{i \alpha \dot\alpha} =
(\sigma^{\mu \nu} \sigma^0)_{\alpha \dot\alpha} v_{\mu \nu} 
$
was used (see relative appendix).
Second line:

\noindent
The first term in the second line 
\be \nonumber
\frac{1}{\sqrt2} {\cal F}''' \lambda^\beta (\sigma^{0 i})_\beta^\gamma
\psi_\gamma \eps_1^\alpha \sigma_{i \alpha \dot\alpha} 
= \frac{1}{2\sqrt2} {\cal F}'''
(\eps_1 \psi \lambda^\alpha \sigma^0_{\alpha \dot\alpha}
- \eps_1 \lambda \psi^\alpha \sigma^0_{\alpha \dot\alpha})
\ee
where the identity
\be \nonumber
\sigma^{0 i \: \gamma}_{\: \beta} \sigma_{i \: \alpha \dot\alpha} =
\frac{1}{2} (\sigma^0_{\: \delta \dot\alpha}
\eps_{\alpha \beta} \eps^{\gamma \delta}  -
\sigma^0_{\: \beta \dot\alpha} \delta^\gamma_\alpha) 
\ee
was used (see Appendix \ref{not}). Combined with the second term in the same 
line we have
\be \label{D1}
- \frac{1}{2\sqrt2} {\cal F}'''
(\eps_1 \psi \lambda^\alpha \sigma^0_{\alpha \dot\alpha}
+ \eps_1 \lambda \psi^\alpha \sigma^0_{\alpha \dot\alpha})
= \frac{1}{2\sqrt2} {\cal F}''' \psi \lambda 
\eps_1^\alpha \sigma^0_{\alpha \dot\alpha}
\ee
where we used the Fierz identity
$
\lambda_\beta \psi^\alpha = - \psi_\beta \lambda^\alpha - \delta_\beta^\alpha
\psi \lambda
$.
Third line:

\noindent
Using similar identities (see the appendix) we can write the first term
in the third line as follows
\be \nonumber
\frac{1}{2\sqrt2} {{\cal F}^\dagger}''' 
(\eps_1 \sigma^0 \bar\psi \bar\lambda_{\dot\alpha}
  - \eps_1 \sigma^0 \bar\lambda \bar\psi_{\dot\alpha})
\ee
which combined with the second term in the same line gives
\be \label{D2}
\frac{1}{2\sqrt2} {{\cal F}^\dagger}''' 
\bar\psi \bar\lambda \eps_1^\alpha \sigma^0_{\alpha \dot\alpha}
\ee
Collecting the terms in (\ref{v}), (\ref{D1}) and (\ref{D2})
we have 
\be
i {\cal I} \sigma^0_{\alpha \dot\alpha} \Delta_1 \lambda^\alpha =
-i {\cal I} (\eps_1 \sigma^{\mu \nu} \sigma^0)_{\dot\alpha} v_{\mu \nu}
+ \frac{1}{2\sqrt2} {\cal F}''' \psi \lambda 
\eps_1^\alpha \sigma^0_{\alpha \dot\alpha}
+ \frac{1}{2\sqrt2} {{\cal F}^\dagger}''' 
\bar\psi \bar\lambda \eps_1^\alpha \sigma^0_{\alpha \dot\alpha}
\ee
which eventually gives the wanted expression (\ref{dl})

\section{$\Delta_2 \psi$ from $Q_2^{I}$}

\noindent
On the one hand
\bea
\Delta_2 \pi^I_{\bar\psi \dot\alpha} (x) &\equiv&
\{\eps_2 Q_2  \; , \;  \pi^I_{\bar\psi \dot\alpha} (x) \}_{-} \nonumber \\
&=& \int d^3 y (\Pi^i(y) \{ \delta_2 v_i(y) \;,\; 
\pi^I_{\bar\psi \dot\alpha} (x) \}_{-} \nonumber \\
&& - \frac{1}{2i} {{\cal F}^\dagger}''(y) v^{* 0 i}(y) 
\{ \eps_2 \sigma_i \bar\psi (y)
\;,\; \pi^I_{\bar\psi \dot\alpha} (x) \}_{-} \nonumber \\
&& + \frac{1}{2\sqrt2} {{\cal F}^\dagger}'''(y) \eps_2 \sigma^0 
\bar\lambda (y) \{ {\bar\psi}^2 (y) \;,\; 
\pi^I_{\bar\psi \dot\alpha} (x) \}_{-} )
\eea
On the other hand
\be
\Delta_2 \pi^I_{\bar\psi \dot\alpha} =
- \frac{1}{\sqrt2} {\cal F}''' \eps_2 \lambda \psi^\alpha 
\sigma^0_{\alpha \dot\alpha} 
+ i {\cal I} \sigma^0_{\alpha \dot\alpha} \Delta_2 \psi^\alpha
\ee
Equating the two expressions, writing explicitly $\Pi^i$ and collecting the 
terms according to the order of the derivative of $\cal F$ we have
\bea
i {\cal I} \sigma^0_{\alpha \dot\alpha} \Delta_2 \psi^\alpha &=& 
- \frac{1}{2} ({\cal F}'' {\hat v}^{0 i} 
- {{\cal F}^{\dagger}}'' {\hat v}^{\dagger 0 i}) 
\eps_2^\alpha \sigma_{i \alpha \dot\alpha}
+ \frac{i}{2} {{\cal F}^\dagger}'' v^{* 0 i}
\eps_2^\alpha \sigma_{i \alpha \dot\alpha} \nonumber \\
&& - \frac{1}{\sqrt2} {\cal F}''' \psi\sigma^{0 i}\lambda
\eps_2^\alpha \sigma_{i \alpha \dot\alpha} 
+ \frac{1}{\sqrt2} {\cal F}''' \eps_2 \lambda \psi^\alpha 
\sigma^0_{\alpha \dot\alpha} \nonumber \\
&& + \frac{1}{\sqrt2} {{\cal F}^\dagger}''' 
     \bar\psi \bar\sigma^{0 i} \bar\lambda
\eps_2^\alpha \sigma_{i \alpha \dot\alpha}
+ \frac{1}{\sqrt2} {{\cal F}^\dagger}''' 
     \eps_2 \sigma^0 \bar\lambda \bar\psi_{\dot\alpha}
\eea
First line:
\be \label{v2}
-\frac{1}{2} ({\cal F}'' - {{\cal F}^{\dagger}}'') {\hat v}^{0 i}
\eps_2^\alpha \sigma_{i \alpha \dot\alpha} 
= -i {\cal I} (\eps_2 \sigma^{\mu \nu} \sigma^0)_{\dot\alpha} v_{\mu \nu}
\ee 
Second line:
\be \label{DD1}
\frac{1}{2\sqrt2} {\cal F}'''
(\eps_2 \lambda \psi^\alpha \sigma^0_{\alpha \dot\alpha}
+ \eps_2 \psi \lambda^\alpha \sigma^0_{\alpha \dot\alpha})
= - \frac{1}{2\sqrt2} {\cal F}''' \psi \lambda 
\eps_2^\alpha \sigma^0_{\alpha \dot\alpha}
\ee
Third line:
\be \label{DD2}
\frac{1}{2\sqrt2} {{\cal F}^\dagger}''' 
(\eps_2 \sigma^0 \bar\psi \bar\lambda_{\dot\alpha}
  + \eps_2 \sigma^0 \bar\lambda \bar\psi_{\dot\alpha})
= - \frac{1}{2\sqrt2} {{\cal F}^\dagger}''' 
\bar\psi \bar\lambda \eps_2^\alpha \sigma^0_{\alpha \dot\alpha}
\ee
Collecting the terms in (\ref{v2}), (\ref{DD1}) and (\ref{DD2})
we have 
\be
i {\cal I} \sigma^0_{\alpha \dot\alpha} \Delta_2 \psi^\alpha =
-i {\cal I} (\eps_2 \sigma^{\mu \nu} \sigma^0)_{\dot\alpha} v_{\mu \nu}
- \frac{1}{2\sqrt2} {\cal F}''' \psi \lambda 
\eps_2^\alpha \sigma^0_{\alpha \dot\alpha}
- \frac{1}{2\sqrt2} {{\cal F}^\dagger}''' 
\bar\psi \bar\lambda \eps_2^\alpha \sigma^0_{\alpha \dot\alpha}
\ee
or
\be 
\Delta_2 \psi^\beta =
- \eps_2^\alpha (\sigma^{\mu \nu})_\alpha^\beta v_{\mu \nu}
- i \eps_2^\beta (-\frac{1}{2\sqrt2}(f \psi \lambda 
+ f^\dagger \bar\psi \bar\lambda)) = \delta_2^{\rm on} \psi^\beta
\ee

\section{The transformations from $Q_1^{II}$}

The charge is given by
\be \label{QII}
\eps_1 Q_1^{II} = 
\int d^3 x {\Big (} \sqrt2 {\cal I} \eps_1 (\not\!\partial A^\dagger) 
  \bar\sigma^0 \psi
+ \Pi^i \delta_1 v_i
- \frac{1}{2i} {\cal F}^{'' \dagger} \epsilon_1 \sigma_i \bar{\lambda} 
  v^{* 0 i}
- \frac{1}{2\sqrt2} {\cal F}^{''' \dagger} \epsilon_1 \sigma^0 \bar{\psi} 
  \bar{\lambda}^2 {\Big )}
\ee
let us write again the momenta
\be
\pi^{II}_A = - I \partial^0 A^{\dagger} 
+ \frac{1}{2} {\cal F}''' (\psi \sigma^0 \bar\psi
+ \lambda \sigma^0 \bar\lambda) \quad
\pi^{II}_{A^\dagger} = \pi_{A^\dagger} \quad
\Pi^{II i} = \Pi^{i} 
\ee
\be
(\pi^{II}_{\bar \psi})_{\dot\alpha} = 0  \quad
(\pi^{II}_{\psi})^{\alpha} =
i{\cal I}\bar\psi_{\dot\alpha}\bar\sigma^{0 \dot\alpha \alpha} \quad
(\pi^{II}_{\bar \lambda})_{\dot \alpha} = 0 \quad
(\pi^{II}_{\lambda})^{\alpha} =
   i{\cal I}\bar\lambda_{\dot\alpha}\bar\sigma^{0 \dot\alpha \alpha}
\ee
The charge re-expressed
\bea
\eps_1 Q_1^{II} &=& 
\int d^3 x {\Big (} \sqrt2 \eps_1 \psi \pi_A^{II} 
- \frac{1}{\sqrt2} {\cal F}''' \eps_1 \psi (\psi \sigma^0 \bar\psi
+ \lambda \sigma^0 \bar\lambda)
+ \sqrt2 {\cal I} \eps_1 \sigma^i \bar\sigma^0 \psi \partial_i A^\dagger 
\nonumber \\
&+& \Pi^i \delta_1 v_i
- \frac{1}{2i} {\cal F}^{'' \dagger} \epsilon_1 \sigma_i \bar{\lambda} 
  v^{* 0 i}
- \frac{1}{2\sqrt2} {\cal F}^{''' \dagger} \epsilon_1 \sigma^0 \bar{\psi} 
  \bar{\lambda}^2 {\Big )}  \label{qII}
\eea
where the first line in (\ref{qII}) corresponds to the first term 
in (\ref{QII}).
Let us call $\Delta_1$ the transformations induced by this charge.

\noindent
{\bf Bosonic transformations.}

\noindent
When we commute the charge (\ref{qII}) with $A$ according to the 
canonical Poisson brackets we only have contribution from the first term 
therefore $\Delta_1 A = \delta_1 A$. Trivially we see that 
$\Delta_1 A^\dagger = 0 = \delta_1 A^\dagger$ and 
$\Delta_1 v_i = \delta_1 v_i$.

\noindent
{\bf Fermionic transformations.}

\noindent
The interesting part is the commutation of the fermions. Let us start 
with $\Delta_1 \psi$. First we commute the last term in (\ref{qII})
that can be written as 
\be 
- \frac{1}{2\sqrt2} {\cal F}^{''' \dagger} \epsilon_1 \sigma^0 \bar{\psi} 
\bar{\lambda}^2 
= \frac{i}{2\sqrt2} f^\dagger \bar{\lambda}^2 \epsilon_1 \pi_\psi^{II} 
\ee
and we see immediately that it is not enough to produce the expression
(\ref{Fdag}) of $E_{\rm on}$, therefore we need also the piece
introduced in the first line to write the canonical momentum for $A$.
The relevant term there is 
\be
- \frac{1}{\sqrt2} {\cal F}''' \eps_1 \psi \psi \sigma^0 \bar\psi
= - \frac{i}{2 \sqrt2} f \psi^2 \eps_1 \pi^{II}_\psi
\ee
Now it is clear that
\bea
\Delta_1 \psi_\alpha  &\equiv& 
\{ \eps_1 Q_1^{II}\;,\; \psi_\alpha  \}_{-} \nonumber \\
&=& \sqrt2 \eps_{1 \alpha}(\frac{i}{4} (f^\dagger {\bar\lambda}^2
- f \psi^2)) = \sqrt2 \eps_{1 \alpha} E_{\rm on} 
= \delta_1 \psi_\alpha 
\eea
Similarly for $\Delta_1 \lambda$ when we consider the terms in the second
line of (\ref{qII}) they are not enough to give the right expression
of $D_{\rm on}$ in (\ref{D}) and also the term 
$
- \frac{1}{\sqrt2} {\cal F}''' \eps_1 \psi \lambda \sigma^0 \bar\lambda
$
in the first line has to be considered. We do not show the explicit
computation being in any respect identical to the one we have done 
with $\eps_1 Q^I$. At this end one could use the independent Poisson
(\ref{fermi}) in both cases. 

\noindent
Let us show in some details what happens for 
the other two fermions $\bar\lambda$ and $\bar\psi$. 
The re-expressed charge (\ref{qII}) has the term 
$
- \frac{1}{\sqrt2} {\cal F}''' \eps_1 \psi \lambda \sigma^0 \bar\lambda
$, therefore we have to commute it
\be
\Delta_1 \pi_\lambda^{\alpha II} 
\equiv \{ \eps_1 Q_1^{II}\;,\;  \pi_\lambda^{\alpha II} \}_{-} 
= \frac{1}{\sqrt2} {\cal F}''' \eps_1 \psi 
(\bar\lambda \bar\sigma^0 )^\alpha
\ee
and
\be
\Delta_1 \pi_\lambda^{II \alpha} =
\frac{1}{\sqrt2} {\cal F}''' \eps_1 \psi (\bar\lambda 
\bar\sigma^0)^\alpha 
+ i {\cal I} \bar\sigma^{0 \dot\alpha \alpha} \Delta_1 
\bar\lambda_{\dot\alpha}
\ee
equating the two expressions we have the wanted 
$\Delta_1 \bar\lambda_{\dot\alpha} = 0 = \delta_1 \bar\lambda_{\dot\alpha}$.

\noindent
Finally $\Delta_1 \bar\psi$. The only contributions come from the first 
line of (\ref{qII})
\bea
\Delta_1 \pi_\psi^{\alpha II} 
&\equiv& \{ \eps_1 Q_1^{II}\;,\;  \pi_\psi^{\alpha II} \}_{-} 
\nonumber \\
&=& \sqrt2 \pi_A^{II} \eps_1^\alpha 
+ \frac{1}{\sqrt2} {\cal F}''' \eps_1 \sigma^0 \bar\psi \psi^\alpha
- \frac{1}{\sqrt2} {\cal F}''' \lambda \sigma^0 \bar\lambda \eps_1^\alpha
+ \sqrt2 {\cal I} (\eps_1 \sigma^i \bar\sigma^0)^\alpha \partial_i A^\dagger
\nonumber \\
\eea
using the Fierz identity $\psi^\alpha \eps_1^\beta = \psi^\beta \eps_1^\alpha 
- \eps^{\alpha \beta} \eps_1 \psi$ we have
\bea
\Delta_1 \pi_\psi^{\alpha II}
&=& \sqrt2 \eps_1^\alpha (\pi_A^{II} 
- \frac{1}{2} {\cal F}''' 
(\psi \sigma^0 \bar\psi + \lambda \sigma^0 \bar\lambda)) 
+ \sqrt2 {\cal I} (\eps_1 \sigma^i \bar\sigma^0)^\alpha \partial_i A^\dagger
\nonumber \\
&& + \frac{1}{\sqrt2} {\cal F}''' \eps_1 \psi (\bar\psi \bar\sigma^0 )^\alpha
\nonumber \\
&=& \sqrt2 {\cal I} 
(\eps_1 \sigma^\mu \bar\sigma^0)^\alpha \partial_\mu A^\dagger
+ \frac{1}{\sqrt2} {\cal F}''' \eps_1 \psi (\bar\psi \bar\sigma^0 )^\alpha
\eea
which combined with the usual
\be
\Delta_1 \pi_\psi^{\alpha II} 
= \frac{1}{\sqrt2} {\cal F}''' \eps_1 \psi 
(\bar\psi \bar\sigma^0 )^\alpha
+ i {\cal I} \bar\sigma^{0 \dot\alpha \alpha} \Delta_1 
\bar\psi_{\dot\alpha}
\ee
gives the wanted
\be
\Delta_1 \bar\psi_{\dot\alpha} = -i \eps_1^\alpha 
\dsl_{\alpha \dot\alpha} A^\dagger = \delta_1 \bar\psi_{\dot\alpha}
\ee
Thus we conclude that $\Delta_1 \equiv \delta_1$ also in the 
{\it ${\cal L}^{II}$-setting} therefore this is a final 
proof that the canonical procedure works even if some labour is needed.
Note that we could not get the right transformations for the spinors if 
we had used the charge in (\ref{QII}).

\section{Transformations of the dummy fields}

\noindent
We want to show here that the transformations of the dummy fields on-shell
can be obtained by the transformations of the fermions. At this end let us 
write again the Euler-Lagrange equations for $E, E^\dag$ and $D$
\bea 
\label{dd}    D & = & - \frac{1}{2 \sqrt 2}
                     ( f \psi \lambda + f^{\dagger}\bar \psi \bar \lambda) \\
\label{ff}    E^\dag &=&- \frac{i}{4} (f \lambda^2 - f^{\dagger}\bar\psi^2)  \\
\label{ffdag} E & = & \frac{i}{4} (f^{\dagger} \bar\lambda^2 - f \psi^2)
\eea
The Euler-Lagrange equations for the fermions, obtained from the Lagrangian 
(\ref{L1}), are given by
\be
{\not\!\bar\partial}^{\dot\alpha \alpha} \psi_{\alpha} = 
\frac{i}{2} f ({\not\!\bar\partial}^{\dot\alpha \alpha}  A) \psi_{\alpha}
- \frac{1}{2} f^{\dagger}
(\frac{1}{\sqrt 2} \bar\sigma^{\mu\nu\!\dot\alpha}_{\quad\dot\beta}
                    \bar\lambda^{\dot\beta} v_{\mu\nu}
+ E \bar\psi^{\dot\alpha}
+ \frac{i}{\sqrt 2} D \bar\lambda^{\dot\alpha})
+ \frac{1}{4} g^{\dagger} \bar\psi^{\dot\alpha} \bar\lambda\bar\lambda
\ee
\be
{\not\!\partial}_{\alpha \dot\alpha} \bar\psi^{\dot\alpha} = 
- \frac{i}{2} f^{\dagger} ({\not\!\partial}_{\alpha \dot\alpha} 
   A^{\dagger}) \bar\psi^{\dot\alpha}
+ \frac{1}{2} f
  (\frac{1}{\sqrt 2} \sigma^{\mu\nu \! \beta}_{\! \alpha}
                    \lambda_{\beta} v_{\mu\nu}
+ E^\dag \psi_{\alpha}
- \frac{i}{\sqrt 2} D \lambda_{\alpha})
- \frac{1}{4} g \psi_{\alpha} \lambda\lambda
\ee
\be
{\not\!\bar\partial}^{\dot\alpha \alpha} \lambda_{\alpha} = 
\frac{i}{2} f ({\not\!\bar\partial}^{\dot\alpha \alpha}  A) \lambda_{\alpha}
- \frac{1}{2} f^{\dagger}
(- \frac{1}{\sqrt 2} \bar\sigma^{\mu\nu\!\dot\alpha}_{\quad\dot\beta}
                    \bar\psi^{\dot\beta} v_{\mu\nu}
+ E^\dag \bar\lambda^{\dot\alpha}
+ \frac{i}{\sqrt 2} D \bar\psi^{\dot\alpha})
+ \frac{1}{4} g^{\dagger} \bar\lambda^{\dot\alpha} \bar\psi\bar\psi
\ee
\be
{\not\!\partial}_{\alpha \dot\alpha} \bar\lambda^{\dot\alpha} = 
- \frac{i}{2} f^{\dagger} ({\not\!\partial}_{\alpha \dot\alpha} 
   A^{\dagger}) \bar\lambda^{\dot\alpha}
+ \frac{1}{2} f
  (-\frac{1}{\sqrt 2} \sigma^{\mu\nu \! \beta}_{\! \alpha}
                    \psi_{\beta} v_{\mu\nu}
+ E \lambda_{\alpha}
- \frac{i}{\sqrt 2} D \psi_{\alpha})
- \frac{1}{4} g \lambda_{\alpha} \psi\psi
\ee
where $f(A,A^{\dagger}) \equiv {\cal F}''' / {\cal I}$ and 
$g(A,A^{\dagger}) \equiv {\cal F}'''' / {\cal I}$.
Note that after integration by parts nothing happens to (\ref{dd}), 
(\ref{ff}) and (\ref{ffdag}), whereas, of course, some of the Euler-Lagrange 
equations for the fermions become meaningless.

\noindent
After a lengthy computation we obtain
\be
\delta_1 E = 0
\ee
\bea
-\frac{i}{\sqrt2} \delta_1 E^\dag &=& 
\eps_1^\alpha {\bigg [} - f^{\dagger}
(\frac{i}{2} (\not\!\partial_{\alpha \dalpha} A^\dag) \bpsi^{\dalpha})
+ \frac{1}{2} \lambda_\alpha [(g-\frac{1}{2i}f^2)\psi\lambda 
- \frac{1}{i4\sqrt2} ff^\dag \bpsi\blambda] \nonumber \\
&+& \frac{1}{2\sqrt2}f
(\sqrt2 \psi_\alpha E^\dag + (\sigma^{\mu\nu}\lambda)_\alpha v_{\mu\nu} 
+ i \lambda_\alpha D) {\bigg ]} 
\eea
and
\bea
-2\sqrt2 \delta_1 D &=& 
\eps_1^\alpha {\bigg [} - f^{\dagger} 
(i\sqrt2 (\not\!\partial_{\alpha \dalpha} A^\dag) \blambda^{\dalpha})
+ \sqrt2 \psi_\alpha [(g-\frac{1}{2i}f^{2})\psi\lambda - 
\frac{1}{2i}f f^{\dagger}\bar\psi\bar\lambda] \nonumber \\
&+& f(\sqrt2 \lambda_\alpha E - (\sigma^{\mu\nu}\psi)_\alpha v_{\mu\nu} 
+ i \psi_\alpha D) {\bigg ]} 
\eea
where we used
\be
\delta f = \delta A (g - \frac{1}{2i} f^{2}) 
+ \delta A^{\dagger} \frac{1}{2i} f f^{\dagger} 
\ee
Comparing these expressions with the Euler-Lagrange equations 
for the fermions we have
\be
\delta_1 E = 0 \quad \delta_1 E^\dag = i \sqrt2 \eps_1\dsl\bpsi \quad
\delta_1 D = - \eps_1\dsl\blambda
\ee
in agreement with the given Susy variations.

\chapter{The SU(2) computations}\label{su2comp}

\noindent
In this Appendix we collect all the formulae and computations relevant for our
analysis of the SW SU(2) effective theory.

\section{Properties of ${\cal F}^{a_1 \cdots a_n}$}

\noindent
Some care is necessary in handling the derivatives of the prepotential
${\cal F}(A^a A^a)$, function of the SU(2) Casimir $A^a A^a$. The first 
four derivatives are given by
\bea
{\cal F}^a &=& 2 A^a {\cal F}' \label{Fder1}\\
{\cal F}^{ab} &=& 2 \delta^{ab}{\cal F}'+ 4 A^a A^b {\cal F}'' \label{Fder2}\\
{\cal F}^{abc} &=& 4 (\delta^{ab} A^c + \delta^{ac} A^b +\delta^{bc} A^a)
{\cal F}''+ 8 A^a A^b A^c {\cal F}''' \label{Fder3} \\
{\cal F}^{abcd} &=& 4 {\cal F}''
(\delta^{ab} \delta^{cd} + \delta^{ac} \delta^{bd} +\delta^{bc} \delta^{ad})
+ 8 {\cal F}'''(A^a A^b \delta^{cd} + A^a A^c \delta^{bd} 
+ A^a A^d \delta^{bc} \nonumber \\
&& + A^b A^c \delta^{ad} + A^b A^d \delta^{ac} 
+ A^c A^d \delta^{ab}) + 16 {\cal F}'''' A^a A^b  A^c A^d \label{Fder4}
\eea
similarly for ${\cal F}^\dag$.

\noindent
Form the expressions (\ref{Fder1})-(\ref{Fder4}) it is easy to obtain the 
following very useful identities: 
\bea
\epsilon^{abc} {\cal F}^{bd} A^c &=& \epsilon^{adc} {\cal F}^c \label{F1} \\ 
\epsilon^{bcd} {\cal F}^{be} A^d &=& - \epsilon^{bed} {\cal F}^{bc} 
A^d \label{F2} \\
{\cal F}^{abc} \epsilon^{cde} A^e &=& 
{\cal F}^{be} \epsilon^{ade} + {\cal F}^{ae} \epsilon^{bde} \label{F3}
\eea
similarly for ${\cal F}^\dag$.

\noindent
Properties (\ref{F1})-(\ref{F3}) are extensively used throughout the SU(2)
computations. As an important example we want to show the explicit computation
of the bosonic coefficients of the spinor terms entering the Gauss constraint 
in the expression (\ref{app}) of the central charge. These terms are given by
\be
i \sqrt2 [i {\cal I}^{be} \eps^{bcd} A^{\dag d}
+ \frac{i}{4} {\cal I}^{be} \eps^{bcd} A^{\dag d}
+ \frac{i}{4} {\cal I}^{bc} \eps^{bed} A^{\dag d}
+ \frac{1}{8} {\cal F}^{ace} \eps^{adb} A^d A^{\dag b}] 
(\psi^e \sigma^0 \bar\psi^c + \lambda^e \sigma^0 \bar\lambda^c)
\ee
By expanding the terms in square brackets we obtain:
\bea
[\frac{1}{2} {\cal F}^{be} \eps^{bcd} A^{\dag d}
&-& \frac{1}{2} {\cal F}^{\dag be} \eps^{bcd} A^{\dag d} \nonumber \\
+ \frac{1}{8} {\cal F}^{be} \eps^{bcd} A^{\dag d}
&-& \frac{1}{8} {\cal F}^{\dag be} \eps^{bcd} A^{\dag d} \nonumber \\
+ \frac{1}{8} {\cal F}^{bc} \eps^{bed} A^{\dag d}
&-& \frac{1}{8} {\cal F}^{\dag bc} \eps^{bed} A^{\dag d} \nonumber \\
- \frac{1}{8} {\cal F}^{cd} \eps^{ebd} A^{\dag b}
&-& \frac{1}{8} {\cal F}^{ed} \eps^{cbd} A^{\dag b} ] 
\eea
where the identity (\ref{F3}) was used to write the last term. By collecting 
similar terms and using the property (\ref{F2}) we end up with
\be
[\frac{1}{2} {\cal F}^{be} \eps^{bcd} A^{\dag d}
- \frac{1}{2} {\cal F}^{\dag be} \eps^{bcd} A^{\dag d}] =
i {\cal I}^{be} \eps^{bcd} A^{\dag d}
\ee
which is the correct coefficient according to the Gauss law (\ref{gauss}).

\section{Computation of the Hamiltonian}

\noindent
First we have to conveniently write $\beps_1 \bar{Q}_1$ in (\ref{bQsu21}) 
introducing $\pi_{A^\dag}$ (see the discussion at the end of Section 3.3.3 
and Appendix \ref{Atrns}). 
The Poisson brackets are then given by 
\bea
\{\eps_1 Q_1 , \bar\eps_1 \bar{Q}_1 \}_{-} &=&
\int d^3 x d^3 y 
\{ \Pi^{a i} \delta_1 v^a_i
+ \delta_1  {\bpsi}^a \pi^a_{\bpsi} 
+ \frac{i}{2}  {\cal F}^{\dagger a b} 
\eps_1 \sigma_i {\blambda}^a v^{* 0 i b} \nonumber \\
&& - \frac{1}{2\sqrt2}  {\cal F}^{\dagger a b c}
\eps_1 \sigma^0 {\bpsi}^a \blambda^b \blambda^c 
+ i  {\cal I}^{a b} \eps_1 \sigma^0 \blambda^b 
\eps^{a c d} A^c A^{d \dagger} \; , \;  \nonumber \\
&& \Pi^{e j} \bar\delta_1 v^e_j
+ \bar\delta_1 A^{e \dagger} \pi^e_{A^\dagger}
+ \frac{1}{\sqrt2} \beps_1 \bpsi^e {\cal F}^{efg \dagger}
(\bpsi^f \bsigma^0 \psi^g + \blambda^f \bsigma^0 \lambda^g) \nonumber \\
&&+ \sqrt2 {\cal I}^{e f} \beps_1 \bsigma^j \sigma^0 \bpsi^f
{\cal D}_j A^e 
+ \frac{i}{2}  {\cal F}^{e f} 
\beps_1 \bsigma_j \lambda^e v^{* 0 j f} \nonumber \\
&&+ \frac{1}{2\sqrt2}  {\cal F}^{efg}
\beps_1 \bsigma^0 \psi^e \lambda^f \lambda^g 
- \beps_1 \pi^e_{\blambda}
\eps^{efg} A^f A^{g \dagger} \}_{-} 
\eea
where $\delta_1 v_i^a = i\eps_1 \sigma_i \blambda^a$, 
$\bar\delta_1 v_j^e = i \beps_1 \bsigma_j \lambda^e$,
$\bar\delta_1 A^{e \dag} = \sqrt2 \beps_1 \bpsi^e$,
$\delta_1 A^b = \sqrt2 \eps_1 \psi^b$,
$\delta_1 \bpsi^a = -i \sqrt2 \eps_1 \not\!{\cal D} A^{a \dag}$ ,and
$\delta_1 \bar\psi^a \pi^a_{\bar\psi} = 
\delta_1 A^b \pi^b_A + 
\sqrt2 {\cal I}^{ab} \epsilon_1 \sigma^i \bar\sigma^0 \psi^b
{\cal D}_i A^{a \dagger}$.
Let us write 
\be
\{\epsilon_1 Q_1 , \bar\epsilon_1 \bar{Q}_1 \}_{-} =  \int d^3 x d^3 y 
{\Big (} {\rm I} + {\rm II} + {\rm III} + {\rm IV} + {\rm V} 
+ {\rm VI} {\Big )}
\ee
where
\bea
{\rm terms \; I} &=& -\Pi^{a i}\Pi^{e j} \{ \eps_1 \sigma_i \blambda^a \; , \;
\beps_1 \bsigma_j \lambda^e \}_{-} \nonumber \\
&& + \frac{i}{\sqrt2} {\cal F}^{efg \dag} \Pi^{ai}\beps_i\bpsi^e
\{ \eps_1 \sigma_i \blambda^a \; , \; \blambda^f \bsigma^0 \lambda^g \}_{-}
\nonumber \\
&& + i \sqrt2 {\cal I}^{ef} \eps_1 \sigma_i \blambda^a
\beps_1 \bsigma^j \sigma^0 \psi^f \{ \Pi^{ai} \; , \; {\cal D}_j A^e \; \}_{-}
\nonumber \\
&& - \frac{1}{2} {\cal F}^{ef} \eps_1 \sigma_i \blambda^a 
\beps_1 \bsigma_j \lambda^e \{ \Pi^{ai} \; , \; v^{* 0j f} \}_{-}
\nonumber \\
&&  - \frac{1}{2} {\cal F}^{ef} \{ \eps_1 \sigma_i \blambda^a \; , \;
\beps_1 \bsigma_j \lambda^e \}_{-} \Pi^{ai}  v^{* 0j f} 
\nonumber \\
&& + \frac{i}{2\sqrt2} {\cal F}^{efg} \beps_1 \bsigma^0 \psi^e \Pi^{ai}
\{ \eps_1 \sigma_i \blambda^a \; , \; \lambda^f \lambda^g \}_{-}
\nonumber \\
&& - i   \eps^{efg} A^f A^{\dag g} \Pi^{ai} 
\{ \eps_1 \sigma_i \blambda^a \; , \; \beps_1 \pi^e_{\blambda}\}_{-}
\eea
\bea
{\rm terms \; II} &=& + i \sqrt2 {\cal I}^{ab} \eps_1\sigma^\mu\bsigma^0\psi^b
\beps_1 \bsigma_j \lambda^e
\{ {\cal D}_\mu A^{\dag a} \; , \Pi^{e j} \; \}_{-}
\nonumber \\
&& + 2 {\cal I}^{ab} \eps_1 \sigma^i \bsigma^0 \psi^b \beps_1 \bpsi^e
\{ {\cal D}_i A^{\dag a} \; , \pi^e_{A^\dag} \; \}_{-}
\nonumber \\
&& - i 2 (\eps_1 \not\!{\cal D} A^{\dag a})_{\dalpha}
\{ \pi^{a \dalpha}_{\bpsi} \; , \; \beps_1 \bpsi^e \}_{-} \pi^e_{A^\dag}
\nonumber \\
&& - i (\eps_1 \not\!{\cal D} A^{\dag a})_{\dalpha}
\{ \pi^{a \dalpha}_{\bpsi} \; , \; 
\beps_1 \bpsi^e \bpsi^f \bsigma^0 \psi^g \}_{-} {\cal F}^{efg \dag}
\nonumber \\
&& - i (\eps_1 \not\!{\cal D} A^{\dag a})_{\dalpha}
\{ \pi^{a \dalpha}_{\bpsi} \; , \; \beps_1 \bpsi^e  \}_{-}
\blambda^f \bsigma^0 \lambda^g {\cal F}^{efg \dag}
\nonumber \\
&& + 2 \eps_1 \psi^a \{ \pi_A^a \; , \; {\cal I}^{ef} \}_{-}
\beps_1 \bsigma^j \sigma^0 \bpsi^f {\cal D}_j A^e
\nonumber \\
&& - i 2 (\eps_1 \not\!{\cal D} A^{\dag a})_{\dalpha}
\{ \pi^{a \dalpha}_{\bpsi} \; , \; \beps_1 \bsigma^j \sigma^0 \bpsi^f \}_{-}
{\cal I}^{ef} {\cal D}_j A^e 
\nonumber \\
&& +2 \eps_1 \psi^a {\cal I}^{ef} \beps_1 \bsigma^j \sigma^0 \bpsi^f
\{ \pi_A^a \; , \; {\cal D}_j A^e \}_{-}
\nonumber \\
&& + \frac{i}{\sqrt2} \eps_1 \psi^a 
\{ \pi_A^a \; , \; {\cal F}^{ef} \}_{-} \beps_1 \bsigma_j \lambda^e 
v^{* 0j f} \nonumber \\
&& + \frac{1}{2} \eps_1 \psi^a 
\{ \pi_A^a \; , \; {\cal F}^{efg} \}_{-}
\beps_1 \bsigma^0 \psi^e \lambda^f \lambda^g
\nonumber \\
&& - \sqrt2   \eps^{efg} \beps_1 \pi^e_{\blambda} A^{\dag g}
\{ \pi^a_A \; , \; A^f \}_{-} \eps_1 \psi^a
\eea
\bea
{\rm terms \; III} &=& -\frac{1}{2} {\cal F}^{\dag ab} \Pi^{ej} v^{* oi b}
\{ \eps_1 \sigma_i \blambda^a \; , \; \beps_1 \bsigma_j \lambda^e \}_{-}
\nonumber \\
&& -\frac{1}{2} {\cal F}^{\dag ab} \{  v^{* oi b} \; , \; \Pi^{ej} \}_{-}
\eps_1 \sigma_i \blambda^a \beps_1 \bsigma_j \lambda^e 
\nonumber \\
&& + \frac{i}{\sqrt2} \beps_1 \bpsi^e 
\{ {\cal F}^{ab \dag} \; , \; \pi^e_{A^\dag} \}_{-}
\eps_1 \sigma_i \blambda^a v^{* 0i b}
\nonumber \\
&& + \frac{i}{2 \sqrt2}  {\cal F}^{ab \dag} \beps_1 \bpsi^e 
\{  \eps_1 \sigma_i \blambda^a \; , \; \blambda^f \bsigma^0 \lambda^g \}_{-}
v^{* 0i b} {\cal F}^{efg \dag}
\nonumber \\
&& -\frac{1}{4} {\cal F}^{\dag ab} {\cal F}^{ef}
v^{* 0i b} v^{* 0j f} 
\{  \eps_1 \sigma_i \blambda^a \; , \; \beps_1 \bsigma_j \lambda^e \}_{-}
\nonumber \\
&& + \frac{i}{4\sqrt2} {\cal F}^{\dag ab} {\cal F}^{efg}
v^{* 0i b} \beps_1 \bsigma^0 \psi^e
\{  \eps_1 \sigma_i \blambda^a \; , \; \lambda^f \lambda^g \}_{-}
\nonumber \\
&& + \frac{i}{2}   {\cal F}^{\dag ab} \eps^{efg} A^f A^{\dag g} v^{* 0i b}
\beps_{1 \dalpha} 
\{ \pi_{\blambda}^{e \dalpha} \; , \; \eps_1 \sigma_i \blambda^a \}_{-}
\eea
\bea
{\rm terms \; IV} &=& -\frac{i}{2\sqrt2}{\cal F}^{\dag abc}\Pi^{ej} 
\eps_1\sigma^0 \bpsi^a
\{ \blambda^b \blambda^c \; , \; \beps_1 \bsigma_j \lambda^e \}_{-}
\nonumber \\
&& - \frac{1}{2} 
\eps_1 \sigma^0 \bpsi^a \blambda^b \blambda^c \beps_1 \bpsi^e
\{ {\cal F}^{\dag abc} \; , \;  \pi^e_{A^\dag} \}_{-}
\nonumber \\
&& -\frac{1}{4} {\cal F}^{\dag abc} {\cal F}^{\dag efg}
\beps_1 \bpsi^e \blambda^b \blambda^c 
\{ \eps_1 \sigma^0 \bpsi^a \; , \; \bpsi^f \bsigma^0 \psi^g \}_{-}
\nonumber \\
&& -\frac{1}{4} {\cal F}^{\dag abc} {\cal F}^{\dag efg}
\beps_1 \bpsi^e \eps_1 \sigma^0 \bpsi^a  
\{  \blambda^b \blambda^c \; , \;  \blambda^f \bsigma^0 \lambda^g \}_{-}
\nonumber \\
&& -\frac{i}{4\sqrt2} {\cal F}^{\dag abc} {\cal F}^{ef} v^{* 0j f}
\eps_1 \sigma^0 \bpsi^a 
\{  \blambda^b \blambda^c \; , \;  \beps_1 \bsigma_j \lambda^e \}_{-}
\nonumber \\
&& - \frac{1}{8} {\cal F}^{\dag abc} {\cal F}^{efg}
\blambda^b \blambda^c \lambda^f \lambda^g
\{ \eps_1 \sigma^0 \bpsi^a \; , \; \beps_1 \bsigma^0 \psi^e \}_{-}
\nonumber \\
&& - \frac{1}{8} {\cal F}^{\dag abc} {\cal F}^{efg}
\{ \blambda^b \blambda^c \; , \; \lambda^f \lambda^g \}_{-}
\eps_1 \sigma^0 \bpsi^a \beps_1 \bsigma^0 \psi^e 
\nonumber \\
&& - \frac{1 }{2\sqrt2} {\cal F}^{\dag abc} \eps_1 \sigma^0 \bpsi^a
\beps_{1 \dalpha} 
\{ \pi^{e \dalpha}_{\blambda} \; , \; \blambda^b \blambda^c \}_{-}
\eps^{efg} A^f A^{\dag g}
\eea
\bea
{\rm terms \; V} &=& - \eps^{acd} {\cal I}^{ab} 
\{ \eps_1 \sigma^0 \blambda^b \; , \; \beps_1 \bsigma_j \lambda^e \}_{-}
\Pi^{e j} A^c A^{\dag d}
\nonumber \\
&& + i \sqrt2   \eps^{acd} \{ {\cal I}^{ab} \; , \; \pi^e_{A^\dag} \}_{-}
\beps_1 \bpsi^e \eps_1 \sigma^0 \blambda^b A^c A^{\dag d}
\nonumber \\
&& + i \sqrt2   \eps^{acd} \eps_1 \sigma^0 \blambda^b \beps_1 \bpsi^e
A^c \{ A^{\dag d} \; , \; \pi^e_{A^\dag} \}_{-} {\cal I}^{ab}
\nonumber \\
&& + \frac{i}{\sqrt2}   \eps^{acd} {\cal I}^{ab} {\cal F}^{\dag efg}
A^c A^{\dag d} \beps_1 \bpsi^e 
\{ \eps_1 \sigma^0 \blambda^b \; , \; \blambda^f \bsigma^0 \lambda^g \}_{-}
\nonumber \\
&& - \frac{1}{2}  \eps^{acd} {\cal I}^{ab} {\cal F}^{ef}
A^c A^{\dag d} v^{* 0j f}
\{ \eps_1 \sigma^0 \blambda^b \; , \; \beps_1 \bsigma_j \lambda^e \}_{-}
\nonumber \\
&& + \frac{i }{2\sqrt2}  \eps^{acd} {\cal I}^{ab} {\cal F}^{efg}
A^c A^{\dag d} \beps_1 \bsigma^0 \psi^e
\{ \eps_1 \sigma^0 \blambda^b \; , \; \lambda^f \lambda^g \}_{-}
\nonumber \\
&& - i      \eps^{acd}\eps^{efg} {\cal I}^{ab} 
A^c A^{\dag d} A^f A^{\dag g} 
\{ \eps_1 \sigma^0 \blambda^b \; , \; \beps_1 \pi^e_{\blambda} \}_{-}
\eea
\bea
{\rm terms \; VI} &=& + i \sqrt2 \eps_1 \psi^a \Pi^{e j} 
\{ \pi_A^a \; , \; \beps_1 \bsigma_j \lambda^e \}_{-}
\nonumber \\
&& + \beps_1 \bpsi^e \eps_1 \psi^a {\cal F}^{\dag efg}
[(\bpsi^f \bsigma^0)^\alpha \{ \pi_A^a \; , \; \psi_\alpha^g \}_{-} +
(\blambda^f \bsigma^0)^\alpha \{ \pi_A^a \; , \; \lambda_\alpha^g \}_{-} ]
\nonumber \\
&& + \frac{i}{\sqrt2} \eps_1 \psi^a {\cal F}^{ef} v^{* 0j f}
(\beps_1 \bsigma_j)^\alpha \{ \pi^a_A \; , \; \lambda^e_\alpha \}_{-}
\nonumber \\
&& + \frac{1}{2} \eps_1 \psi^a {\cal F}^{efg}
[(\beps_1 \bsigma^0)^\alpha \{ \pi^a_A \; , \; \psi^e_\alpha \}_{-}
\lambda^f \lambda^g 
+ \beps_1 \bsigma^0 \psi^e \{ \pi^a_A \; , \; \lambda^f \lambda^g \}_{-} ]
\nonumber \\
\eea
where we kept explicitly the non trivial terms VI.
It is now matter to explicitly compute the Poisson brackets. 
At this end let us write the following useful formulae
\bea
\{ \pi^a_A \; , \; \psi^b_\alpha \}_{-} 
&=& -\frac{i}{2} ({\cal I}^{bc})^{-1} {\cal F}^{cad} \psi^d_\alpha \\
\{ \pi^a_{A^\dag} \; , \; \psi^b_\alpha \}_{-} 
&=& +\frac{i}{2} ({\cal I}^{bc})^{-1} {\cal F}^{\dag cad} \psi^d_\alpha
\eea
same for $\lambda$ (these are responsible for terms VI)
\be
\{ \bpsi^a_{\dalpha} \; , \; \psi^b_\alpha \}_{+}
= \{ \blambda^a_{\dalpha} \; , \; \lambda^b_\alpha \}_{+} 
= -i ({\cal I}^{ab})^{-1} \sigma^0_{\alpha \dalpha}
\ee
\bea
\{ \eps_1 \sigma_i \blambda^a \; , \; \beps_1 \bsigma_j \lambda^e \}_{-}
&=& i ({\cal I}^{ae})^{-1} \eps_1 \sigma_i \bsigma^0 \sigma_j \beps_1 \\
\{ \eps_1 \sigma_i \blambda^a \; , \; \blambda^f \bsigma^0 \lambda^g \}_{-}
&=& i ({\cal I}^{ag})^{-1} \eps_1 \sigma_i \blambda^f \\
\{ \Pi^{ai} \; , \; {\cal D}_j A^e \}_{-} 
&=&   \eps^{aeh} \delta^i_j A^h \\
\{ \Pi^{ai} (x) \; , \; v^{* 0j f} (y) \}_{-} 
&=& - 2 \eps^{0ijk} (\delta^{af}  \partial^y_k +   \eps^{afh} v^h_k(y))
\delta^{(3)} (\vec{x} - \vec{y})  \\
\{ \eps_1 \sigma_i \blambda^a \; , \; \lambda^f \lambda^g \}_{-}
&=& -i({\cal I}^{af})^{-1} \eps_1 \sigma_i \bsigma^0 \lambda^g 
+ (f \leftrightarrow g) \\
\{ \eps_1 \sigma_i \blambda^a  \; , \; \beps_1 \pi^e_{\blambda} \}_{-}
&=& \delta^{ae} \beps_1 \bsigma_i \eps_1 \\
\{ {\cal D}_\mu A^{\dag a} (x) \; , \; \pi^e_{A^\dag} (y) \}_{-}
&=& (\delta^{ae} \partial^x_i +   \eps^{ade} v_i^d (x)) \delta^{(3)} 
(\vec{x} - \vec{y}) \\
\{ \pi^{a \dalpha}_{\bpsi} \; , \; 
\beps_1 \bpsi^e \bpsi^f \bsigma^0 \psi^g \}_{-} 
&=& \beps_1^{\dalpha} \delta^{ae} \bpsi^f \bsigma^0 \psi^g 
+ \beps_1 \bpsi^e (\bsigma^0 \psi^g)^{\dalpha} \delta^{af} \\
\{ \blambda^b \blambda^c \; , \; \beps_1 \bsigma_j \lambda^e \}_{-}
&=& i({\cal I}^{ec})^{-1} \beps_1 \bsigma_j \sigma^0 \blambda^b 
+ (b \leftrightarrow c) \\
\{ \eps_1 \sigma^0 \bpsi^a \; , \; \bpsi^f \bsigma^0 \psi^g \}_{-}
&=& i ({\cal I}^{ag})^{-1} \eps_1 \sigma^0 \bpsi^f \\
\{ \blambda^b \blambda^c \; , \; \blambda^f \bsigma^0 \lambda^g \}_{-}
&=& i ({\cal I}^{gc})^{-1} \blambda^f \blambda^b +(b \leftrightarrow c) \\
\{ \blambda^b \blambda^c \; , \; \lambda^f \lambda^g \}_{-}
&=&  - i ({\cal I}^{cf})^{-1} \blambda^b \bsigma^0 \lambda^g
- i ({\cal I}^{bg})^{-1} \blambda^c \bsigma^0 \lambda^f \nonumber \\
&& + (f \leftrightarrow g) 
\eea
After commutation the terms above given become
\bea
{\rm terms \; I} &=& - i ({\cal I}^{ae})^{-1}
\Pi^{a i} \Pi^{e j} \eps_1 \sigma_i \bsigma^0 \sigma_j \beps_1  \nonumber \\
&& - \frac{1}{\sqrt2} ({\cal I}^{ag})^{-1}
{\cal F}^{efg \dag} \Pi^{ai} \beps_i\bpsi^e
\eps_1 \sigma_i \blambda^f \nonumber \\
&& - i \sqrt2   \eps^{eah} {\cal I}^{ef} A^h \eps_1 \sigma_i \blambda^a
\beps_1 \bsigma^i \sigma^0 \bpsi^f \nonumber \\
&& + \int d^3x d^3y 
[{\cal F}^{ef}(y) \eps_1\sigma_i\blambda^a(x) 
\beps_1\bsigma_j\lambda^e(y) \eps^{0ijk} \nonumber \\
&& \quad \quad \times(\delta^{af} \partial^y_k +   \eps^{afh} v^h_k(y))
\delta^{(3)} (\vec{x} - \vec{y})]  \nonumber \\
&&  - \frac{i}{2} ({\cal I}^{ae})^{-1}
{\cal F}^{ef}  \Pi^{ai}  v^{* 0j f} 
\eps_1 \sigma_i \bsigma^0 \sigma_j \beps_1 \nonumber \\
&& + \frac{1}{\sqrt2} {\cal F}^{efg} ({\cal I}^{af})^{-1}
\Pi^{ai} \beps_1 \bsigma^0 \psi^e
\eps_1 \sigma_i \bsigma^0 \lambda^g \nonumber \\
&& - i   \eps^{efg} A^f A^{\dag g} \Pi^{ei} 
\beps_1 \bsigma_i \eps_1
\eea
\bea
{\rm terms \; II} &=& + i \sqrt2   \eps^{aeh} {\cal I}^{ab} 
\eps_1 \sigma^i \bsigma^0 \psi^b
\beps_1 \bsigma_i \lambda^e A^{\dag h} \nonumber \\
&& + \int d^3x d^3y 
[2 {\cal I}^{ab}(x) \eps_1\sigma^i\bsigma^0\psi^b(x)
\beps_1\bpsi^e(y) \nonumber \\
&& \quad \quad \times (\delta^{ae} \partial^x_i +   \eps^{ade} v_i^d(x)) 
\delta^{(3)}(\vec{x} - \vec{y})] \nonumber  \\
&& - i2 \eps_1 \sigma^\mu \beps_1 \pi^a_{A^\dag} {\cal D}_\mu A^{\dag a} 
\nonumber \\
&& - i {\cal F}^{aeg \dag} ({\cal D}_\mu A^{\dag a})
[ \eps_1 \sigma^\mu \beps_1 
(\bpsi^e \bsigma^0 \psi^g + \blambda^e \bsigma^0 \lambda^g) \nonumber \\
&& \quad \quad + \eps_1\sigma^\mu\bsigma^0 \psi^g \beps_1\bpsi^e] \nonumber \\
&& + i \eps_1 \psi^a {\cal F}^{aef} 
\beps_1 \bsigma^j \sigma^0 \bpsi^f {\cal D}_j A^e \nonumber \\
&& - i 2 \eps_1 \sigma^\mu \bsigma^0 \sigma^j \beps_1 {\cal I}^{ea}
({\cal D}_\mu A^{\dag e}) ({\cal D}_j A^a) \nonumber \\
&& - \int d^3x d^3y
[2 \eps_1 \psi^a (x) {\cal I}^{ef} (y) 
\beps_1 \bsigma^j \sigma^0 \bpsi^f (y) \nonumber \\
&& \quad \quad \times (\delta^{ae} \partial^y_j +   \eps^{eda} v_j^d(y)) 
\delta^{(3)} (\vec{x} - \vec{y})] \nonumber \\
&& - \frac{i}{\sqrt2} \eps_1 \psi^a 
{\cal F}^{aef} \beps_1 \bsigma_j \lambda^e v^{* 0j f} \nonumber \\
&& - \frac{1}{2} \eps_1 \psi^a 
{\cal F}^{aefg} \beps_1 \bsigma^0 \psi^e \lambda^f \lambda^g \nonumber \\
&& + i \sqrt2   {\cal I}^{ed} \eps^{efg} A^{\dag g}
\beps_1 \bsigma^0 \lambda^d  \eps_1 \psi^f
\eea
\bea
{\rm terms \; III} &=& -\frac{i}{2} ({\cal I}^{ae})^{-1}
{\cal F}^{\dag ab} \Pi^{ej} v^{* oi b}
\eps_1 \sigma_i \bsigma^0 \sigma_j \beps_1\nonumber  \\
&& - \int d^3x d^3y
[{\cal F}^{\dag ab}(x) \eps_1\sigma_i\blambda^a(x) 
\beps_1\bsigma_j\lambda^e(y)\eps^{0jik} \nonumber \\
&& \quad \quad \times (\delta^{eb}\partial^x_k +  \eps^{ebh} v^h_k(x))
\delta^{(3)}(\vec{x} - \vec{y})] \nonumber \\
&& + \frac{i}{\sqrt2} {\cal F}^{abe \dag}
\beps_1 \bpsi^e \eps_1 \sigma_i \blambda^a v^{* 0i b} \nonumber \\
&& - \frac{1}{2 \sqrt2}  ({\cal I}^{ag})^{-1}
{\cal F}^{ab \dag} \beps_1 \bpsi^e 
\eps_1 \sigma_i \blambda^f v^{* 0i b} {\cal F}^{efg \dag} \nonumber \\
&& -\frac{i}{4}  ({\cal I}^{ae})^{-1}
{\cal F}^{\dag ab} {\cal F}^{ef} v^{* 0i b} v^{* 0j f} 
\eps_1 \sigma_i \bsigma^0 \sigma_j \beps_1 \nonumber \\
&& + \frac{1}{2\sqrt2} ({\cal I}^{af})^{-1}
{\cal F}^{\dag ab} {\cal F}^{efg} v^{* 0i b} 
\beps_1 \bsigma^0 \psi^e
\eps_1 \sigma_i \bsigma^0 \lambda^g \nonumber \\
&& - \frac{i}{2}   {\cal F}^{\dag ab} \eps^{efg} A^f A^{\dag g} v^{* 0i b}
\beps_1 \bsigma_i \eps_1
\eea
\bea
{\rm terms \; IV} &=& \frac{1}{\sqrt2} ({\cal I}^{ec})^{-1}
{\cal F}^{\dag abc} \Pi^{ej} \eps_1 \sigma^0 \bpsi^a
\beps_1 \bsigma_j \sigma^0 \blambda^b \nonumber \\
&& - \frac{1}{2} {\cal F}^{\dag abce} \beps_1 \bpsi^e
\eps_1 \sigma^0 \bpsi^a \blambda^b \blambda^c  \nonumber \\
&& -\frac{i}{4} {\cal F}^{\dag abc} {\cal F}^{\dag efg}
({\cal I}^{ag})^{-1}
\beps_1 \bpsi^e \blambda^b \blambda^c 
\eps_1 \sigma^0 \bpsi^f \nonumber \\
&& -\frac{i}{2} {\cal F}^{\dag abc} {\cal F}^{\dag efg}
({\cal I}^{gc})^{-1}
\beps_1 \bpsi^e \eps_1 \sigma^0 \bpsi^a  
\blambda^f \blambda^b \nonumber \\
&& +\frac{1}{2\sqrt2} {\cal F}^{\dag abc} {\cal F}^{ef} 
({\cal I}^{ec})^{-1} v^{* 0j f} 
\eps_1 \sigma^0 \bpsi^a 
\beps_1 \bsigma_j \sigma^0 \blambda^b \nonumber \\
&& - \frac{i}{8} {\cal F}^{\dag abc} {\cal F}^{efg}
({\cal I}^{ae})^{-1}
\blambda^b \blambda^c \lambda^f \lambda^g
\eps_1 \sigma^0 \beps_1 \nonumber \\
&& + \frac{i}{2} {\cal F}^{\dag abc} {\cal F}^{efg}
({\cal I}^{cf})^{-1}
\blambda^b \bsigma^0 \lambda^g 
\eps_1 \sigma^0 \bpsi^a \beps_1 \bsigma^0 \psi^e \nonumber \\
&& - \frac{1}{\sqrt2} {\cal F}^{\dag abc} \eps_1 \sigma^0 \bpsi^a
\beps_1 \blambda^c \eps^{bfg} A^f A^{\dag g} 
\eea
\bea
{\rm terms \; V} &=& - i \eps^{acd} {\cal I}^{ab} ({\cal I}^{be})^{-1} 
\eps_1 \sigma_j \beps_1 \Pi^{e j} A^c A^{\dag d} \nonumber \\
&& - \frac{1}{\sqrt2}   
\eps^{acd} {\cal F}^{abe \dag} \beps_1 \bpsi^e 
\eps_1 \sigma^0 \blambda^b A^c A^{\dag d} \nonumber \\
&& + i \sqrt2 \eps^{acd} \eps_1 \sigma^0 \blambda^b 
\beps_1 \bpsi^d A^c {\cal I}^{ab} \nonumber \\
&& - \frac{1}{\sqrt2}   \eps^{acd} {\cal I}^{ab} ({\cal I}^{bg})^{-1}
{\cal F}^{\dag efg} A^c A^{\dag d} \beps_1 \bpsi^e 
\eps_1 \sigma^0 \blambda^f \nonumber \\
&& - i \frac{1}{2}  \eps^{acd} {\cal I}^{ab} ({\cal I}^{be})^{-1} 
{\cal F}^{ef} A^c A^{\dag d} v^{* 0j f} \eps_1 \sigma_j \beps_1 \nonumber \\
&& + \frac{1}{\sqrt2} \eps^{acd} {\cal I}^{ab} ({\cal I}^{bf})^{-1} 
{\cal F}^{efg} A^c A^{\dag d} \beps_1 \bsigma^0 \psi^e \eps_1 \lambda^g 
\nonumber\\
&& + i \eps^{acd} \eps^{bfg} {\cal I}^{ab} 
A^c A^{\dag d} A^f A^{\dag g} \eps_1 \sigma^0 \beps_1 
\eea
\bea
{\rm terms \; VI} &=& \frac{1}{\sqrt2}  ({\cal I}^{ec})^{-1} {\cal F}^{cad}
\eps_1\psi^a \Pi^{e j} \beps_1\bsigma_j\lambda^d \nonumber \\
&& -\frac{i}{2} ({\cal I}^{gc})^{-1} {\cal F}^{cad}{\cal F}^{\dag efg}
\beps_1 \bpsi^e \eps_1 \psi^a 
[\bpsi^f \bsigma^0 \psi^d  + \blambda^f \bsigma^0 \lambda^d ] \nonumber \\
&& + \frac{1}{2\sqrt2} ({\cal I}^{ec})^{-1} {\cal F}^{cad}
{\cal F}^{ef} v^{* 0j f} \eps_1\psi^a \beps_1\bsigma_j\lambda^d \nonumber \\
&& -\frac{i}{4} ({\cal I}^{ec})^{-1} {\cal F}^{cad}
{\cal F}^{efg} \eps_1\psi^a
\beps_1\bsigma^0\psi^d \lambda^f\lambda^g \nonumber \\ 
&& -\frac{i}{2} ({\cal I}^{ec})^{-1} {\cal F}^{cad}
{\cal F}^{efg} \eps_1\psi^a  \beps_1\bsigma^0\psi^f \lambda^d\lambda^g ]
\eea
\noindent
We now get rid of $\eps_1$ and $\beps_1$ by using the property 
$\{ \eps_1 Q_1 \; , \; \beps_1 \bar{Q}_1 \}_{-}
= \eps_1^\alpha \beps_1^{\dalpha} 
\{ Q_{1 \alpha} \; , \; \bar{Q}_{1 \dalpha} \}_{+} $ and after that we take
the trace with $\bsigma^{0 \dalpha \alpha}$ ($\bsigma^0 = - \bsigma_0$). 
The non zero terms are all collected in the following. Note that we have 
still to divide by a factor $4i$ and note also that, for instance, II(6) 
stands for the sixth term in the group II and so forth.

\centerline{\bf ``Classical'' terms}

\centerline{\it Kinetic terms for the e.m. field\footnote{Classical test on 
the e.m. kinetic piece: 
${\cal I}^{-1} \Pi^2 + {\cal I}^{-1} {\cal R} \Pi v^*
+\frac{1}{4} {\cal I}^{-1} ({\cal R}^2 + {\cal I}^2) v^{* 2}
= {\cal I} (E^2 + B^2)$ 
where $v^* = 2 B$ and $\Pi = -({\cal I}E + {\cal R}B)$.}}
\bea
{\rm I(1)} &\quad& -2i ({\cal I}^{ab})^{-1} \Pi^{ai} \Pi^{bi}  \\
{\rm I(5)}+ {\rm III(1)} &\quad& -2i ({\cal I}^{ab})^{-1} {\cal R}^{bc}
\Pi^{ai} v^{* 0i c} \\
{\rm III(5)} &\quad& 
- \frac{i}{2} ({\cal I}^{ae})^{-1} {\cal F}^{\dag ab} {\cal F}^{ef}
v^{* 0i b} v^{* 0i f}
\eea
\centerline{\it Kinetic terms for the scalar fields}
\bea
{\rm II(3)}+ {\rm II(4)} &\quad& -4i ({\cal I}^{ab})^{-1}
\pi^a_A [\pi^b_{A^\dag} + \frac{1}{2} {\cal F}^{\dag bcd}
(\bpsi^c \bsigma^0 \psi^d + \blambda^c \bsigma^0 \lambda^d)] \nonumber \\
&&= -4i ({\cal I}^{ab})^{-1} \pi^a_A (\pi^b_A)^\dag \\
{\rm II(6)} &\quad& 4i {\cal I}^{ab} ({\cal D}^i A^a)({\cal D}^i A^{\dag b})
\eea

\centerline{\it Kinetic terms for the spinors}

for $\lambda$: 
\bea
{\rm I(4)} &\quad& 
2i {\cal F}^{ab} \lambda^a \sigma^i {\cal D}_i \blambda^b \\
{\rm III(2)} &\quad& 
-2i {\cal F}^{\dag ab} \blambda^a \bsigma^i {\cal D}_i \lambda^b 
\eea

for $\psi$: 
\bea
{\rm II(2)} &\quad& 
-2 {\cal I}^{ab} \psi^a \sigma^i {\cal D}_i \bpsi^b  \\
{\rm II(7)} &\quad& 
-2 {\cal I}^{ab} \bpsi^a \bsigma^i {\cal D}_i \psi^b  \\ 
{\rm II(4)} + {\rm III(5)} &\quad& 
2 \bpsi^e \bsigma^i \psi^g 
[\partial_i {\cal I}^{ab} +   \eps^{ebc} v_i^b {\cal I}^{gc} 
+   \eps^{gbc} v_i^b {\cal I}^{ec}] 
\eea
to write the last term (purely quantum) we used the properties given in the 
first Section of this Appendix. Integrating by parts we have 
\bea
- 4 {\cal I}^{ab} \lambda^a \sigma^i {\cal D}_i \blambda^b
+ 2i (\partial {\cal F}^{\dag ab}) \blambda^a \bsigma^i \lambda^b
+ \partial_i (- 2i {\cal F}^{\dag ab} \blambda^a \bsigma^i \lambda^b) \\
- 4 {\cal I}^{ab} \psi^a \sigma^i {\cal D}_i \bpsi^b
+ \partial_i ( -2 {\cal I}^{ab} \bpsi^a \bsigma^i \psi^b)
\eea

\centerline{\it Yukawa potential}

\bea
{\rm I(3)} &\quad& 
-i 3 \sqrt2   \eps^{eah} {\cal I}^{ef} A^h \bpsi^f \blambda^a \\
{\rm V(3)} &\quad& 
-i \sqrt2   \eps^{acd} {\cal I}^{ad} A^c \bpsi^d \blambda^b \\
{\rm IV(8)} + {\rm V(2)} + {\rm V(4)}  &\quad&
- \frac{3}{\sqrt2}   \eps^{bfg} {\cal F}^{\dag abc} A^f A^{\dag g}
\bpsi^a \blambda^c \\
{\rm II(1)} &\quad& 
- 3 i \sqrt2   \eps^{aeh} {\cal I}^{ab} A^{\dag h} \psi^b \lambda^e \\
{\rm II(10)} &\quad& 
- i \sqrt2   \eps^{efg} {\cal I}^{ed} A^{\dag g} \psi^f \lambda^d \\
{\rm V(6)} &\quad&
-\frac{1}{\sqrt2}   \eps^{acd} {\cal F}^{eag} A^c A^{\dag d}
\psi^e \lambda^g
\eea
By using the properties of $\eps^{abc} {\cal F}^{ade}$ listed in the previous 
Section of this Appendix we can recast these terms into
\be
-i 2 \sqrt2   \eps^{abc} {\cal I}^{ad} 
(A^c \bpsi^d \blambda^b + A^{\dag c} \psi^d \lambda^b)
\ee

\centerline{\it Higgs potential}

\be
{\rm V(7)} \quad 2i \eps^{acd} {\cal I}^{ab} \eps^{bfg}
A^c A^{\dag d} A^f A^{\dag g}
\ee

\centerline{\bf Purely quantum corrections}

\centerline{\it Terms that will contribute to 
${\cal F}^{abc} \sigma_{\mu \nu} v^{\mu \nu}$}

\centerline{ \it and to dummy fields on shell via the two fermions piece 
$\Pi_{\rm F}$}

\bea
{\rm I(2)}+ {\rm IV(1)} &\quad& 2\sqrt2 ({\cal I}^{ec})^{-1}
{\cal F}^{\dag abc} \Pi^{ei} \bpsi^a \bsigma_{i0} \blambda^b \\
{\rm I(6)}+ {\rm VI(1)} &\quad& - 2\sqrt2 ({\cal I}^{af})^{-1}
{\cal F}^{feg} \Pi^{ai} \psi^e \sigma_{i0} \lambda^g \\
{\rm III(4)}+ {\rm IV(5)} &\quad& \sqrt2 ({\cal I}^{ag})^{-1} {\cal R}^{ab}
{\cal F}^{\dag efg} v^{* 0i b} \bpsi^e \bsigma_{i0} \blambda^f \\
{\rm III(6)}+ {\rm VI(3)} &\quad& - \sqrt2 ({\cal I}^{af})^{-1} {\cal R}^{ab}
{\cal F}^{efg} v^{* 0i b} \psi^e \sigma_{i0} \lambda^g \\
{\rm III(3)}&\quad& 
i \sqrt2 {\cal F}^{\dag abc} v^{* 0i a} \bpsi^b \bsigma_{i0} \blambda^c \\
{\rm II(8)}&\quad& 
i \sqrt2 {\cal F}^{abc} v^{* 0i a} \psi^b \sigma_{i0} \lambda^c 
\eea
Summing them up we obtain\footnote{
$\Pi + \frac{1}{2}{\cal F}'' v^{*} = -\frac{1}{2i}{\cal I}\hat{v}^\dag
+ \Pi_{\rm F}$ and 
$\Pi + \frac{1}{2}{{\cal F}^\dag}'' v^{*} = -\frac{1}{2i}{\cal I}\hat{v}
+ \Pi_{\rm F}$. Also $\hat{v} = E + i B$ and $\hat{v}^\dag = E - i B$.}
\bea
&&-2 \sqrt2 ({\cal I}^{af})^{-1} {\cal F}^{feg} \psi^e\sigma_{i0}\lambda^g 
(\Pi^{ia}+ {\cal F}^{\dag ab} B^{ib}) \\
&& 2 \sqrt2 ({\cal I}^{ec})^{-1}{\cal F}^{\dag abc} 
\bpsi^a\bsigma_{i0}\blambda^b (\Pi^{ie}+ {\cal F}^{ed} B^{id}) 
\eea

\centerline{\it Terms that contribute to the dummy fields on-shell only}

\bea
{\rm VI(2)} &\quad& 
\frac{i}{4} {\cal F}^{\dag efg} {\cal F}^{cad} ({\cal I}^{gc})^{-1}
\bpsi^e\bpsi^f \psi^a\psi^c \\
{\rm VI(4)} &\quad& 
\frac{3i}{4} {\cal F}^{bec} {\cal F}^{efg} ({\cal I}^{ab})^{-1}
\psi^a \psi^c \lambda^f \lambda^g \\
{\rm IV(3)} + {\rm VII(4)}&\quad& 
\frac{3i}{4} {\cal F}^{\dag bec} {\cal F}^{\dag efg} ({\cal I}^{ab})^{-1}
\bpsi^a \bpsi^c \blambda^f \blambda^g \\
{\rm IV(6)} &\quad& 
+\frac{i}{4} {\cal F}^{\dag abc} {\cal F}^{efg} ({\cal I}^{ae})^{-1}
\blambda^b \blambda^c \lambda^f \lambda^g 
\eea

\centerline{\it ${\cal F}^{abcd}$-type of terms}

\be
{\rm II(9)} + {\rm IV(2)} \quad
-\frac{1}{2} ({\cal F}^{abcd} \psi^a \psi^b \lambda^c \lambda^d
- {\cal F}^{\dag abcd} \bpsi^a \bpsi^b \blambda^c \blambda^d)
\ee
Collecting all these terms and dividing by $4i$ we obtain the Hamiltonian 
given in Chapter 4.

\section{Tests on the Lagrangian}

\noindent
Let us first rewrite the four fermions terms in the SU(2) Lagrangian
\bea
{\cal L}^{\rm SU(2)}_{\rm F} &=& \frac{3}{16} ({\cal I}^{ab})^{-1} {\Big [}
{\cal F}^{acd} {\cal F}^{bef} 
(\psi^d\psi^f \lambda^c\lambda^e - \psi^d\lambda^e \lambda^c\psi^f)
- {\cal F}^{gbc} {\cal F}^{gef} \psi^a\psi^c \lambda^e\lambda^f \nonumber \\ 
&+&{\cal F}^{\dag acd} {\cal F}^{\dag bef}
(\bpsi^d\bpsi^f \blambda^c\blambda^e - \bpsi^d\blambda^e \blambda^c\bpsi^f)
- {\cal F}^{\dag gbc} {\cal F}^{\dag gef}
\bpsi^a\bpsi^c \blambda^e\blambda^f \nonumber \\
&+&{\cal F}^{\dag acd} {\cal F}^{\dag bef}
(\lambda^c\sigma^0\bpsi^f \psi^d\sigma^0\blambda^e 
- \lambda^c\sigma^0\blambda^e \psi^d\sigma^0\bpsi^f ) {\Big ]} \nonumber \\
&-&\frac{1}{16} ({\cal I}^{ab})^{-1} {\cal F}^{\dag acd} {\cal F}^{bef} 
(\bpsi^c\bpsi^d \psi^e\psi^f +  
\blambda^c\blambda^d \lambda^e\lambda^f) \nonumber \\
&+&\frac{1}{2i} 
(\frac{1}{4}{\cal F}^{abcd} \psi^a\psi^b \lambda^c\lambda^d
- \frac{1}{4}{\cal F}^{\dag abcd} \bpsi^a\bpsi^b \blambda^c\blambda^d) 
\eea
We now want to compare these terms with the U(1) correspondent ones. At this 
end we write here the four fermions contributions to the U(1) effective 
Lagrangian obtained after elimination of the dummy fields 
\bea
{\cal L}^{\rm U(1)}_{\rm F} &=& \frac{3}{32} {\cal I}^{-1} ({\cal F}''')^2
\psi^2 \lambda^2 
+ \frac{3}{32} {\cal I}^{-1} ({{\cal F}^\dag}''')^2 \bpsi^2 \blambda^2 
\nonumber \\
&-&\frac{1}{16} {\cal I}^{-1} {{\cal F}^\dag}''' {\cal F}''' 
(\bpsi^2 \psi^2 + \blambda^2 \lambda^2 + 2 \psi\lambda \bpsi\blambda) 
\nonumber \\
&+&\frac{1}{2i} (\frac{1}{4}{\cal F}'''' \psi^2 \lambda^2
- \frac{1}{4}{{\cal F}^\dag}'''' \bpsi^2 \blambda^2) 
\eea
We see immediately that the ${\cal F}''''$ terms, the 
${{\cal F}^\dag}''' {\cal F}''' \psi^2\bpsi^2$ and 
${{\cal F}^\dag}''' {\cal F}''' \lambda^2\blambda^2$ have the correct factors. 
For the $({\cal F}''')^2 \psi^2\lambda^2$ terms we have simply to notice that 
in the Abelian limit 
\be
\frac{3}{16} ({\cal I}^{ab})^{-1} {\cal F}^{acd} {\cal F}^{bef} 
(\psi^d\psi^f \lambda^c\lambda^e - \psi^d\lambda^e \lambda^c\psi^f)
- {\cal F}^{gbc} {\cal F}^{gef} \psi^a\psi^c \lambda^e\lambda^f 
\ee
reduces to 
\be
- \frac{3}{16}{\cal I}^{-1} ({\cal F}''')^2 \psi\lambda \lambda\psi
\ee
and by using the Fierz identities given in Appendix \ref{not}, 
$\psi\lambda \lambda\psi = -\frac{1}{2} \psi^2 \lambda^2$, we obtain 
the correct factor $\frac{3}{32}$.
Similarly for the $({{\cal F}^\dag}''')^2 \bpsi^2\blambda^2$ terms.
For the ${{\cal F}^\dag}''' {\cal F}''' \psi\lambda \bpsi\lambda$ terms we 
cannot recast them into a proper form. Nevertheless there is no other terms
to combine them with and we conclude that they must give the right factors.

\noindent
Note also that
\bea
{\cal L}^{\rm SU(2)}_{\rm e.m.} &=& \frac{1}{2i} {\Big [} 
- \frac{1}{4} {\cal F}^{ab} v^{a \mu \nu} \hat{v}^b_{\mu \nu}
+ \frac{1}{4} {\cal F}^{\dag ab} v^{a \mu \nu} \hat{v}^{\dag b}_{\mu \nu}
{\Big ]} \nonumber \\
&=& \frac{1}{2i} {\Big [} 
- \frac{1}{4} {\cal F}^{ab} 
(2 v^{a 0 i} v^b_{0 i} + i v^{a 0 i} v^{* b}_{0 i}
+ v^{a i j} v^b_{i j} + \frac{i}{2} v^{a i j} v^{* b}_{i j}) \nonumber \\
&& + \frac{1}{4} {\cal F}^{\dag ab} 
(2 v^{a 0 i} v^b_{0 i} - i v^{a 0 i} v^{* b}_{0 i}
+ v^{a i j} v^b_{i j} - \frac{i}{2} v^{a i j} v^{* b}_{i j})
{\Big ]} \nonumber \\
&=& - \frac{1}{2} {\cal I}^{ab} E^{a i} E^{b i}
- \frac{1}{2} {\cal R}^{ab} E^{a i} B^{b i}
+ \frac{1}{2} {\cal I}^{ab} B^{a i} B^{b i}
- \frac{1}{2} {\cal R}^{ab} E^{a i} B^{b i} \nonumber \\
\eea
The term ${\cal R} E B$ is the effective version of the CP violating 
$\theta$ term.

\end{document}